\documentclass[a4paper]
{IEEEtran}
\usepackage{cite}
\usepackage{graphicx}
\usepackage{tcolorbox}
\usepackage{psfrag}
\usepackage{caption}
\usepackage{subfigure}
\usepackage{url}
\usepackage{amsmath}
\usepackage{amsthm}
\usepackage{array}
\usepackage{amssymb}
\usepackage{amsfonts}
\usepackage{float}
\newtheorem{proposition}{Proposition}
\newtheorem*{conjecture*}{Conjecture}
\newtheorem{lemma}{Lemma}

\newtheorem{remark}{Remark}
\newtheorem{definition}{Definition}
\newtheorem{theorem}{Theorem}
\newtheorem{example}{Example}

\title{Optimal Index Codes via a Duality between Index Coding and Network Coding}
\begin{document}
\author{
\IEEEauthorblockN{Ashok Choudhary, Vamsi Krishna Gummadi, Prasad Krishnan\\}
\IEEEauthorblockA{Signal Processing and Communications Research Centre,\\
International Institute of Information Technology, Hyderabad.\\
Email: \{ashok.choudhary@research,vamsi.gummadi@research,prasad.krishnan@\}iiit.ac.in\\}
}

\vspace{-0.7cm}

\date{\today}
\maketitle

\thispagestyle{empty}	
\pagestyle{empty}

\begin{abstract}
In Index Coding, the goal is to use a broadcast channel as efficiently as possible to communicate information from a source to multiple receivers which can possess some of the information symbols at the source as side-information. In this work, we present a duality relationship between index coding (IC) and multiple-unicast network coding (NC). It is known that the IC problem can be represented using a side-information graph $G$ (with number of vertices $n$ equal to the number of source symbols). The size of the maximum acyclic induced subgraph, denoted by $MAIS$ is a lower bound on the \textit{broadcast rate}. For IC problems with $MAIS=n-1$ and $MAIS=n-2$, prior work has shown that binary (over ${\mathbb F}_2$) linear index codes achieve the $MAIS$ lower bound for the broadcast rate and thus are optimal. In this work, we use the duality relationship between NC and IC to show that for a class of IC problems with $MAIS=n-3$, binary linear index codes achieve the $MAIS$ lower bound on the broadcast rate. In contrast, it is known that there exists IC problems with $MAIS=n-3$ and optimal broadcast rate strictly greater than $MAIS$.
\end{abstract}
\section{Introduction}
Index Coding (IC), introduced in \cite{BiK}, considers the problem of efficiently broadcasting a number of messages available at a source, to receivers that already possess some prior knowledge of the messages. 
Index coding problems where each receiver demands a unique message are called \textit{single unicast} and are the most widely studied class. In this paper, we consider single unicast index coding.

Formally, the single-unicast IC problem (over some finite field $\mathbb{F}$) consists of a broadcast channel which can carry symbols from ${\mathbb F}$, along with a source with $n$ messages denoted by ${\cal X}=\{x_i, i\in[n]\triangleq \{1,...,n\}\}$ and $n$ receivers each demanding an unique message. The messages are modelled as $t$-length vectors over ${\mathbb F}$. For $j\in[n]$, receiver $j$ has a subset $S(j)\subset {\cal X}$ as prior side-information messages. 

For a message vector $\boldsymbol{x}\in{\mathbb F}^{nt}$, the source transmits a $l$-length codeword ${\mathbb E}(\boldsymbol{x})$ (the function $\mathbb E:{\mathbb F}^{nt}\rightarrow {\mathbb F}^l$, is known as the \textit{index code}), such that all the receivers can recover their demands. The quantity $l$ is known as the \textit{length of the code} $\mathbb E$. The \textit{transmission rate} of the code is defined as $\frac{l}{t}$. If $t=1$, then the index code is known as a \textit{scalar index code}, else it is known as a \textit{vector index code}. A linear encoding function ${\mathbb E}$ is also called a \textit{linear index code}. The goal of index coding is to find optimal index codes, i.e., those with the minimum possible transmission rate. For an index coding problem ${\cal I}$ (over ${\mathbb F}_q$) with $t$-length messages, let $\beta_q(t,{\cal I})$ denote the length of an optimal vector index code. The \textit{broadcast rate} \cite{BKL2} is then $\beta_q({\cal I})=\lim_{t\rightarrow \infty}\frac{\beta_q(t,{\cal I})}{t}$. 

A given single unicast index coding problem ${\cal I}$ can be modelled using a directed graph called the \textit{side-information graph} \cite{BBJK}, denoted by ${G}_{SI}({\cal V}_{SI},{\cal E}_{SI})$, where the set of vertices ${\cal V}_{SI}=[n]$, identified with the set of message symbols, represents also the set of receivers (each demanding an unique message). A directed edge  $(j,i)$ in ${\cal E}_{SI}$ indicates the availability of the message symbol $x_j$ as side-information at the receiver $i$ (which demands message symbol $x_i$). It was shown in \cite{BBJK} that the length of any optimal scalar linear index code (over ${\mathbb F}_q$) is equal to a property of the graph ${G}_{SI}$ called the minrank, denoted by $mrk_q({G}_{SI})$. While computing $mrk_q({G}_{SI})$ is known to be NP-hard \cite{Pee} in general, several authors have given lower bounds and upper bounds for the quantity, as well as specific graph structures for which the bounds are met with equality (see for example, \cite{BKL2,BBJK,DSC}). From \cite{BKL2,BBJK,DSC}, we know that given a single unicast IC problem ${\cal I}$ on ${G}_{SI}$ (with $n$ message vertices), we have the following.
\begin{align}
\label{eqn5}
MAIS=n-\tau\hspace{-0.05cm}\leq \hspace{-0.05cm}\beta_q({\cal I})\hspace{-0.05cm} \leq mrk_q({G}_{SI})\hspace{-0.05cm}\leq n-\nu,
\end{align}
where $MAIS$ is the size of the maximum acyclic induced subgraph of $G_{SI}$, $\tau$ is the minimum number of vertices that have to be removed to make $G_{SI}$ acyclic, and $\nu$ is the maximum number of disjoint directed cycles of $G_{SI}$. All the above quantities are NP-hard to compute for general graphs \cite{Pee,Kar}. It has been shown in \cite{Ong3} that for the case of side-information graphs with $\tau\in\{1,2\}$, optimal binary scalar linear index codes of length $n-\tau$ can be constructed. 

In this work we consider the case of constructing binary scalar linear index codes of length $n-3$ for a subclass of index coding problems with $\tau = 3$. Towards this end, we first show using a matroid theoretic argument that a linear index code for a given IC problem is dual to linear network code for an associated multiple unicast NC problem. Dual relationships of IC problems and distributed storage were discussed in \cite{Maz,ShD}, while the connection of network coding and index coding were discussed in \cite{RSG,ShD,ERL}. Though the duality presented in this paper is closely related to the results in \cite{Maz,ShD} and can be interpreted as being loosely connected to the ideas of \cite{ShD},  the explicit connection to network coding as given in this paper is novel and enables us to show our results on optimal binary codes. Using this connection, we show that for a class of IC problems with $\tau=3$, there exist linear codes over ${\mathbb F}_2$ of length equal to $MAIS=n-3$, which can be easily generalized to any field. In contrast, it has been shown in \cite{BKL1} that for the undirected side-information graph (in which an edge $\{i,j\}$ indicates directed edges $(i,j)$ and $(j,i)$) which is a $5$-cycle, we have $\tau=3$ but $\beta_q=2.5$ (for any ${\mathbb F}_q$) which is strictly greater than $MAIS=2$. 

The organization of this paper is as follows. In Section \ref{sec2}, we establish the duality properties of matroids associated with index codes, and in Section \ref{sec3}, we use these properties to show that a feasible network code on a particular graph is dual to a valid index code for a given side-information graph. Using this connection, we show in Section \ref{sec4} and Section \ref{sec5} that for IC problems with $n$ messages and  $MAIS=n-3$ satisfying some properties (we call such IC problems as \textit{Class} ${\mathbb I}$\textit{a} \textit{problems}), optimal linear index codes over ${\mathbb F}_2$ of length $MAIS$ can be constructed. The index code constructed can also be easily seen to be independent of the field size. This is accomplished by listing out all possible configurations within Class $\mathbb I$a IC problems (Section \ref{sec5}). 
 

\subsection{Notations and Terminology}
For sets $A$ and $B$, $A\backslash B$ is the set of elements in $A$ but not in $B$. $\mathbb{F}$ denotes a finite field. For a positive integer $m$, let $[m]=\{1,...,m\}.$ The vertex and edge sets of a directed graph $\cal G$ are denoted by ${\cal V}({\cal G})$ and ${\cal E}({\cal G})$ respectively. An edge from vertex $i$ to $j$ in a directed graph is denoted as $(i,j)$. For an edge $e=(i,j)$, $head(e)=j$ and $tail(e)=i$.  Throughout this paper, only simple directed graphs are used. On a directed graph $G,$ a path $p$ of length $(l-1)$ from vertices $v_1$ to $v_l$, denoted by $v_1\xrightarrow{p}v_l$ (thus identifying the start and end vertices), is a ordered $l$-tuple of distinct vertices $(v_1,...,v_{l-1},v_l)$ such that there is a directed edge between every two consecutive vertices. Subpaths of such paths can be written as $v_i\xrightarrow{p}v_j,$ for any $1\leq i<j\leq l$. We denote some path from $v$ to $w$ as $v\rightarrow w$. For two paths $p,q$, the set of vertices that are common to $p$ and $q$ is denoted by $p\cap q$. Let $u\xrightarrow{p}v$ and $v\xrightarrow{q}w$ be two paths with the only common vertex being $v$. Then the natural union of these paths can be written as $u\xrightarrow{p\cup q}w=u\xrightarrow{p}v\xrightarrow{q}w$. We can also extend this beyond two paths naturally. A vertex $v$ is said to be in the \textit{upstream} (equivalently, \textit{downstream}) of vertex $w$ if there exists some path from $v$ to $w$ (from $w$ to $v$). 
A cycle of length $l$ in a directed graph has the same definition for a path except that start and end vertices are identical. A trail can be a cycle or a path. 

\section{Preliminaries}
\label{sec2}
\subsection{Matroids}
We provide some basic definitions and results from matroid theory. For more details, the reader is referred to \cite{Oxl}.
\begin{definition}[Matroids]
\label{matroiddefnindp}
Let $E$ be a finite set. A \textit{matroid} $\cal{M}$ on $E$ (called the ground set of ${\cal M}$) is an ordered pair $(E,\cal{I}),$ where the set $\cal{I}$ is a collection of subsets (called independent sets) of $E$ satisfying the following three conditions
\begin{enumerate}
\item $\phi \in \cal{I}.$ 
\item If $X \in \cal{I}$ and $X' \subseteq X,$ then $X'  \in \cal{I}.$
\item If $X_1$ and $X_2$ are in $\cal{I}$ and $|X_1|<|X_2|,$ then there is an element $e$ of $X_2-X_1$ such that $X_1 \cup e\in \cal{I}.$
\end{enumerate}
For $X\subseteq E$, the rank function $r(.)$ of a matroid is a function which associates to $X$ an integer which is the size of a maximal independent subset of $X$. An set $X \in {\cal I}$ such that $r(X)=r(E)$ is called a basis for ${\cal M}$.
\end{definition}
\begin{definition}[Vector matroid]
Let $A$ be a matrix over some field $\mathbb{F}$. The ordered pair $(E,\cal{I})$ where $E$ consists of the set of column labels of $A,$ and $\cal{I}$ consists of all the subsets of $E$ which index columns that are linearly independent over $\mathbb{F}$ is a matroid called the \textit{vector matroid} associated with $A$.
\end{definition}
\begin{definition}[Dual matroid]
Let ${\cal M}=(E,{\cal I})$ be a matroid. Then the set $\left\{E-B:B~\text{is a basis of}~{\cal M}\right\}$ forms the set of bases of a matroid on $E({\cal M}),$ defined as the \textit{dual matroid} of ${\cal M}$ denoted as ${\cal M}^*$ (note that $({\cal M}^*)^*={\cal M}$). Let $r^*$ be the rank function of ${\cal M}^*$. It can be shown that, for any $X\subseteq E$, 
\begin{align}
\label{eqn2}
r^*(X)=|X|-r(E)+r(E-X).
\end{align}
The dual matroid of a vector matroid of a matrix $A$ (over some field $\mathbb F$) is also a vector matroid of a matrix over the same field $\mathbb F$. In particular, if $A$ is a $k\times n$ matrix with rank $k$, the dual matroid is a vector matroid of a $(n-k)\times n$ matrix with rank $n-k$.
\end{definition}

\subsection{Multiple unicast network coding}
A multiple unicast network coding problem on a acyclic network (see \cite{LiL}, for example) consists of an acyclic graph $G_{NC}$ representing the network, a set of $n$ source vertices generating one message (a symbol from $\mathbb F$ each, denoted by $\{x_i:i\in[n]\}$) and a set of $n$ receiver vertices demanding the corresponding symbol generated by the source. The edges in the network have capacity of one symbol from $\mathbb F$ per unit time. In the linear network coding framework \cite{LiY}, the vertices of the network are allowed to transmit linear combinations of the incoming symbols on their outgoing edges. A linear network code can be defined by the coefficients (called \textit{local encoding coefficients}) of these linear combinations taken at the network vertices. 

Given the set of all local encoding coefficients, we can associate with every edge $e$ a column-vector $\boldsymbol{f_e}$ (called the \textit{global encoding vector} of $e$) such that if $\boldsymbol{x}$ is the row-vector of messages then the symbol flowing on edge $e$ can be obtained as $\boldsymbol{x}\boldsymbol{f_e}$. A linear network code can also be specified by set of all \textit{global encoding vectors}. Furthermore, we can associate the standard basis vectors of $\mathbb{F}^n$ with the $n$ source symbols. As the symbols on any edge $e$ are linear combinations of those in the incoming edges of $tail(e)$, it is clear that the following conditions hold.
\begin{itemize}
\item \textit{Flow conservation condition: }$\boldsymbol{f_e}$ is a linear combination of $\{\boldsymbol{f_{e'}}:head(e')=tail(e)\}.$
\end{itemize}
A network code is said to be \textit{feasible} if all the receivers can decode their respective demand. For linear multiple unicast network codes, a receiver demanding $x_i$ can decode successfully if a linear combination of the corresponding incoming global encoding vectors gives the $i^{th}$ standard basis from $\mathbb{F}^n$.
\subsection{The Matroid Theoretic Dual of an Index Code}
The following lemma is well known in different forms from prior work (for instance, \cite{MCJ}). It is essentially a decodability criterion. The phrasing here is however in terms of the rank of specific sets in the vector matroid associated with a index code for ${G}_{SI}$. We skip the proof as it is straightforward. 
\begin{theorem}
\label{thmMatroidIC}
Let $B$ be denote a $k\times n$ ($k\leq n$) matrix over $\mathbb F$, where $n=|{\cal V}({G}_{SI})|$ . Let $r(.)$ be the rank function of ${\cal M}[B]$. Then the following statements are equivalent (equating the ground set of ${\cal M}[B]$ with ${\cal V}({G}_{SI})$).
\begin{enumerate}
\item $B$ is a valid index coding matrix for ${G}_{SI}$.
\item For all $v\in {\cal V}({G}_{SI})$, we have 
\begin{align}
\nonumber
r({\cal V}({G}_{SI})\backslash S(v))&=r(v)+r({\cal V}({G}_{SI})\backslash\{S(v)\cup v\})\\
\label{eqn1}
&=1+r({\cal V}({G}_{SI})\backslash\{S(v)\cup v\}).
\end{align}
\end{enumerate}
\end{theorem}
We now give a result which gives the condition for a given vector matroid to be an index code, based on its dual matroid of this code. This result can be obtained from \cite{Maz,ShD}. However we give a simple proof here based on matroid theory.
\begin{theorem}
\label{dual:thm}
Let $C$ be a $(n-k)\times n$ matrix over ${\mathbb F}$ with rank $n-k$ (with its columns indexed by ${\cal V}({G}_{SI})$. Let $r^*$ be the rank function of ${\cal M}[C]$.  Let $B_{k\times n}$ be a matrix of rank $k$ such that ${\cal M}[B]$ is the dual matroid of ${\cal M}[C]$. Then $B$ denotes a valid index code for ${G}_{SI}$ if and only if 
\begin{align}
\label{eqn3}
r^*(v\cup S(v))=r^*(S(v)),\forall v\in {\cal V}({G}_{SI}).
\end{align}
\end{theorem}
\begin{IEEEproof}
Suppose (\ref{eqn3}) holds. Let $r$ be the rank function of ${\cal M}[B]$. It is sufficient then to prove that (\ref{eqn1}) holds in Theorem \ref{thmMatroidIC}. We have
\begin{align*}
r&({\cal V}({G}_{SI})\backslash S(v))\\
&\stackrel{\tiny\text{by (\ref{eqn2})}}{=}|{\cal V}({G}_{SI})\backslash S(v)|-r^*({\cal V}({G}_{SI}))+r^*(S(v)).\\
&\stackrel{\tiny\text{by (\ref{eqn3})}}{=}|{\cal V}({G}_{SI})\backslash\{S(v)\cup v\}|+1-r^*({\cal V}({G}_{SI}))+r^*(S(v)\cup v)\\
&\stackrel{\tiny\text{by (\ref{eqn2})}}{=}1+r({\cal V}({G}_{SI})\backslash\{S(v)\cup v\}).
\end{align*}
In the same way, we can use (\ref{eqn1}) and (\ref{eqn2}) to show that (\ref{eqn3}) holds. We leave this part to the reader.
\end{IEEEproof}
\section{A relationship of Index Coding with Network Coding}
\label{sec3}
For a given single-unicast index coding problem (with $MAIS=(n-\tau)$ being the lower bound on the rate), we associate a multiple unicast network coding problem and show that a scalar linear network code for the NC problem corresponds to a scalar linear index code  of length = $MAIS$ (and hence an optimal index code) for given IC problem. 

We first choose some set of $\tau$ vertices$, {\cal V}_\tau\subset{\cal V}({{G}_{SI}}),$ such that removing ${\cal V}_\tau$ makes ${G}_{SI}$ acyclic. We proceed by generating an acyclic graph $G_{NC}$ and an associated multiple unicast problem on $G_{NC}$ corresponding to the side information graph ${G}_{SI}$ as follows.
\begin{itemize}
\item For each vertex $v\in {\cal V}({{G}_{SI}})$, a pair of vertices, denoted by $v$ and $v'$, is created in $G_{NC}$. The vertices $v$ and $v'$ in $G_{NC}$ are connected by a edge (which we call a `coding edge'). For any $w\in{\cal V}_\tau$, the vertices $w\in {\cal V}(G_{NC})$ are assigned as the \textit{source nodes} of $G_{NC}$, each of which generates a unique message (from $\mathbb F$) independently of other sources. 
\item For all $w\notin {\cal V}_{\tau}$, for all $v\in S(w)\subset{\cal V}({{G}_{SI}})$, we create an edge $(v',w)$ in $G_{NC}$. Such edges are called `forwarding' edges.
\item For each $w\in{\cal V}_\tau$, we also create vertices $D_w$ and $D_{w'}$ in $G_{NC}$ (with $D_{w'}$ denoting the receiver vertex demanding the message generated at source vertex $w$), and connect them by a (coding) edge $(D_w,D_{w'})$.
\item Finally, for any  $w\in {\cal V}_\tau$, for any $v\in S(w)\subset{\cal V}({{G}_{SI}})$, we create a (forwarding) edge from $(v',D_w)$. 
\end{itemize}
\begin{remark}
In the forthcoming discussions, we identify the nodes $v \in {G}_{SI}$ with the node $v\in G_{NC}$. The meaning will be clear from the context. Furthermore, we call the vertices $\{v,D_v : v\in {\cal V}(G_{SI})\}$ as \textit{undashed vertices} of $G_{NC}$ and the other vertices of $G_{NC}$ as the \textit{dashed vertices}.
\end{remark}

Fig. \ref{fig:IC_tau2} shows an example $G_{SI}$ and Fig. \ref{fig:IC_converted} shows the corresponding $G_{NC}$. It is clear that trails in $G_{NC}$ have corresponding trails in $G_{SI}$.
%
The following theorem puts down some simple facts about these corresponding trails in $G_{NC}$ and $G_{SI}$. 
\begin{theorem}
\label{statements}
The following statements, along with the converses of statements $1$-$5$ are true.
\begin{enumerate}
\item For some $v\in{\cal V}_\tau$ and $p_{NC}=v\rightarrow D_{v'}$, the trail $p_{SI}$ is a cycle passing through the corresponding vertex $v$ in ${G}_{SI}$ (and no other vertex in ${\cal V}_\tau$).
\item If paths $p_{NC}=v\rightarrow D_{v'}$ and $q_{NC}=w\rightarrow D_{w'}$ for distinct $v,w\in{\cal V}_\tau$ are edge-disjoint in $G_{NC}$, then the cycles $p_{SI}$ and $q_{SI}$ in ${G}_{SI}$ are node-disjoint. 
\item If paths $p_{NC}=v\rightarrow D_{v'}$ and $q_{NC}=w\rightarrow D_{w'}$ for distinct $v,w\in{\cal V}_\tau$  intersect at the edge $(u,u'),$ then the cycles $p_{SI}$ and $q_{SI}$ in ${G}_{SI}$ intersect at the corresponding vertex $u\in{\cal V}({G}_{SI})$. Moreover any such intersecting node $u\notin {\cal V}_\tau.$ 
\item For $\{v_i\in{\cal V}_\tau,i=1,...,r\}$ ($r\leq \tau$), suppose there are paths $\{p_{i,NC}=v_i\rightarrow D_{v_{i+1}'}: i=1,...,r-1\}$ and $p_{r,NC}=v_{r}\rightarrow D_{v_1'}$, such that they are all edge-disjoint in $G_{NC}$. Then the union $\cup_{i=1}^r p_{i,SI}$ is a cycle in ${G}_{SI}$ passing through the vertices $v_i\in{\cal V}_\tau, i=1,...,r,$ and no other vertices in ${\cal V}_\tau$. 
\item Every path $p$ in $G_{NC}$ in which an undashed vertex $v$ is an intermediate vertex (neither the start vertex nor the end vertex of $p$) also passes through $v'$, the corresponding dashed vertex.
\item The graph $G_{NC}$ associated with the side information graph ${G}_{SI}$ is acyclic. 
\end{enumerate}
\end{theorem}
\begin{IEEEproof}
We prove only the last statement, the others are follow from the definition of $G_{NC}$. In ${G}_{SI},$ all the cycles pass through the vertices in ${\cal V}_\tau$. From (converse of) statement (1), any cycle which passes through only one vertex $v\in {\cal V}_\tau$ has a corresponding path $v\rightarrow D_{v'}$ in $G_{NC}$. If a cycle in ${G}_{SI}$ passes through multiple vertices $v_1,v_2,...,v_l\in{\cal V}_\tau$ in ${G}_{SI}$ in succession, then by (converse of) statement (4), we have paths $v_1\rightarrow D_{v_2'},$ $v_2\rightarrow D_{v_3'}, ..., v_{l-1}\rightarrow D_{v_l'}, v_{l}\rightarrow D_{v_1'}.$ Thus any cycle in ${G}_{SI}$ (which involves at least one vertex from ${\cal V}_\tau$) is converted into one or many paths in $G_{NC}$. 


If there is any other cycle in $G_{NC}$ (which necessarily has to be a cycle not containing any vertex from ${\cal V}_\tau$), then clearly there should be a cycle in ${G}_{SI}$ not containing any vertex from ${\cal V}_\tau$. By definition of ${\cal V}_\tau$ this is not possible. Hence $G_{NC}$ is acyclic.
\end{IEEEproof}
The following theorem shows the dual relationship between a NC solution in $G_{NC}$ and IC solution to $G_{SI}$.
\begin{theorem}
\label{ICNCdualitythm}
Over any field $\mathbb F$, there exists a optimal feasible linear index code for ${G}_{SI}$ of length $n-\tau$ if and only if a feasible linear network code exists for the multiple unicast network $G_{NC}$.
\end{theorem}
\begin{IEEEproof}
We prove the theorem by construction. Suppose there is a feasible network code for $G_{NC}$.

Consider the matrix $A$ formed by concatenating column-wise the global encoding vectors $\boldsymbol{f_e}$, for all coding edges except $(D_v,D_{v'}):v\in{\cal V}_\tau$.  This matrix therefore has $n=|{\cal V}({G}_{SI})|$ columns. We assume that the column $\boldsymbol{f}_{(v,v')}$ is indexed by the matroid element $v$ in the vector matroid of $A$. By construction of $G_{NC}$ and because the given network code is feasible, it is clear that for any $v\in {\cal V}({G}_{SI})\backslash {\cal V}_{\tau},$ $\boldsymbol{f}_{(v,v')}$ is linearly dependent on $\{\boldsymbol{f}_{(w,w')}:w\in S(v)\}$. Hence we have $r^*(v)=r^*(S(v)\cup v)$, where $r^*(.)$ is the rank function associated with the vector matroid of $A$. 

For any $v\in {\cal V}_{\tau}$, because of the construction of $G_{NC}$ and since the network code is feasible, the vector $\boldsymbol{f}_{(D_v,D_{v'})}$ should be dependent on $\{\boldsymbol{f}_{(w,w')}:w\in SI(v)\}$. Since the network code is feasible, we must have for some constant $c\neq 0$, $\boldsymbol{f}_{(v,v')}=c\boldsymbol{f}_{(D_v,D_{v'})}$.  Hence, again we have $r^*(v)=r^*(S(v)\cup v)$ for all $v\in {\cal V}_\tau$. Thus, by Theorem \ref{dual:thm}, the dual matrix of $A$ represents a feasible index coding matrix for ${G}_{SI}$.

The only if part follows in a similar way.
\end{IEEEproof}
\begin{example}
Consider the index coding problem represented by the side information graph in Fig. \ref{fig:IC_tau2}. This represents a problem for which $\tau=2$. WLOG, we may assume ${\cal V}_\tau=\{1,2\}$. 

We convert this into a multiple unicast network coding problem on a network shown in Fig. \ref{fig:IC_converted}. The network is obtained according to the procedure described Section \ref{sec3}. Note that a feasible multiple unicast code for this network can be obtained by fixing the global encoding vectors of the edges on the path $1\rightarrow 4\rightarrow 6\rightarrow 9\rightarrow 10\rightarrow D_{1'}$ as $(1~~0)^T$ and along the edges of $2\rightarrow 11\rightarrow 5\rightarrow 7\rightarrow 8\rightarrow D_{2'}$ as $(0~~1)^T$.  This can be done as the two paths are edge-disjoint. Other edges not on these paths are assigned zero vector. This leads to a optimal index code for Fig. \ref{fig:IC_tau2} by Theorem \ref{ICNCdualitythm}. 

\begin{figure}
\centering
  \includegraphics[width=0.5\linewidth]{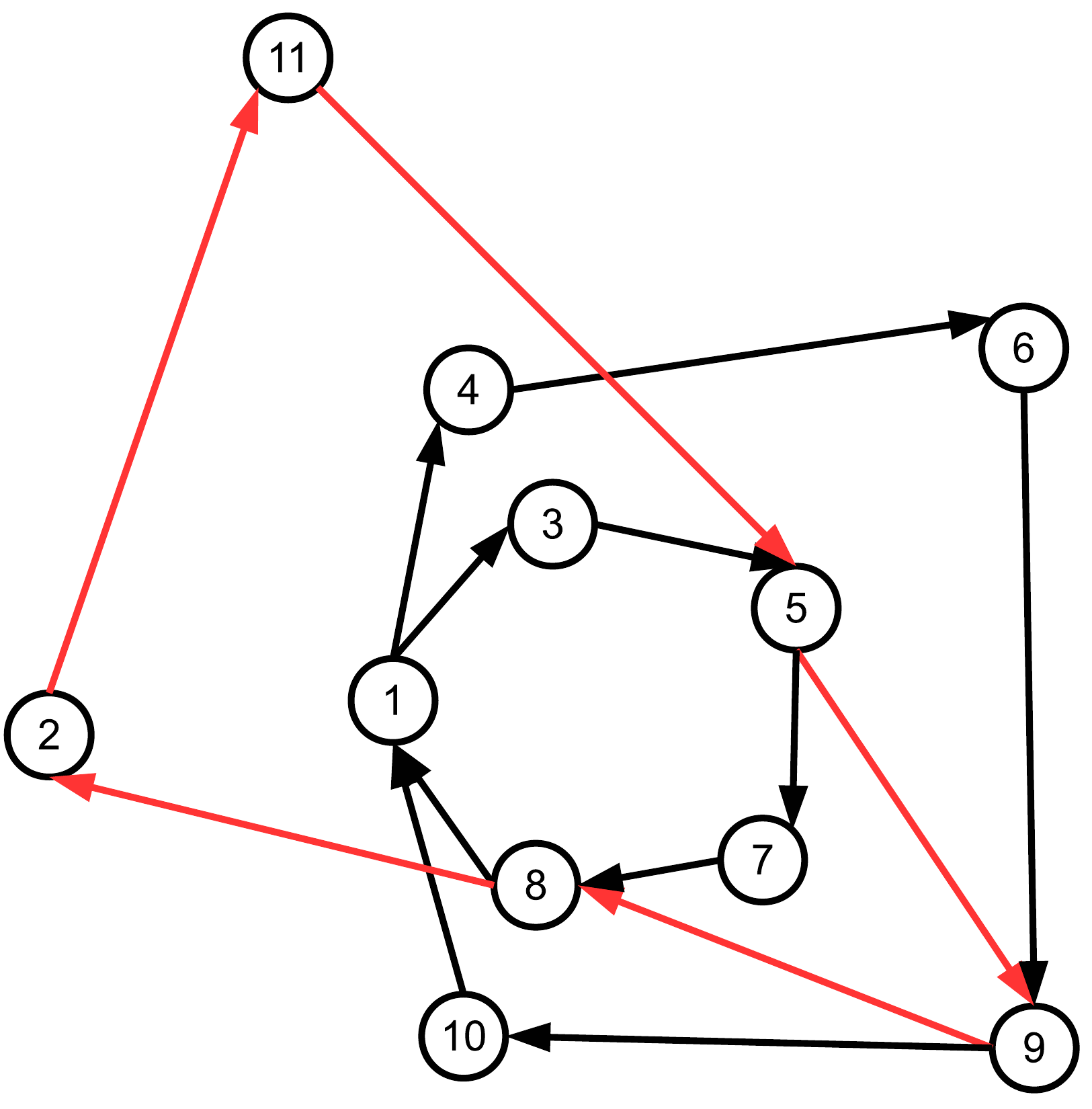}
  \caption{\small Side information graph ${G}_{SI}$ of an example index coding problem with $\tau=2$. The cycles in black are two cycles passing through vertex $1$. The cycle in red denotes a cycle passing through vertex $2$.}
	\label{fig:IC_tau2}
	\hrule
\end{figure}
\begin{figure}
  \includegraphics[width=\linewidth]{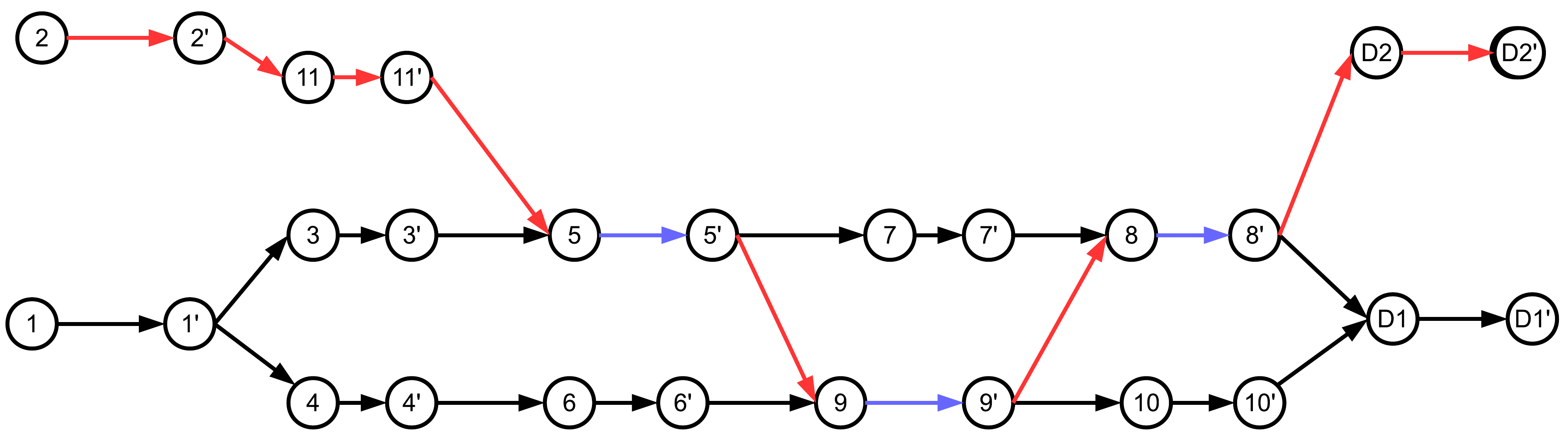}
  \caption{\small The network coding equivalent of the index coding problem in Fig. \ref{fig:IC_tau2}. The cycles passing through vertices $1$ and $2$ in ${G}_{SI}$ become paths from $1$ to $D_1,$ and $2$ to $D_2$ respectively, in $G_{NC}$. 
	} 
  \label{fig:IC_converted}
	\hrule
\end{figure}
\end{example}
\section{Index Coding Problems on Side-information graph with $\tau=3$}
\label{sec4}
We first give some terminology for our use in this paper.
\begin{definition}
We refer to a cycle in $G_{SI}$ that passes through only one vertex in ${\cal V}_\tau$ as a unicycle, and if that vertex is $v\in{\cal V}_\tau$, we denote this unicycle as a $v$-cycle. We call the path $v\rightarrow D_{v'}$ in $G_{NC}$ corresponding to a $v$-cycle as a $v$-unipath. A cycle in $G_{SI}$ passing through $\{v_i\in{\cal V}_\tau, i\in[r]\}$ (and not through any other vertices of ${\cal V}_\tau$) is called a bicycle if $r=2$ and a tricycle if $r=3$. We denote such a cycle as $(v_i:i\in[r])$-cycle. For $r\geq 2$ and $i\in[r]$ the paths $p_{i,NC}$ in $G_{NC}$ associated with the $(v_i:i\in[r])$-cycle of $G_{SI}$ are respectively called the $(v_i,v_{i+1})$-crosspaths ($i\in[r-1]$) and the $(v_r,v_1)$-crosspath in $G_{NC}$.
\end{definition}
In \cite{Ong3}, it has been shown that if $\tau\leq 2$, then the IC problem on $G_{SI}$ can be solved with a binary linear index code of length $n-\tau$. Furthermore, the index code generalizes over any finite field naturally. By Theorem \ref{ICNCdualitythm}, this means that the corresponding $G_{NC}$ with $\tau$-unicast ($\tau\leq 2$) has a feasible network code. In this paper we look at the case IC problems on side-information graphs with $\tau=3$.

For an $IC$ problem on $G_{SI}$ with $\tau=3$, the NC problem on the corresponding $G_{NC}$ is a $3$-unicast NC problem. It is not difficult to see that, in this case, if there exists one unipath in $G_{NC}$ that is edge-disjoint to the unipaths from the two other sources to the respective receivers, then we can view the NC problem on $G_{NC}$ as a disjoint union of a $1$-unicast problem (on a subgraph $H_{NC}$ containing the source-sink pair and the edge-disjoint unipath) and a $2$-unicast problem (with the subgraph $G_{NC}\backslash {\cal E}(H_{NC})$). These two problems can be solved separately and merged together again with minor changes (adding an extra zero appropriately in the global encoding vectors), leading to a $3$-length network code for $G_{NC}$, and hence a $(n-3)$-length optimal index code for $G_{SI}$. 

Thus the novel IC problems with $\tau=3$ are those in which we cannot find any one unipath which is edge-disjoint to the other unipaths. We consider a class of such problems, henceforth which satisfy the following condition (WLOG we assume that ${\cal V}_\tau=\{1,2,3\}$).
\begin{tcolorbox}[title= Definition: Class $\mathbb I$ IC problem and associated $G_{NC}$]
A Class $\mathbb I$ IC problem is one (with $MAIS=n-3$) in which there exists a non-zero set of vertices ${\cal V}_I$ which lie in the intersection of all unicycles of ${\cal G}_{SI}$. Thus, in the associated NC graph $G_{NC}$, if $P^1_r, P^2_s,$ and $P^3_t$ denote the $r^{th}$ $1$-unipath, $s^{th}$ $2$-unipath, and $t^{th}$ $3$-unipath respectively, then the set of vertices
\[
\left(\cap_{r} P^1_r\right)\bigcap\left(\cap_{s} P^2_s\right)\bigcap\left(\cap_{t} P^3_t\right)
\]
must be non-empty. We call the class of $G_{NC}$ networks satisfying the above conditions as Class $\mathbb I$ networks.
\end{tcolorbox}
In Section \ref{sec5}, we will consider a subclass of Class $\mathbb I$ IC problems and construct optimal index codes of length $n-3$ for such problems by constructing feasible network codes for the associated Class $\mathbb I$ $G_{NC}$ networks. In the rest of this section, we establish some structural properties of Class $\mathbb I$ networks.

We need a simple lemma which we will repeatedly use throughout the paper henceforth. The proof can be seen from a simple pictorial argument and so is skipped. Due to Lemma \ref{contiintersection}, intersections between two subpaths of $G_{NC}$ can always assumed to be contiguous.
\begin{lemma}
\label{contiintersection}
Consider paths $p_{ab}$ and $p_{cd}$ for (not necessarily distinct) vertices $a,b,c,d$ (dashed or undashed) in $G_{NC}$. If $p_{ab}$ and $p_{cd}$ intersect in some vertices or edges, we can find a path from $c$ to $d$, say $q_{cd}$, such that $p_{ab}$ and $q_{cd}$ intersect contiguously (i.e., $p_{ab}\cap q_{cd}$ is itself a single subpath). For this reason, we can assume WLOG that if there exists at least one pair of $a\rightarrow b$ and $c\rightarrow d$ paths which have a non-empty intersection, then there exist $a\rightarrow b$ and $c\rightarrow d$ paths having contiguous intersections . 
\end{lemma}
\subsection{Skeleton-network configurations and crosspaths within Class $\mathbb I$}
\begin{figure}[ht]
  \centering
  \vspace{0.0in}
  \subfigure[Skeleton Network A. The subpaths (for instance, $1\rightarrow v_{12}$) may contain several vertices and edges.]{\label{skeleton_style_A}	
	\includegraphics[width=0.3\textwidth]{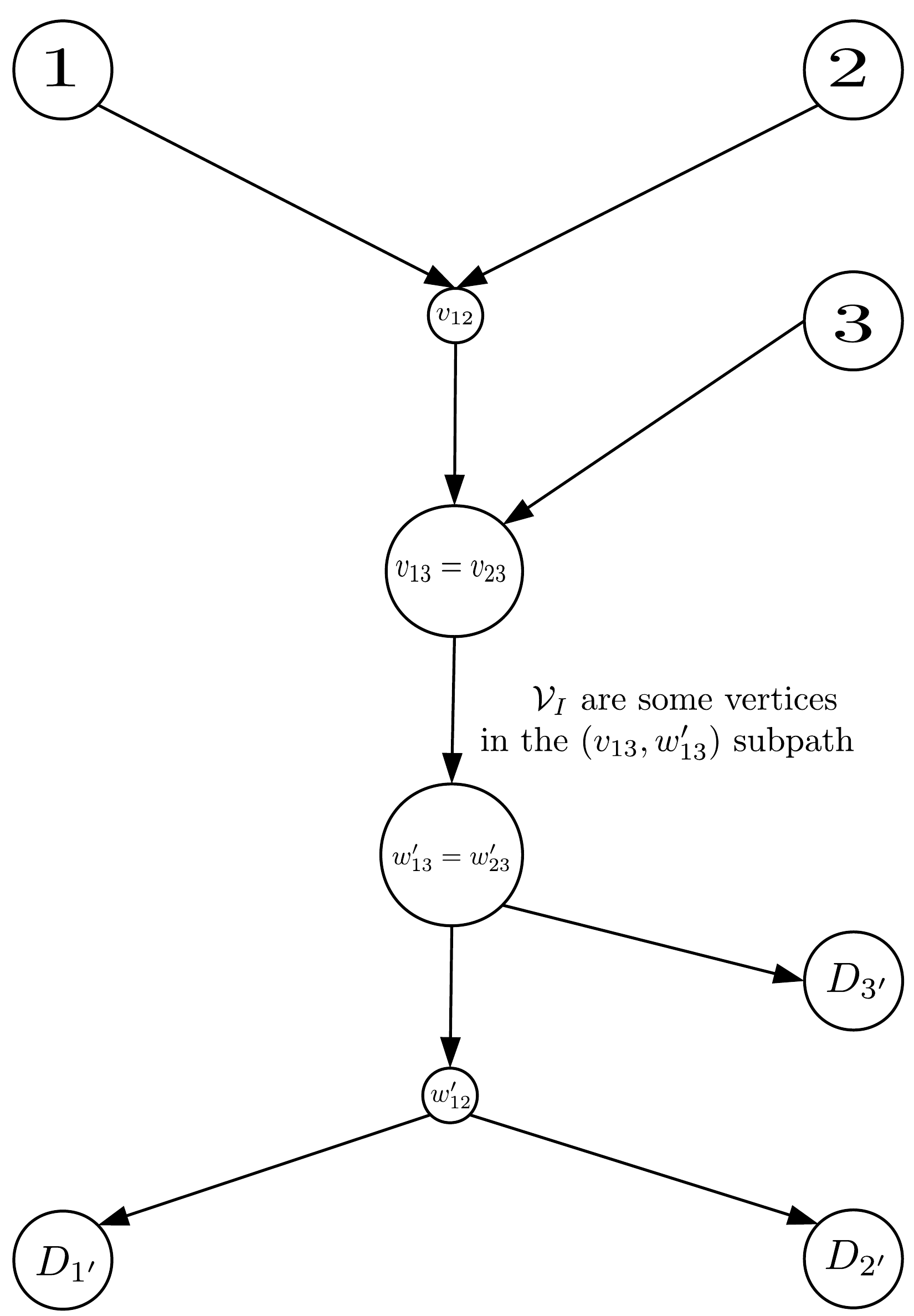}} 
  \hrule\subfigure[Skeleton Network B]{\label{skeleton_style_B}
	\includegraphics[width=0.3\textwidth]{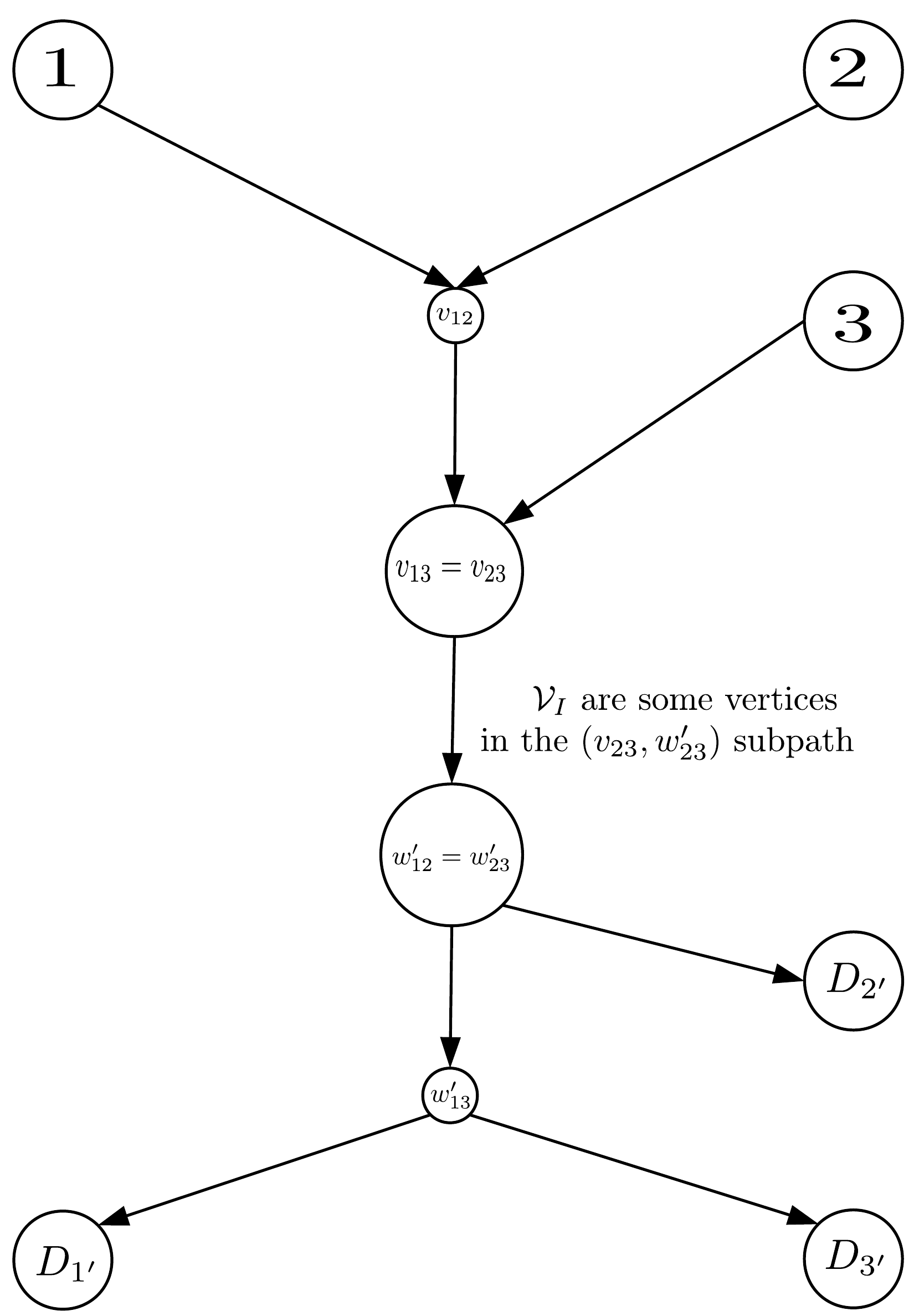}}
	\hrule
	\caption{\small Skeleton networks of $G_{NC}$}
\end{figure}
	
As mentioned before, WLOG we can assume ${\cal V}_\tau=\{1,2,3\}$. Consider a subgraph of $G_{NC}$ consisting of three unipaths (such unipaths must exist by definition of ${\cal V}_\tau$ and Theorem \ref{statements}), one from each source to each receiver. Since Class $\mathbb I$ property have to be satisfied and since we can consider any intersection between unipaths to be a contiguous intersection by Lemma \ref{contiintersection} and by Theorem \ref{statements}, it is not hard to see that this subgraph can be essentially of two types (upto a permutation of ${\cal V}_\tau$) as shown in Fig. \ref{skeleton_style_A} and Fig. \ref{skeleton_style_B}. We call these two configurations as \textit{skeleton network A} and \textit{skeleton network B} (note that bolded lines with arrowheads indicate subpaths in figures henceforth, not just an edge).

For distinct $i,j\in\{1,2,3\}$, we denote by the undashed vertex $v_{ij}$ the first vertex in the skeleton subnetwork in the intersection of $i$-unipath and $j$-unipath and $w'_{ij}$ denote the last vertex (which is necessarily dashed by statement (5) of Theorem \ref{statements}) in the intersection (note that $v_{ij}=v_{ji}$ and $w'_{ij}=w'_{ji}$). 

We now show that $G_{NC}$ must necessarily have other paths apart from the skeleton network, else $\tau<3$.  
\begin{lemma}
\label{allcrosspaths}
\begin{enumerate}
\item In $G_{SI}$, there exists an $ij$-bicycle for each pair of distinct $i,j\in[3]$. 
\item In $G_{NC}$, there exists an $ij$-crosspath for each distinct pair of $i,j\in[3]$. 
\item Also, for distinct $i,j$, for any given $i$-unipath $P$ and $j$-unipath $P'$, there exists an $ij$-crosspath that does not intersect $P\cap P'$. 
\end{enumerate}
\end{lemma}
\begin{IEEEproof}
We show the first part by contradiction. Assume that there is no $12$-bicycle in $G_{SI}$. The possible other cycles of $G_{SI}$ are the unicycles, the $13$-bicycles, the $23$-bicycles, and the $123$-tricycles. Suppose we delete any one vertex from ${\cal V}_I$ and vertex $3$ (and the associated edges), the graph $G_{SI}$ becomes acyclic. Thus $\tau\leq 2$, which is a contradiction as $\tau=3$. This completes the proof of the first part. The second part follows from the first part and Theorem \ref{statements} (statement (4)).

Now to prove the third part. Suppose every $ij$-crosspath passes through all vertices in ${\cal V}_I$. Then removing any vertex in ${\cal V}_I$ makes ${\cal G}_{SI}$ acyclic (by statement (4) of Theorem \ref{statements}), but we have $\tau=3$. Thus there should exist $ij$-crosspaths which don't pass through all vertices in ${\cal V}_I$. Consider such a $ij$-crosspath labelled $q$, and also an $i$-unipath $P$ and a $j$-unipath $P'$.  We will now prove by contradiction that $q$ does not pass through any vertex in $P\cap P'$ either.


Consider that some vertex $w\in {\cal V}_I$ does not lie in $q$ (such a vertex should exist by the previous arguments), but there is some $v\in P\cap P'$ which lies upstream of $w$ and also in $q$. Then the $j$-unipath $j\xrightarrow{P'}v\xrightarrow{q}D_{j'}$ exists. But this unipath does not contain $w\in {\cal V}_I$, which is a contradiction to the assumption of Class $\mathbb I$ networks. Similarly, if there is some $v\in P\cap P'\cap q$ which lies downstream of $w$, we can construct a $i$-unipath which does not contain any $w$. But our assumption is that each unipath contains ${\cal V}_I$, and hence $q\cap P\cap P'$ must be empty. This concludes the proof.
\end{IEEEproof}
\section{Style A and Style B networks}
\label{sec5}
\begin{figure}
\centering
  \includegraphics[width=0.7\linewidth]{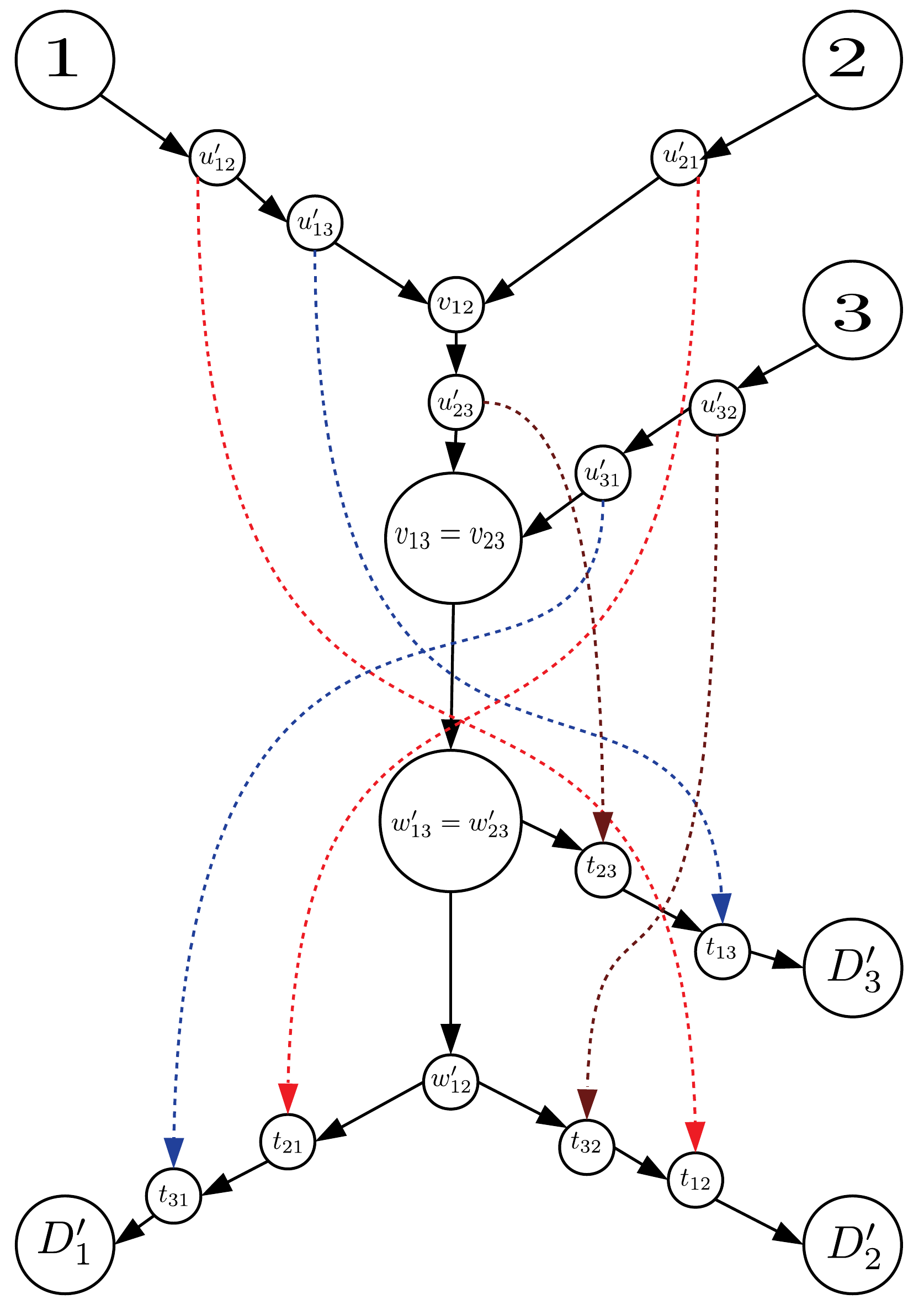}
  \caption{\small Style A network example showing the skeleton network A with all cross paths.  The crosspaths are colored and dotted. This is Configuration 3 as shown in Fig. \ref{styleAallconfigs}.}	
	\hrule
	\label{styleAexample}
\end{figure}

Lemma \ref{allcrosspaths} results in $G_{NC}$ graphs with different configurations. In this work, we look at a subclass of Class $\mathbb I$ $G_{NC}$ networks which we call as \textit{Class} $\mathbb I$\textit{a networks}. 
\begin{tcolorbox}[title=Definition: Class $\mathbb I$a networks]
Consider a Class $\mathbb I$ $G_{NC}$ and let $P^m, m\in [3]$ be $m$-unipaths. Consider the subgraph $H$ consisting of $P^1, P^2, P^3$ and at least one $ij$-crosspath chosen for each $i,j\in[3]$ according to Statement (3) of Lemma \ref{allcrosspaths}. We call $G_{NC}$ as a Class $\mathbb I$a $G_{NC}$, if in $H$ the $ij$-crosspath (for each distinct pair $i,j\in[3]$) does not intersect with any of the subpaths of $H$ except in the vertices of the 
subpaths $i\rightarrow v_{ij}$ and $w'_{ij}\rightarrow D_{j'}$.
\end{tcolorbox}
Within the set of all Class $\mathbb I$a networks, clearly each has a subgraph that is either a skeleton network A or a skeleton network B. We call these two as Style A and Style B networks. An example of a Style A network (with skeleton network A) is shown in Fig. \ref{styleAexample}.
The main result of this paper is as follows.
\begin{theorem}
\label{maintheorem}
Let $G_{SI}$ represent an IC problem with $\tau=3$ (hence $MAIS=n-3$), such that the network $G_{NC}$ corresponding to $G_{SI}$ is a Class $\mathbb I$a network. Then $G_{NC}$ has a feasible network code over ${\mathbb F}_2$ and hence (by Theorem \ref{ICNCdualitythm}) the given IC problem has a binary optimal linear index code of length $n-3$.
\end{theorem}
To prove Theorem \ref{maintheorem}, we need to obtain a feasible network code for $G_{NC}$. We do this by listing out all possible subgraph structures (of both Style A and Style B) that can possibly exist in $G_{NC}$ if it has to satisfy the Class $\mathbb I$a properties and then obtain feasible network codes for these subgraphs (given in Proposition \ref{mainthmStyleA} in Section \ref{styleA} and Proposition \ref{mainthmStyleB} in Section \ref{styleB}). 

We need a few more notations. As shown in Fig. \ref{styleAexample}, for Style A and B networks, we denote the last (dashed) vertex in the intersection of $i$-unipath and the $ij$-crosspath as $u'_{ij}$, and the first (dashed) vertex in the intersection of the $j$-unipath and $ij$-crosspath as $t_{ij}$ ($u'_{ij}$ and $t_{ij}$ are clearly well defined). We now prove some conditions on the positions of these crosspath intersection vertices in the following lemma.
\begin{lemma}
\label{upstreamdownstreamlemma}
Let $\{i,j,k\}=\{1,2,3\}$ (with some permutation). Then the following statements are true.
\begin{enumerate}
\item $u'_{ij}$ has to be upstream of $v_{ij}$, and $t_{ij}$ must be downstream of $w'_{ij}$. 
\item If $u'_{ij}$ is downstream of $v_{ik}$, then $t_{ij}$ cannot be	 upstream of $w'_{jk}$.
\end{enumerate}
\end{lemma}
\begin{IEEEproof}
Statement (1) is a direct consequence of Statement (3) of Lemma \ref{allcrosspaths}.

We now prove statement (2). Suppose $u'_{ij}$ is downstream of $v_{ik}$ (Note that this possibility can happen only if $v_{ik}$ is upstream of $v_{ij}$, else Statement 1) will be violated), and $t_{ij}$ is upstream of $w'_{jk}$. Let the $ij$-crosspath through $u'_{ij}$ and $t_{ij}$ be denoted as $p$, and $q$ denote the $k$-unipath. Consider the $k$-unipath constructed as follows. 
\[
k\xrightarrow{q}v_{ik}\xrightarrow{q}u'_{ij}\xrightarrow{p}t_{ij}\xrightarrow{p}w'_{jk}\xrightarrow{q}D_{k'}.
\]
First note that as $u'_{ij}$ is upstream of $v_{ij}$, the subpath $k\xrightarrow{q}v_{ik}\xrightarrow{q}u'_{ij}$ does not contain any $v\in{\cal V}_I$. The crosspath $p$ does not contain any vertex in ${\cal V}_I$. As $t_{ij}$ is downstream of $w'_{ij}$, $t_{ij}\xrightarrow{p}w'_{jk}\xrightarrow{q}D_{k'}$ contains no vertex from ${\cal V}_I$. Thus, the $k$-unipath constructed above does not contain any vertices from ${\cal V}_I$. This is a contradiction as ${\cal V}_I$ lies on every unipath. This proves statement (2). 
\end{IEEEproof}

Based on Lemma \ref{upstreamdownstreamlemma}, for $\{i,j,k\}=\{1,2,3\}$, the $ij$- crosspath can be in one of the three configurations (note that statement (1) of Lemma \ref{upstreamdownstreamlemma} is true always).
\begin{enumerate}
\item[]\textbf{Type 1:}~$u'_{ij}$ is not downstream of $v_{ik}$ (i.e., there is no path from $v_{ik}$ to $u'_{ij}$) and $t_{ij}$ is not upstream of $w'_{jk}$. 
\item[]\textbf{Type 2:}~$u'_{ij}$ is not downstream of $v_{ik}$ and $t_{ij}$ is upstream of $w'_{jk}$. 
\item[]\textbf{Type 3:}~ $u'_{ij}$ is downstream of $v_{ik}$, in which case $t_{ij}$ is not upstream of $w'_{jk}$ (i.e., there is no path from $t_{ij}$ to $w'_{jk}$.
\end{enumerate}
If we apply the above possibilities to each distinct $i,j\in[3]$ of Style A and Style B networks, we get all possible Style A and Style B networks, which we shall see henceforth.
\subsection{Style A networks}
\label{styleA}
\begin{figure}
\centering
  \includegraphics[width=0.9\linewidth]{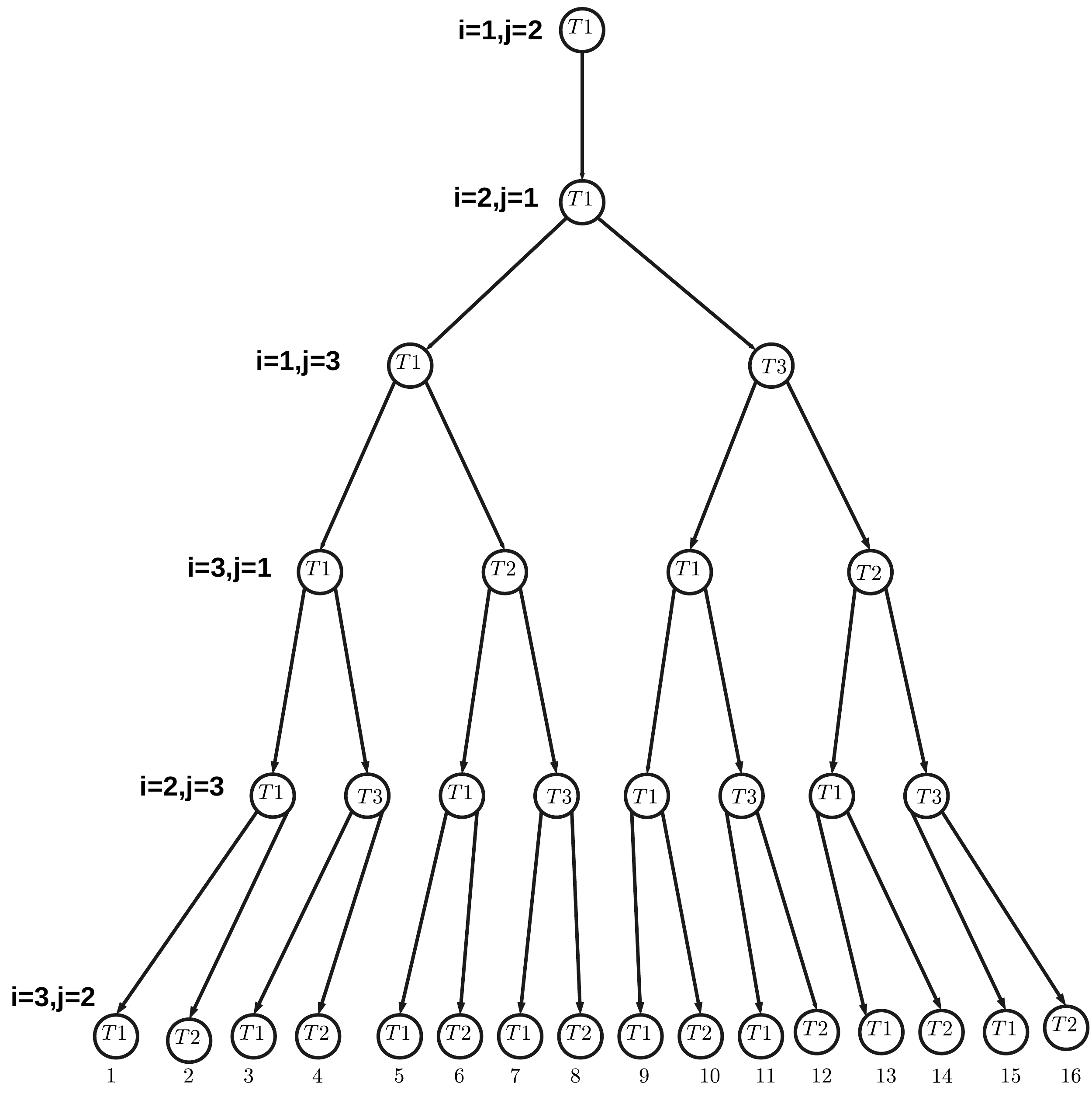}
  \caption{\small The sixteen configurations of Style A networks according to the possibilities of Lemma \ref{upstreamdownstreamlemma}, as the paths from root to leaves of this tree. $T1$ means a Type 1 crosspath, etc. Table \ref{tab1:reductionsstyleA} gives the details.}	
	\hrule
	\label{styleAallconfigs}
\end{figure}

Style A networks are based on skeleton network A shown in Fig. \ref{skeleton_style_A}. In this skeleton network A, we note the following. 
\begin{itemize}
\item The $12$-crosspath and $21$-crosspath must necessarily be of Type 1. Type 2 is not possible as $w'_{12}$ is downstream of $w'_{13}(=w'_{23})$. Type 3 is not possible as $v_{12}$ is upstream of $v_{13} (=v_{23})$.
\item For the $13$-crosspath, Type 2 is not possible as $t_{13}$ cannot be upstream of $w'_{12}$ as if $t_{13}\in w'_{13}\rightarrow w'_{12}$, then $t_{13}\rightarrow D_{3'}$ path cannot exist.
\item For the $31$-crosspath, Type 3 is not possible. This is because if $u'_{13}\in v_{12}\rightarrow v_{13}$ then $3\rightarrow u'_{13}$ path cannot exist. 
\item Similarly we have that for the $23$-crosspath Type 2 is not possible, and for the $32$-crosspath, Type 3 is not possible.
\end{itemize}
With the above constraints, we have 16 different configurations, as shown in Fig. \ref{styleAallconfigs} in the form of the paths from the root to the leaves of the tree. For instance, the Configuration 1 in Fig. \ref{styleAallconfigs} has the crosspath types Type 1 for all the crosspaths, while the Configuration 16 has Type 1 for $12$-crosspath, $21$-crosspath, Type 3 for $13$-crosspath and $23$-crosspath, and Type 2 for $31$-crosspath and $32$-crosspath.

\begin{figure*}[ht]
  \centering
  \vspace{0.0in}
  \subfigure[Final Config S21 of Style A]{\label{Aconfig21}\includegraphics[width=0.23\textwidth]{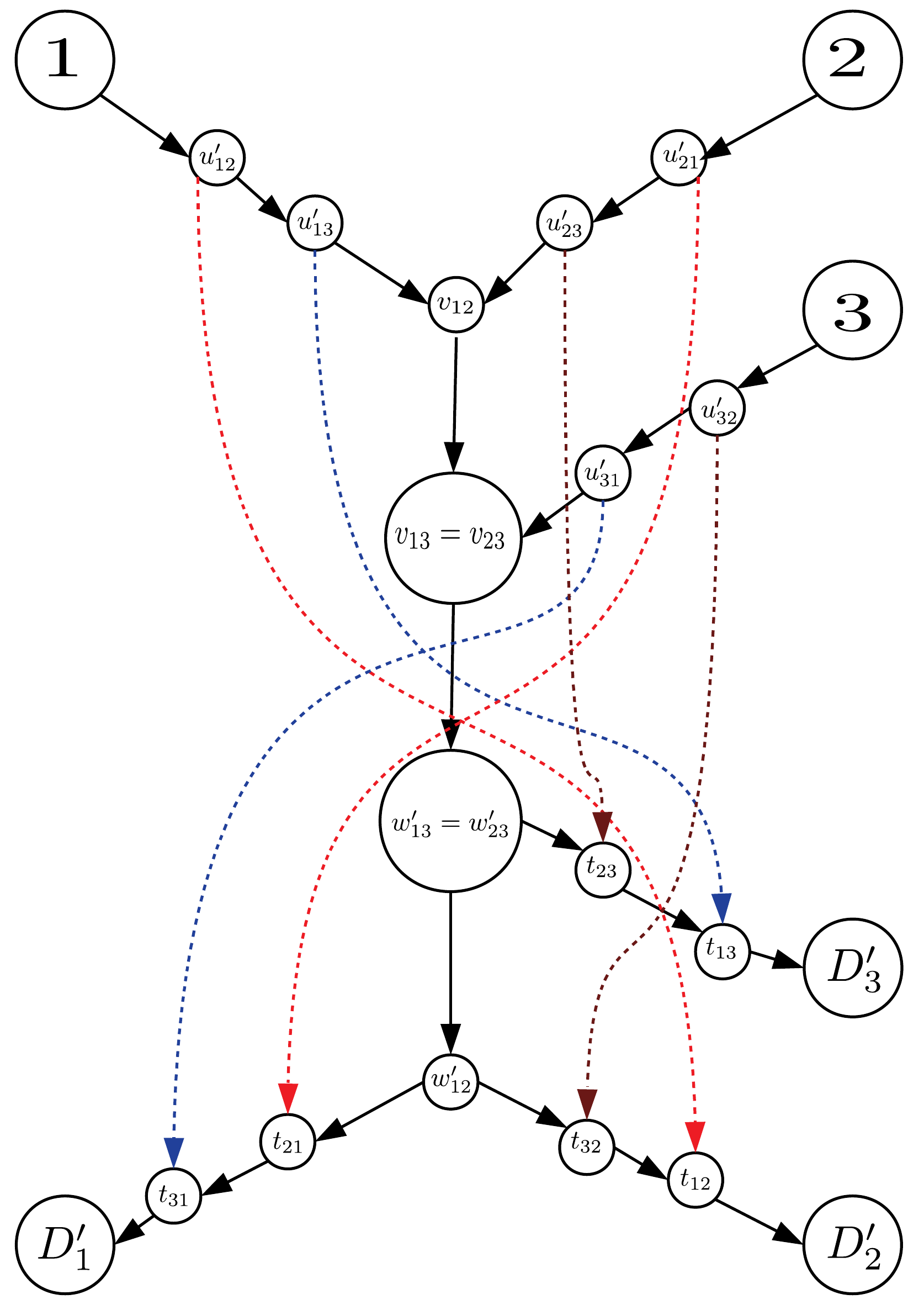}} 
\subfigure[Final Config S22 of Style A]{\label{Aconfig23}\includegraphics[width=0.23\textwidth]{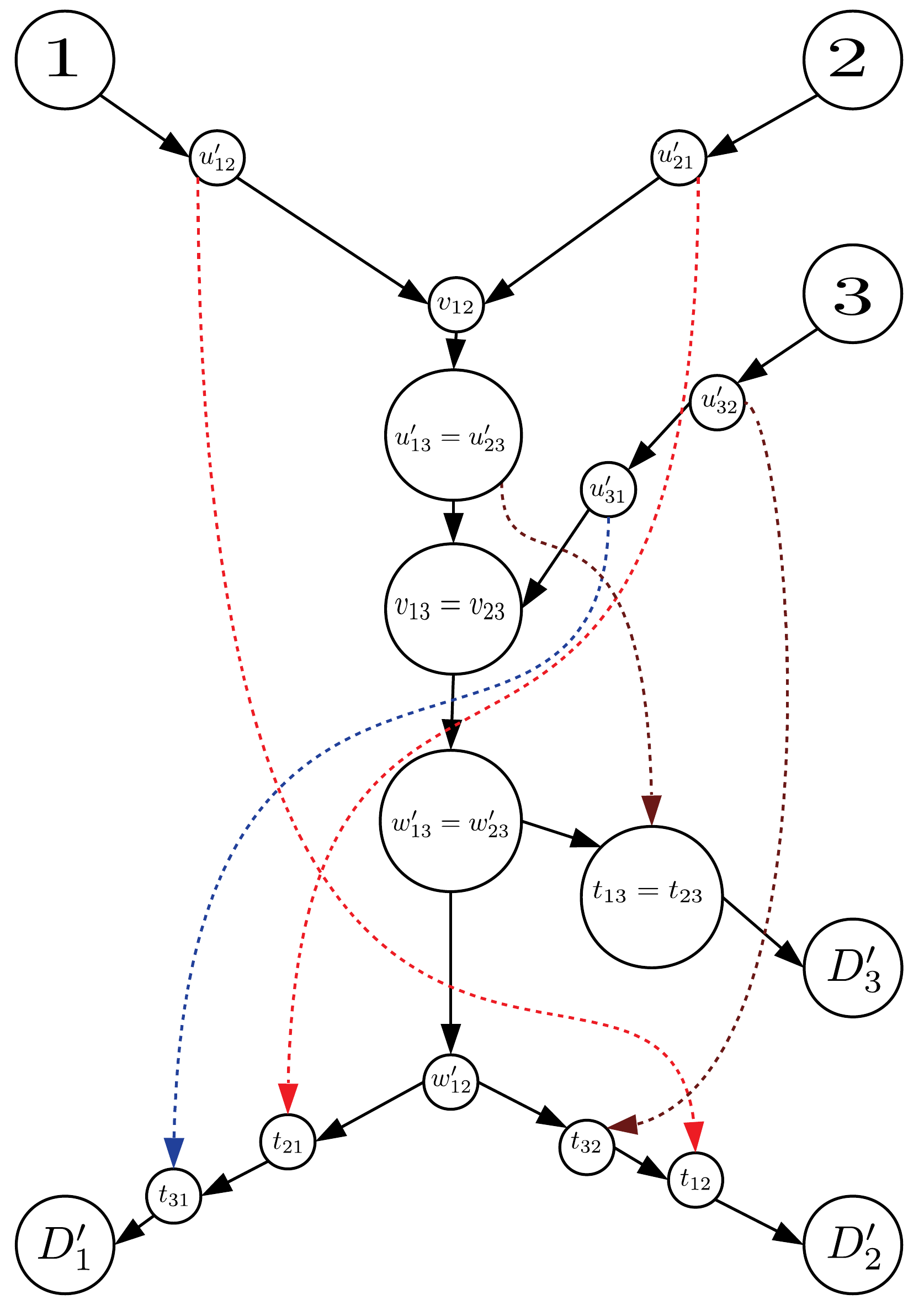}}
\subfigure[Final Config S23 of Style A]{\label{Aconfig22}\includegraphics[width=0.23\textwidth]{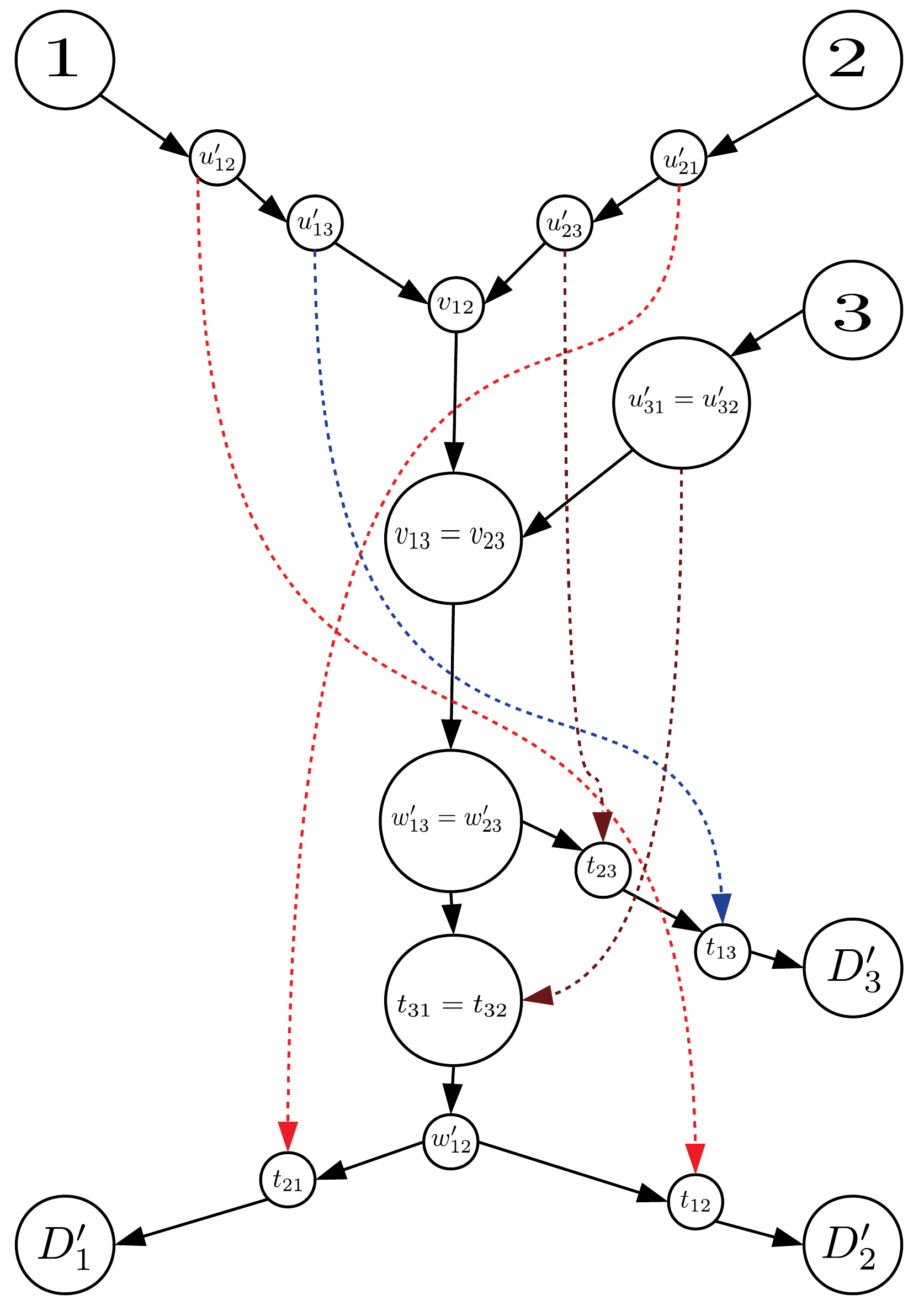}}
 \subfigure[Final Config S24 of Style A]{\label{Aconfig24}\includegraphics[width=0.23\textwidth]{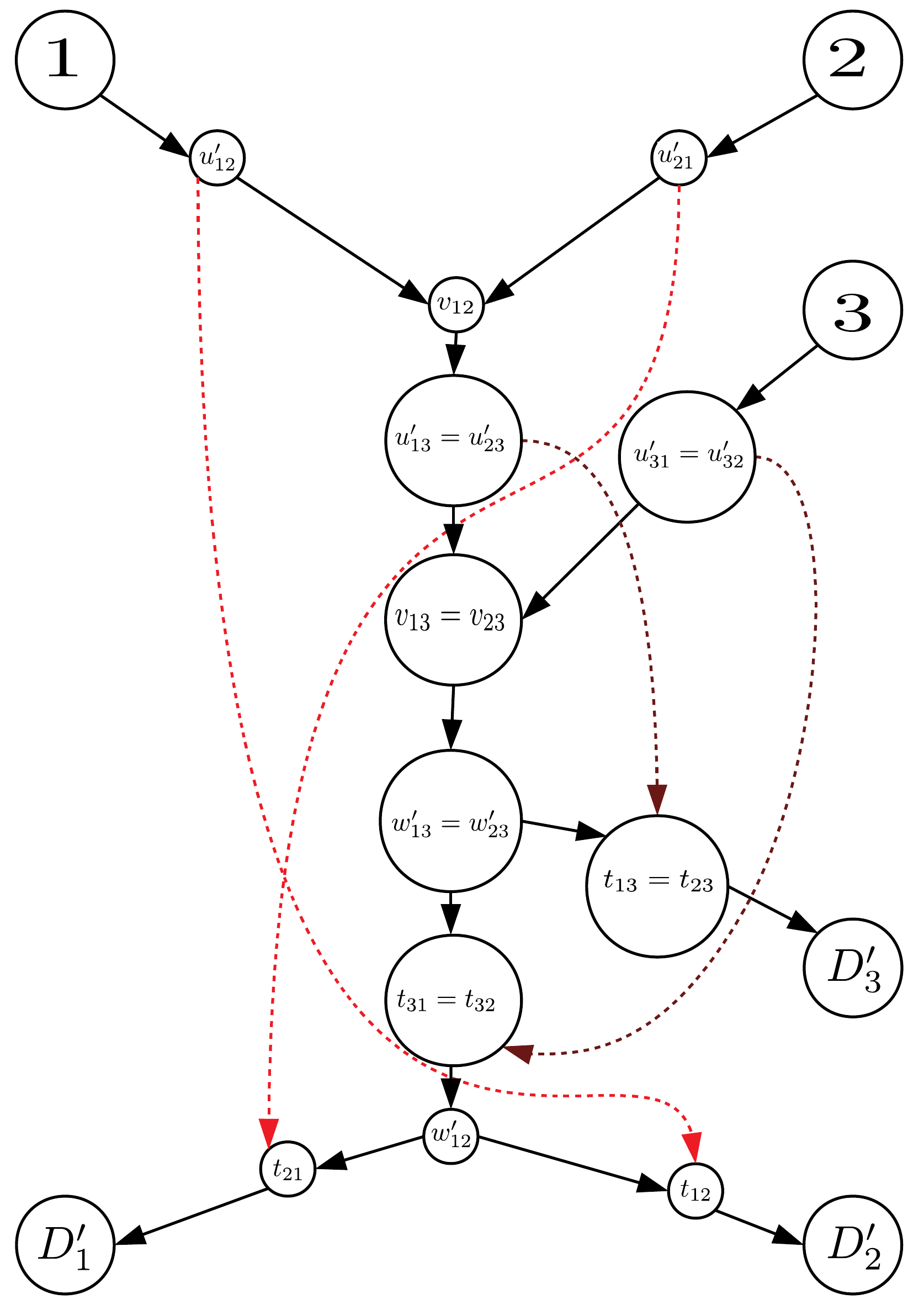}}
	\hrule
	\centering
	\caption{\small Final configurations of Style A}
	\label{styleAfinal}
		\captionsetup{justification=centering}
\end{figure*}

Now, if we examine these configurations carefully, we will notice that many of these configurations have more than one $ij$-crosspaths for some $i,j$. For instance, consider the Style A configuration 3 network shown in Fig. \ref{styleAexample}. In this configuration, $u'_{13}\rightarrow t_{13}$ is a valid $13$-crosspath, but so is $u'_{23}\rightarrow t_{23}$. Hence this configuration can be reduced to another configuration in which $u'_{13}\rightarrow t_{13}$ as shown in Fig. \ref{styleAexample} does not exist (without disturbing requirements of Class $\mathbb I$a conditions). We get this same reduced configuration from Configuration 11 of Fig. \ref{styleAallconfigs}, in which $13$-crosspath can be removed without disturbing Class $\mathbb I$a conditions. Thus, Configuration $3$ and $11$ boil down to the same reduced configuration, which we term as configuration S12 in Table \ref{tab1:reductionsstyleA}. The set of all such unique reduced configurations (by deleting extra crosspaths) are called the Stage I reduced configurations, with the new titles  S1j,~~$j\in[9]$. The Stage I reduced configurations and the deleted crosspaths to get Stage I reduced configurations are shown in column $2$ and $3$ of Table \ref{tab1:reductionsstyleA}. Furthermore, in Stage I reduced configurations, it can be observed that some of the crosspaths in two different configurations can be identified as one (because of the network topology) to reduce the configurations further. For instance, the $23$-crosspath of configuration S12 and the $13$-crosspath of S13 have the exact same structure (both are Type 3) and hence can be identified with each other. This gives only $4$ unique configurations among Stage I, which we call the Stage II reduced configurations, shown in Fig. \ref{styleAfinal}. These duplicate crosspaths and the new labels for the final $4$ unique configurations in Style A are given in the last two columns of Table \ref{tab1:reductionsstyleA}. The complete figures for the 16 original configurations and their reductions are shown in the end of this paper (page 11 onwards).

\begin{table*}[htbp]
\centering
\begin{tabular}{|c|c|c|c|c|}
\hline
\textbf{Stage I Reduction:}       & \textbf{New title for}   & \textbf{Extra $ij$-crosspaths}  & \textbf{Duplicate Crosspaths}    & \textbf{New title}    \\
\textbf{Original Configs clubbed} & \textbf{Stage I}         & \textbf{deleted from original}  & \textbf{identified}              & \textbf{for Stage II} \\
\textbf{as one by deleting}       & \textbf{Reduced Configs} & \textbf{configs to get Stage I} & \textbf{as one to}               & \textbf{Reduced}      \\
\textbf{extra crosspaths}         & \textbf{}                & \textbf{reduced configs}        & \textbf{get Stage II}            & \textbf{Configs}      \\ \hline
Configuration 1                   & Config. S11              & (NA: Not applicable)            & NA                               & Config. S21           \\ \hline
3,11                              & S12                      & $13$-crosspath from Config 3,   & $23$ crosspath of  S12 and       &                       \\
                                  &                          & $13$-crosspath from Config 11   & $13$ crosspath from S13          & S22                   \\ \cline{1-3}
9                                 & S13                      & $23$-crosspath                  &                                  &                       \\ \hline
2,6                               & S14                      & $31$-crosspaths from            & $32$ crosspath of S14            &                       \\
                                  &                          & both configs                    & and $31$ crosspath from S15      & S23                   \\ \cline{1-3}
5                                 & S15                      & $32$-crosspath                  &                                  &                       \\ \hline
4                                 & S16                      & $13$ and $31$ crosspaths.       &                                  &                       \\ \cline{1-3}
7,8,15,16                         & S17                      & $13$ and $32$ crosspaths        &                                  &                       \\
                                  &                          & from all configs.               & $32$ crosspath of S16 and S18,   &                       \\ \cline{1-3}
10,12,14                          & S18                      & $31$ and $23$ crosspaths        & $31$ crosspath from S17 and S19. & S24                   \\
                                  &                          & in all configs                  &                                  &                       \\ \cline{1-3}
13                                & S19                      & $23$ and $32$ crosspaths        &                                  &                       \\ \hline
\end{tabular}
~\\\caption{Reductions of Style A network configurations. From the original 16, we get 9 configurations in Stage I reductions (first three columns), by deleting extra crosspaths without disturbing Class $\mathbb I$a conditions. The title for the Stage I reduced configurations are shown in column 2. The last two columns give the Stage II reductions.}
\hrule
\label{tab1:reductionsstyleA}
\end{table*}
Finally, we have the following proposition regarding the feasible network codes for Style A configurations.
\begin{proposition}
\label{mainthmStyleA}
Style A networks (which have to be in one of the 4 configurations shown in Fig. \ref{styleAfinal}) have feasible binary linear network codes.
\end{proposition}
\begin{IEEEproof}
We proceed by assigning valid global encoding vectors for the main subpaths in each final Stage II configuration of Fig. \ref{styleAfinal}. All other local encoding coefficients are assumed to be $1$. The reader can easily check that these assignments give a feasible network code. 

\textit{Configuration S21 (Fig. \ref{Aconfig21}):} The subpath $v_{12}\rightarrow v_{13}$ gets $(1~1~0)^T$, and subpath $v_{13}\rightarrow w'_{12}$ gets $(1~1~1)^T$. For $i\in[3]$, the crosspaths originating from $i$ get the $i^{th}$ standard basis vector.

\textit{Configuration S22 (Fig. \ref{Aconfig22}):} The subpath $v_{12}\rightarrow v_{13}$ gets $(1~1~0)^T$, and subpath $v_{13}\rightarrow w'_{12}$ gets $(1~1~1)^T$.	 The crosspaths originating from $3$ get $(0~0~1)^T$, the $12$-crosspath gets $(1~0~0)^T$, the $21$-crosspath gets $(0~1~0)^T$. Finally, the $13$-crosspath gets $(1~1~0)^T$. 

\textit{Configuration S23 (Fig. \ref{Aconfig23}):} The subpath $v_{12}\rightarrow v_{13}$ gets $(1~1~0)^T$, the subpath $v_{13}\rightarrow t_{31}$ gets $(1~1~1)^T$, and $t_{31}\rightarrow w'_{12}$ gets $(1~1~0)^T$. The crosspaths originating from $i$ get the $i^{th}$ standard basis vector.

\textit{Configuration S24 (Fig. \ref{Aconfig24}):} The subpath $v_{12}\rightarrow v_{13}$ gets $(1~1~0)^T$, the subpath $v_{13}\rightarrow t_{31}$ gets $(1~1~1)^T$, and $t_{31}\rightarrow w'_{12}$ gets $(1~1~0)^T$. the $12$-crosspath gets $(1~0~0)^T$, the $21$-crosspath gets $(0~1~0)^T$, the $13$-crosspath gets $(1~1~0)^T$, and $u'_{31}\rightarrow t_{31}$ is $(0~0~1)^T$.
\end{IEEEproof}
\subsection{Style B networks}
\label{styleB}
\begin{figure}
\centering
  \includegraphics[width=0.7\linewidth]{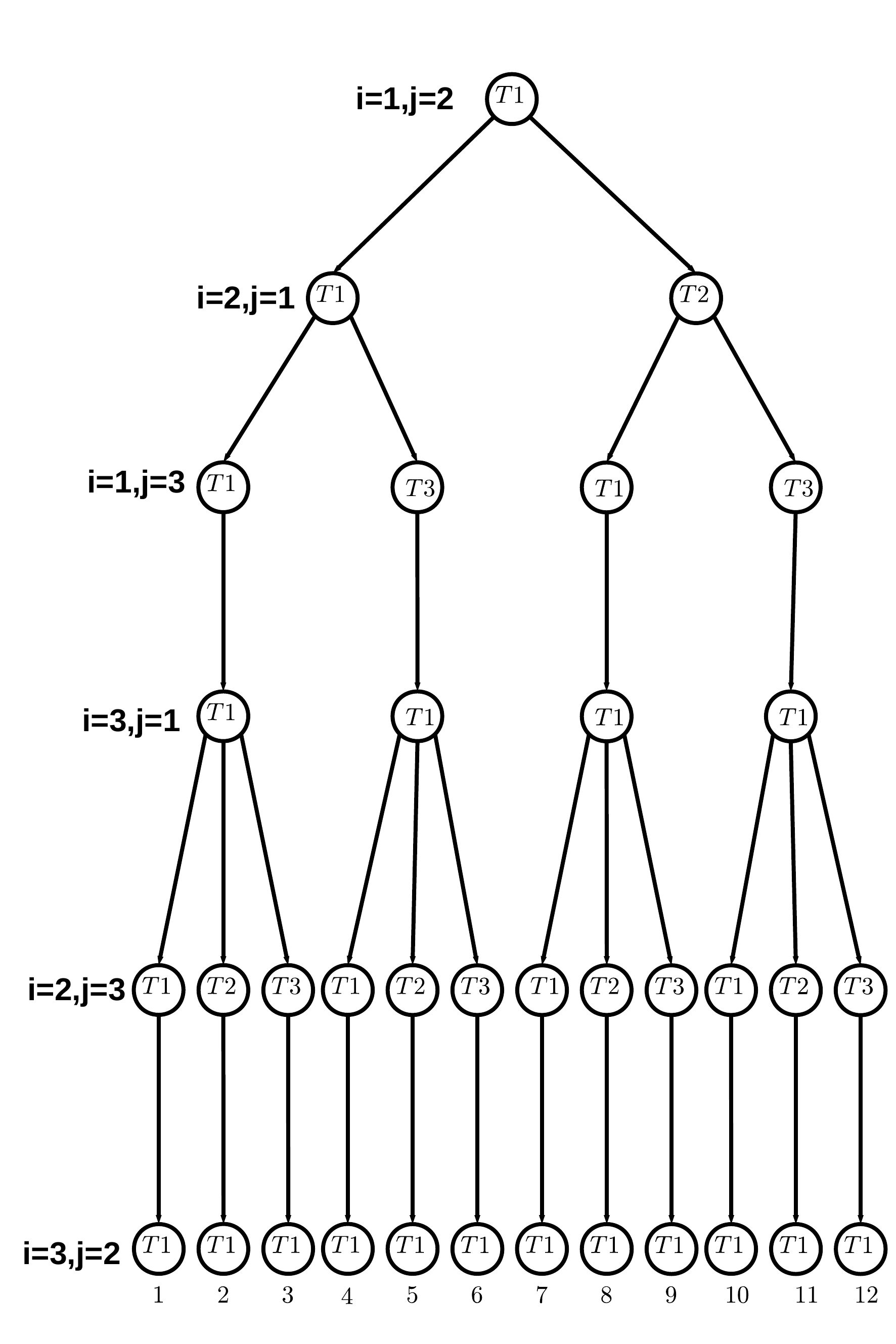}
  \caption{\small The twelve configurations of Style B networks according to the possibilities of Lemma \ref{upstreamdownstreamlemma}, shown as the paths from the root to the leaves of this tree.}	
	\label{styleBallconfigs}
\end{figure}

\begin{figure}[ht]
  \centering
  \vspace{0.0in}
\subfigure[Configuration 10]{\label{Bconfig10}\includegraphics[width=0.23\textwidth]{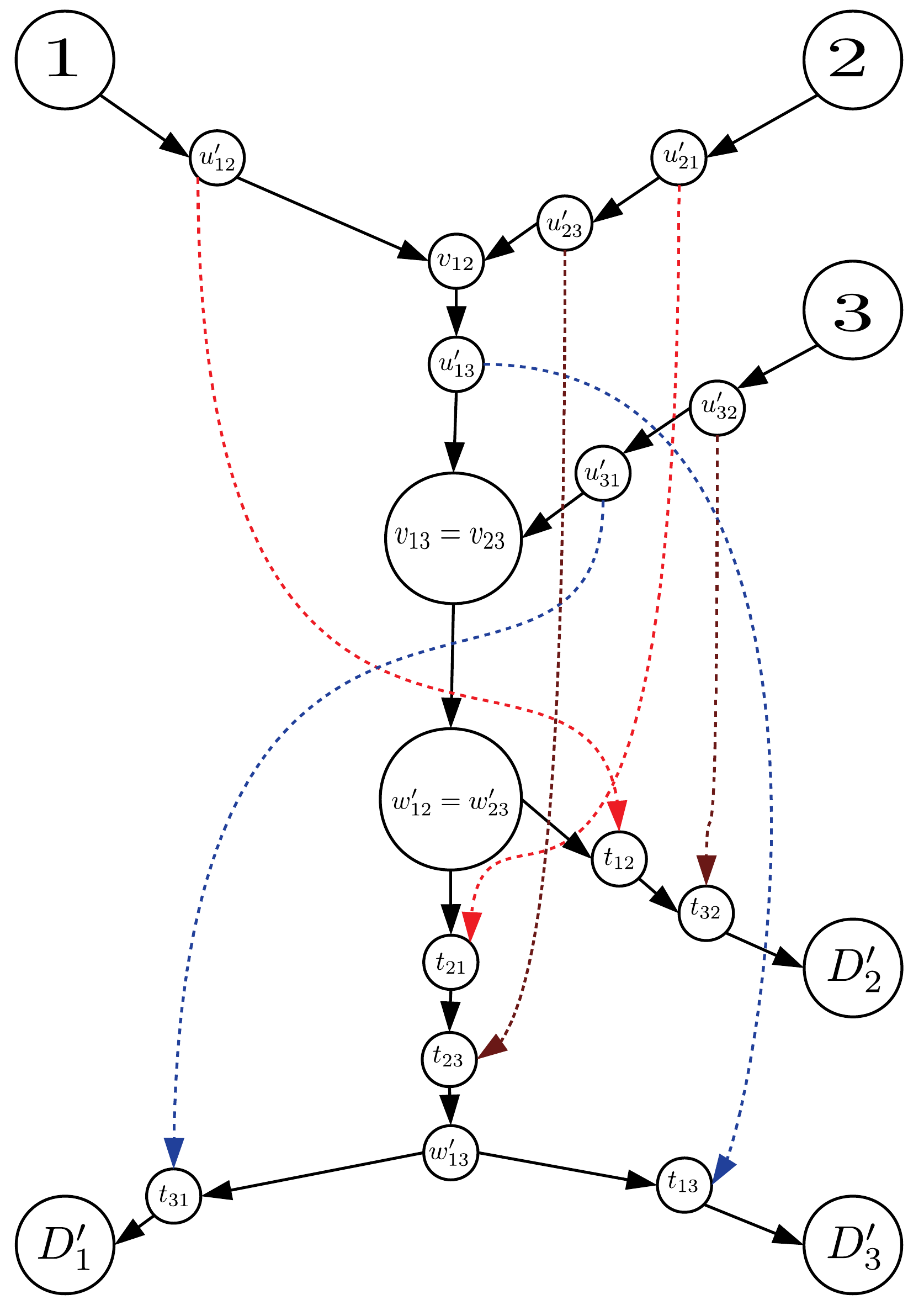}}
 \subfigure[Configuration 12]{\label{Bconfig12}\includegraphics[width=0.23\textwidth]{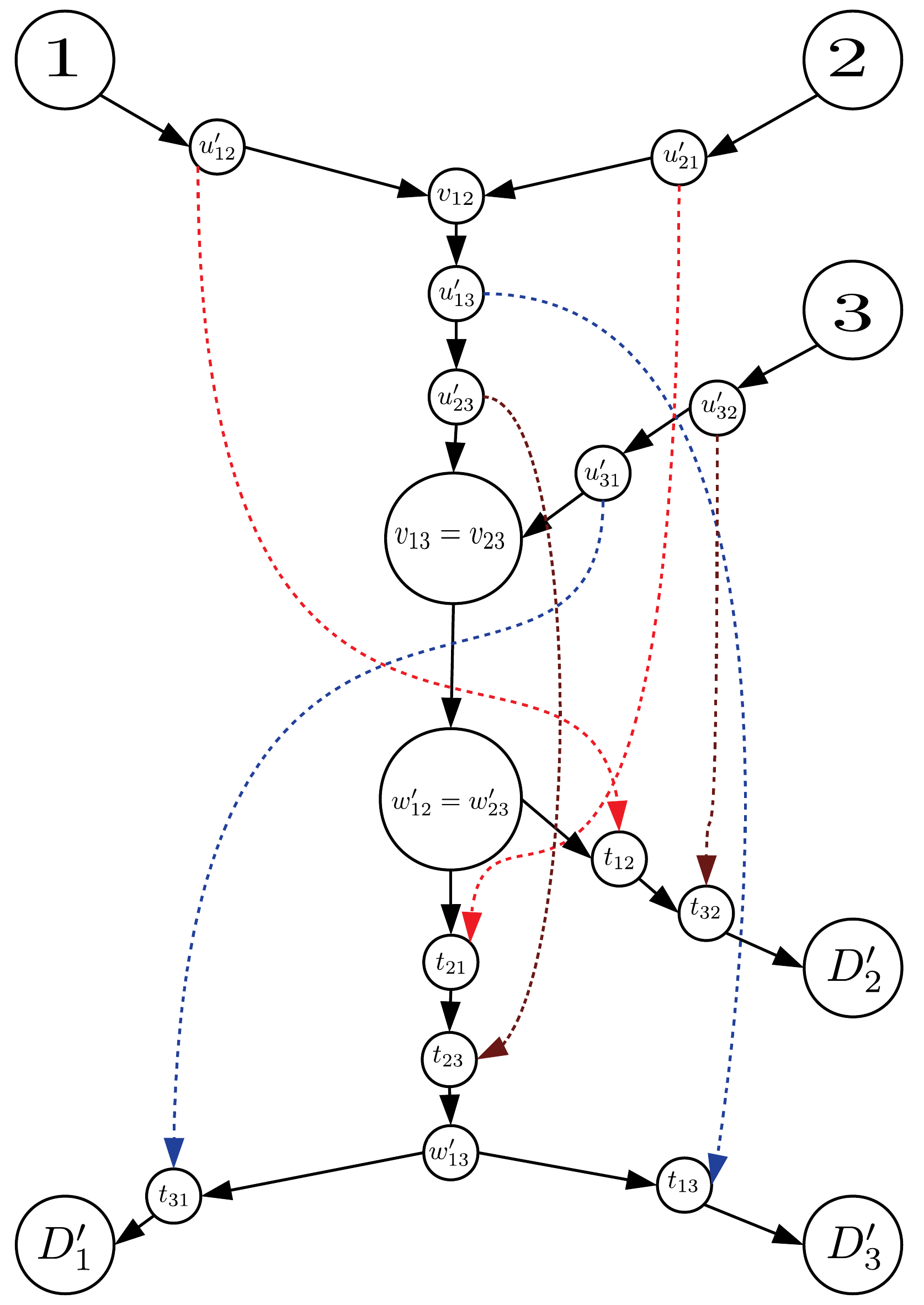}}
	\hrule
	\centering
	\caption{\small Illegitimate ($\tau=2$) configurations of Style B. To ensure that $\tau=3$ holds, addition crosspaths must exist, in which case we will get other legitimate configurations}
	\label{illegalStyleB}
		\captionsetup{justification=centering}
\end{figure}

We now move to Style B networks which have the skeleton network B as the basic subgraph as shown in Fig. \ref{skeleton_style_B}. By a similar approach to Style A networks, we get $12$ configurations, represented by the paths from the root to the leaves in Fig. \ref{styleBallconfigs}. Thus, a Style B network has to necessarily be in one of the $12$ configurations. However, unlike Style A, there exist \textit{illegitimate} configurations, i.e., configurations which do not satisfy the condition of $\tau=3$. Among the $12$, Configurations $10$ and $12$, shown in Fig. \ref{illegalStyleB} have $\tau=2$. In Configuration 10 (Fig. \ref{Bconfig10}), note that removing $u'_{13}$ and $t_{23}$ is sufficient to disconnect the all unipaths (corresponding to unicycles of $G_{SI}$) and one of each crosspath-pair of $G_{NC}$ (a crosspath pair would be $ij$-crosspath and a $ji$-crosspath, corresponding to a $ij$-bicycle of $G_{SI}$), which in other words means that removing $u_{13}$ and $t_{23}$ from $G_{SI}$ is sufficient to make $G_{SI}$ acyclic. Thus $\tau=2$ for Configuration $10$. Similarly, one can observe that removing $u_{13}$ and $t_{23}$ is sufficient to make $G_{SI}$ acyclic, if the corresponding $G_{NC}$ is in configuration $12$ (Fig. \ref{Bconfig12}). The illegitimate configurations are an artifact of the result in Lemma \ref{allcrosspaths} which says that the existence of the crosspaths is only a necessary condition for $\tau=3$, not a sufficient condition. Thus, in order to `legitimize' Configurations 10 and 12, we would have to consider that other crosspaths than those considered already must exist so as to make the condition $\tau=3$ valid again. It is not difficult to see from a pictorial argument that introducing these new crosspaths will result in Configurations 10 and 12 becoming equivalent to some of the other $10$ legitimate configurations of Style B (we omit the proof of this claim here). 

\begin{figure*}[htbp]
  \centering
  \vspace{0.0in}
  \subfigure[Final Config S21 of Style B]{\label{Bconfig21}\includegraphics[width=0.23\textwidth]{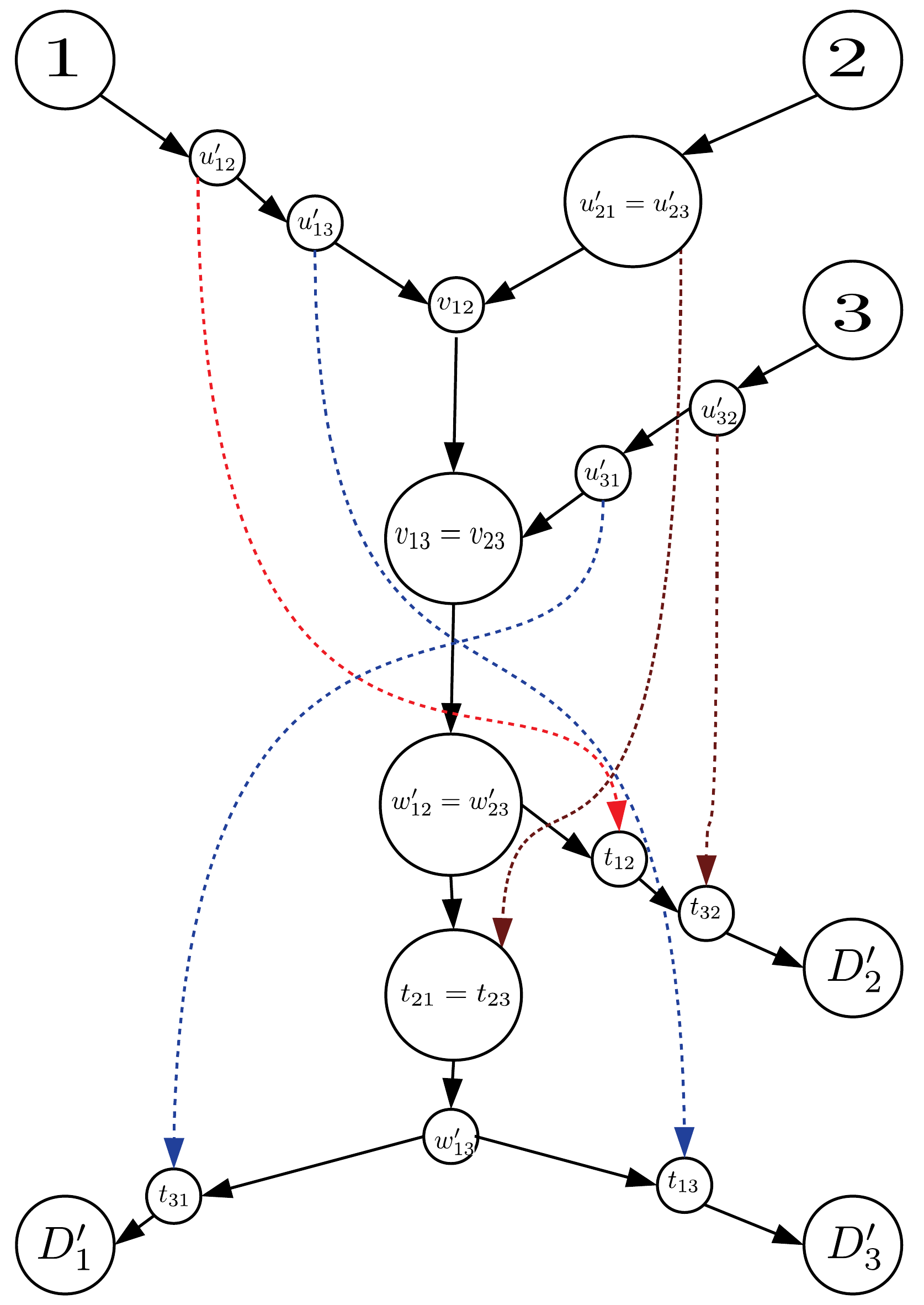}} 
\subfigure[Final Config S22 of Style B]{\label{Bconfig23}\includegraphics[width=0.23\textwidth]{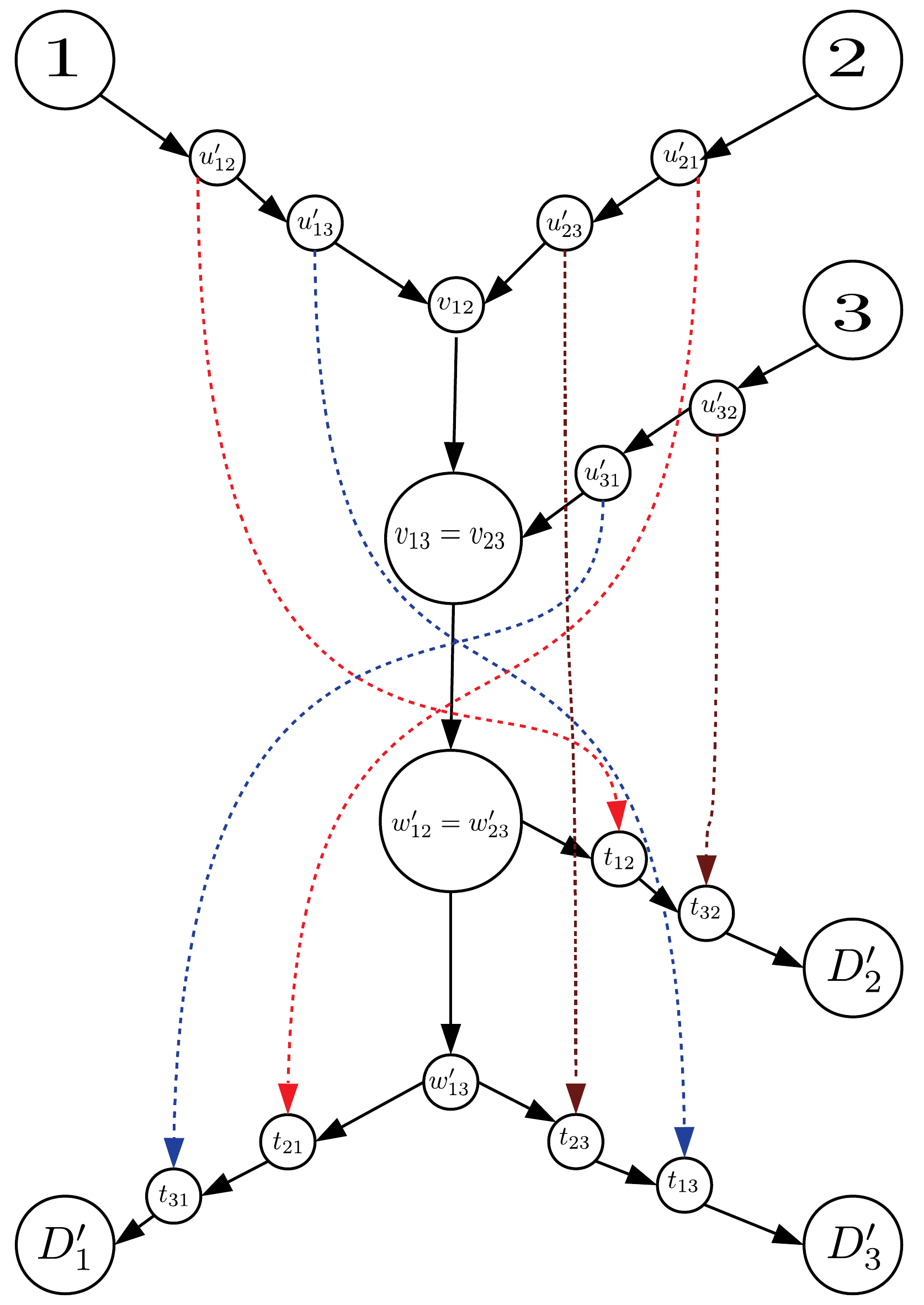}}
\subfigure[Final Config S23 of Style B]{\label{Bconfig22}\includegraphics[width=0.23\textwidth]{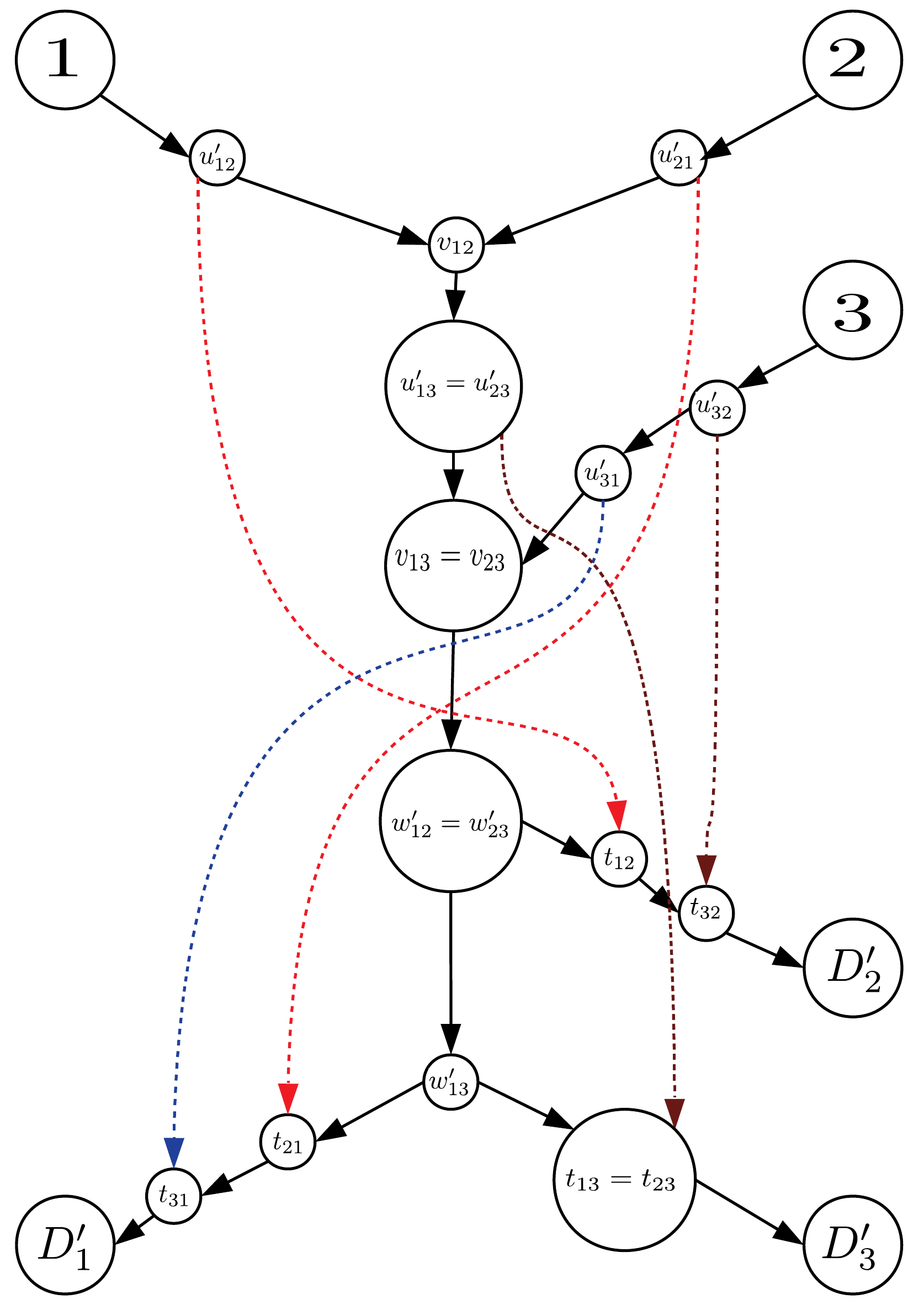}}
 \subfigure[Final Config S24 of Style B]{\label{Bconfig24}\includegraphics[width=0.23\textwidth]{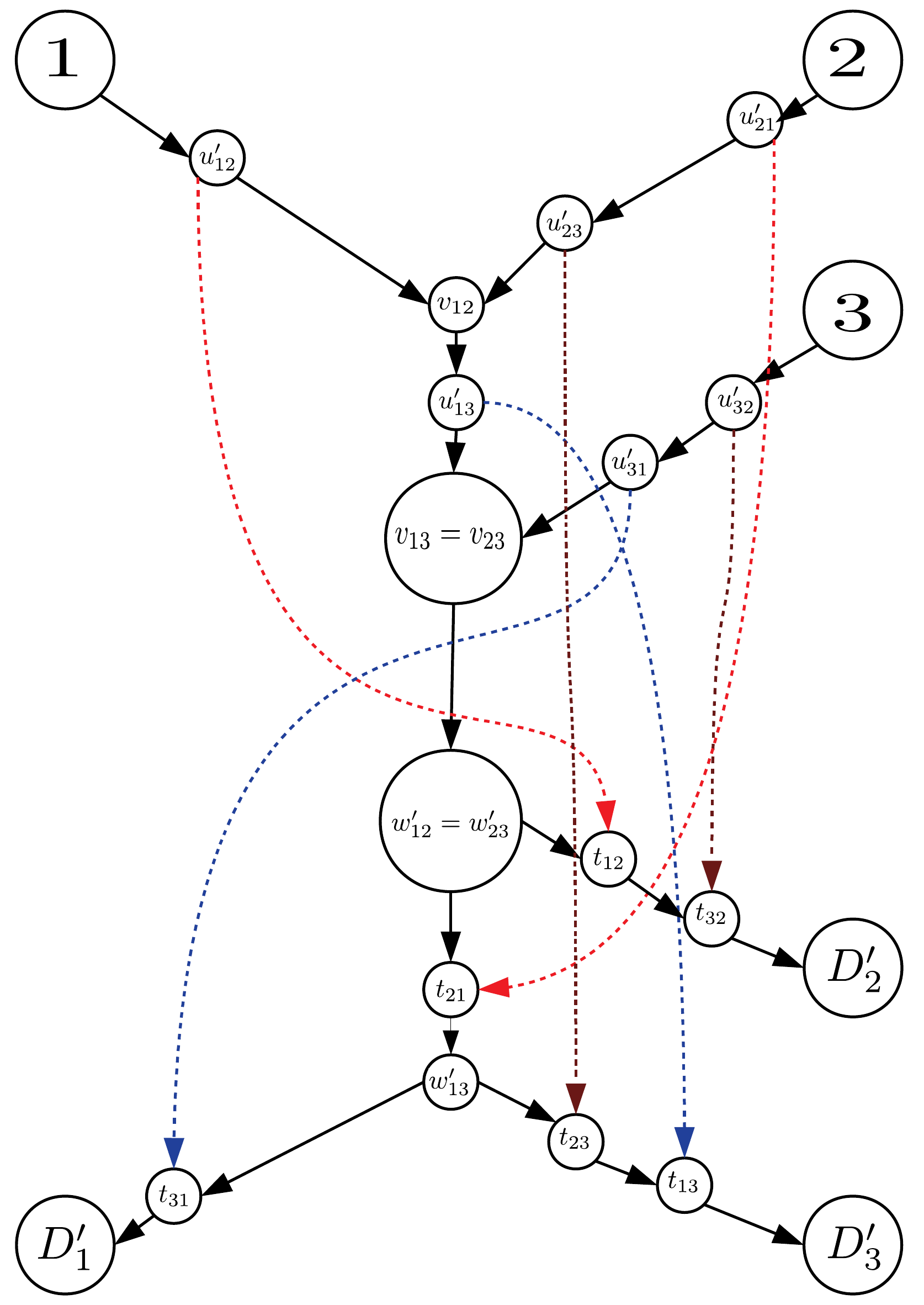}}
	\centering
	\caption{\small Final configurations of Style B}
	\label{styleBfinal}
		\captionsetup{justification=centering}
		\hrule
\end{figure*}

Hence we proceed ahead using the following simple lemma which we have established through above arguments.
\begin{lemma}
Each $G_{NC}$ network in Style B must contain at least one of the configurations in Fig. \ref{styleBallconfigs} amongst Configurations 1-9 and Configuration 11.
\end{lemma}
Thus it is sufficient to establish network coding scheme for these Configurations. As with Style A, Table \ref{tab:styleBreduced} gives the reduction of these 10 original Style B configurations based on deleting additional crosspaths (Reduction Stage I, as given by first 3 columns of Table \ref{tab:styleBreduced}) and further identifying crosspaths in different configurations as one (last 2 columns of Table \ref{tab:styleBreduced}).  Fig. \ref{styleBfinal} shows the four final valid Style B configurations. Complete figures for all configurations of Style B and their reductions are available at the end of this paper.

The following proposition shows that each of the four final Style B configurations have valid 3-unicast network codes.
\begin{proposition}
\label{mainthmStyleB}
Style B networks (which have to be in one of the 4 configurations shown in Fig. \ref{styleBfinal}) have feasible binary linear network codes.
\end{proposition}
\begin{IEEEproof}
We proceed the same way as for Style A, by assigning valid global encoding vectors for the main subpaths in each configuration of Fig. \ref{styleBfinal}. All other local encoding coefficients are assumed to be $1$. It is left to the reader to check that these assignments give a feasible network code for style B configurations. 

\textit{Configuration S21 (Fig. \ref{Bconfig21}):} The subpath $v_{12}\rightarrow v_{13}$ gets $(1~1~0)^T$, and subpath $v_{13}\rightarrow t_{21}$ gets $(1~1~1)^T$. The crosspaths originating from $3$ get $(0~0~1)^T$, while those originating from vertex $1$ get $(1~0~0)^T$, and the $21$-crosspath gets $(0~1~0)^T$. The Subpath $t_{21}\rightarrow w'_{13}$ gets $(1~0~1)^T$.

\textit{Configuration S22 (Fig. \ref{Bconfig23}):} The subpath $v_{12}\rightarrow v_{13}$ gets $(1~1~0)^T$, and subpath $v_{13}\rightarrow w'_{13}$ gets $(1~1~1)^T$. For $i\in[3]$, the crosspaths originating from $i$ get the $i^{th}$ standard basis vector.

\textit{Configuration S23 (Fig. \ref{Bconfig22}):} The subpath $v_{12}\rightarrow v_{13}$ gets $(1~1~0)^T$, and subpath $v_{13}\rightarrow w'_{13}$ gets $(1~1~1)^T$.	 The crosspaths originating from $3$ get $(0~0~1)^T$, the $12$-crosspath gets $(1~0~0)^T$, the $21$-crosspath gets $(0~1~0)^T$. Finally, The Subpath $u'_{13}\rightarrow t_{13}$ gets $(1~1~0)^T$.

\textit{Configuration S24 (Fig. \ref{Bconfig24}):} The subpath $v_{12}\rightarrow v_{13}$ gets $(1~1~0)^T$, the subpath $v_{13}\rightarrow t_{21}$ gets $(1~1~1)^T$, and $t_{21}\rightarrow w'_{13}$ gets $(1~0~1)^T$. the $12$-crosspath gets $(1~0~0)^T$,the crosspaths starting from $3$ get $(0~0~1)^T$,the crosspaths starting from $2$ get $(0~1~0)^T$, and the subpath $u'_{13}\rightarrow t_{13}$ is $(1~1~0)^T$.

\end{IEEEproof}
\subsection*{Proof of Theorem \ref{maintheorem}}
Using Proposition \ref{mainthmStyleA} and Proposition \ref{mainthmStyleB}, Theorem \ref{maintheorem} follows. It is also easy to check that all the index codes obtained in this paper can be extended to non-binary fields. 
\section{Discussion}
In this work, we have shown that for a class of single-unicast IC problems with $n$ messages and $\tau=3$ (i.e., $MAIS=n-\tau$), we have $\beta_2=\beta_q=n-3$. This is in contrast with known IC problems \cite{BKL1} with $\tau=3$ for which $\beta_2>n-3$. The precise subclasses of IC problems with $\tau=3$ which have $\beta_q=n-3$, and those for which $\beta_q>n-3$ are still to be found. We are currently in the process of identifying optimal broadcast rate and constructing codes for more such subclasses. We also believe that the IC-NC duality approach can be used to find or bound optimal broadcast rates for other classes of IC problems. 
\begin{table*}[htbp]
\centering
\caption{Reduced Configurations of Style B in Stage I and Stage II (final)}
\label{tab:styleBreduced}
\begin{tabular}{|c|cccc|}
\hline
\textbf{Stage I Reduction:}       & \multicolumn{1}{c|}{\textbf{New title for}}   & \multicolumn{1}{c|}{\textbf{Extra $ij$-crosspaths}}  & \multicolumn{1}{c|}{\textbf{Duplicate Crosspaths}} & \textbf{New title}    \\
\textbf{Original Configs clubbed} & \multicolumn{1}{c|}{\textbf{Stage I}}         & \multicolumn{1}{c|}{\textbf{deleted from original}}  & \multicolumn{1}{c|}{\textbf{identified}}           & \textbf{for Stage II} \\
\textbf{as one by deleting}       & \multicolumn{1}{c|}{\textbf{Reduced Configs}} & \multicolumn{1}{c|}{\textbf{configs to get Stage I}} & \multicolumn{1}{c|}{\textbf{as one to}}            & \textbf{Reduced}      \\
\textbf{extra crosspaths}         & \multicolumn{1}{c|}{\textbf{}}                & \multicolumn{1}{c|}{\textbf{reduced configs}}        & \multicolumn{1}{c|}{\textbf{get Stage II}}         & \textbf{Configs}      \\ \hline
1                                 & \multicolumn{1}{c|}{S11}                      & \multicolumn{1}{l|}{$21$-crosspath from Config 1}    & \multicolumn{1}{l|}{$23$ crosspath of S11 and}     &                       \\ \cline{1-2}
7,8,9                             & \multicolumn{1}{c|}{S15}                      & \multicolumn{1}{c|}{$23$-crosspath from all}         & \multicolumn{1}{l|}{$21$ crosspath from S15}       & S21                   \\
\multicolumn{1}{|l|}{}            & \multicolumn{1}{l|}{}                         & \multicolumn{1}{c|}{three configs}                   & \multicolumn{1}{l|}{}                              & \multicolumn{1}{l|}{} \\ \hline
2                                 & \multicolumn{1}{c|}{S12}                      & \multicolumn{1}{c|}{(NA: Not applicable)}            & \multicolumn{1}{c|}{NA}                            & S22                   \\ \hline
3                                 & \multicolumn{1}{c|}{S13}                      & \multicolumn{1}{c|}{$13$-crosspath from Config 3}    & \multicolumn{1}{c|}{$23$ crosspath of S13 and}     &                       \\ \cline{1-3}
4,5,6                             & \multicolumn{1}{c|}{S14}                      & \multicolumn{1}{c|}{$23$-crosspath from all}         & \multicolumn{1}{c|}{$13$ crosspath from S14}       & S23                   \\
                                  & \multicolumn{1}{c|}{}                         & \multicolumn{1}{c|}{three configs}                   & \multicolumn{1}{c|}{}                              &                       \\ \hline
11                                & \multicolumn{1}{c|}{S16}                      & \multicolumn{1}{c|}{NA}                              & \multicolumn{1}{c|}{NA}                            & S24                   \\ \hline
10                                &                                               & illegitmate as we get                         &                                                    &                       \\ \cline{1-1}
12                                &                                               & $\tau$=2                                &                                                    &                       \\ \hline
\end{tabular}
\end{table*}

\newpage
\begin{figure*}[htbp]
  \centering
 \subfigure[Network style A original Config 1 (same for reduced configs S11 and S21)]{\includegraphics[height=3.2in]{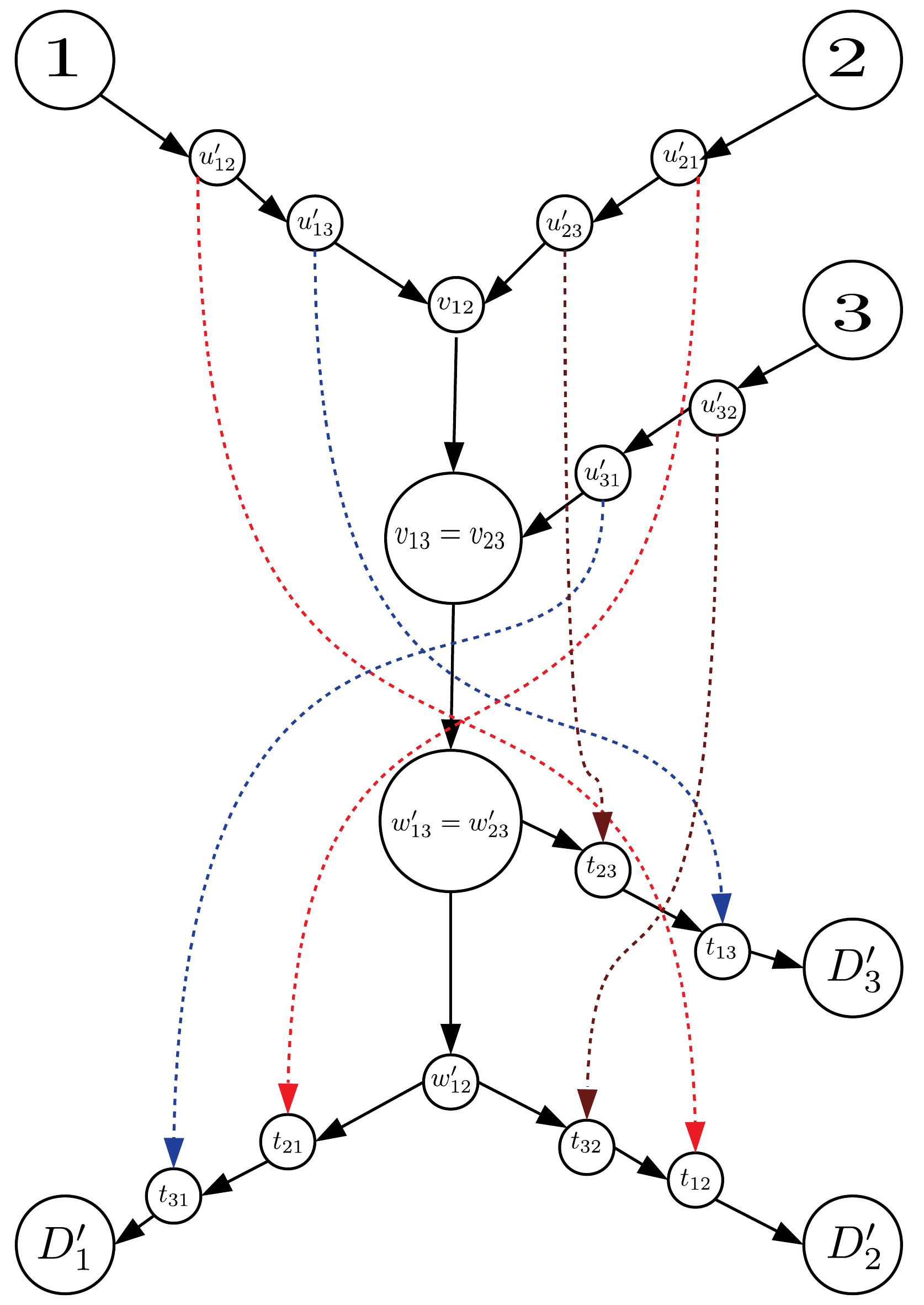}
  \label{network style A-1}}
	\hrule
	\caption{}
    \end{figure*}
\begin{figure*}[htbp]
\centering
  \subfigure[Network style A original Config  2]{\label{network style A-2}\includegraphics[height=2.0in]{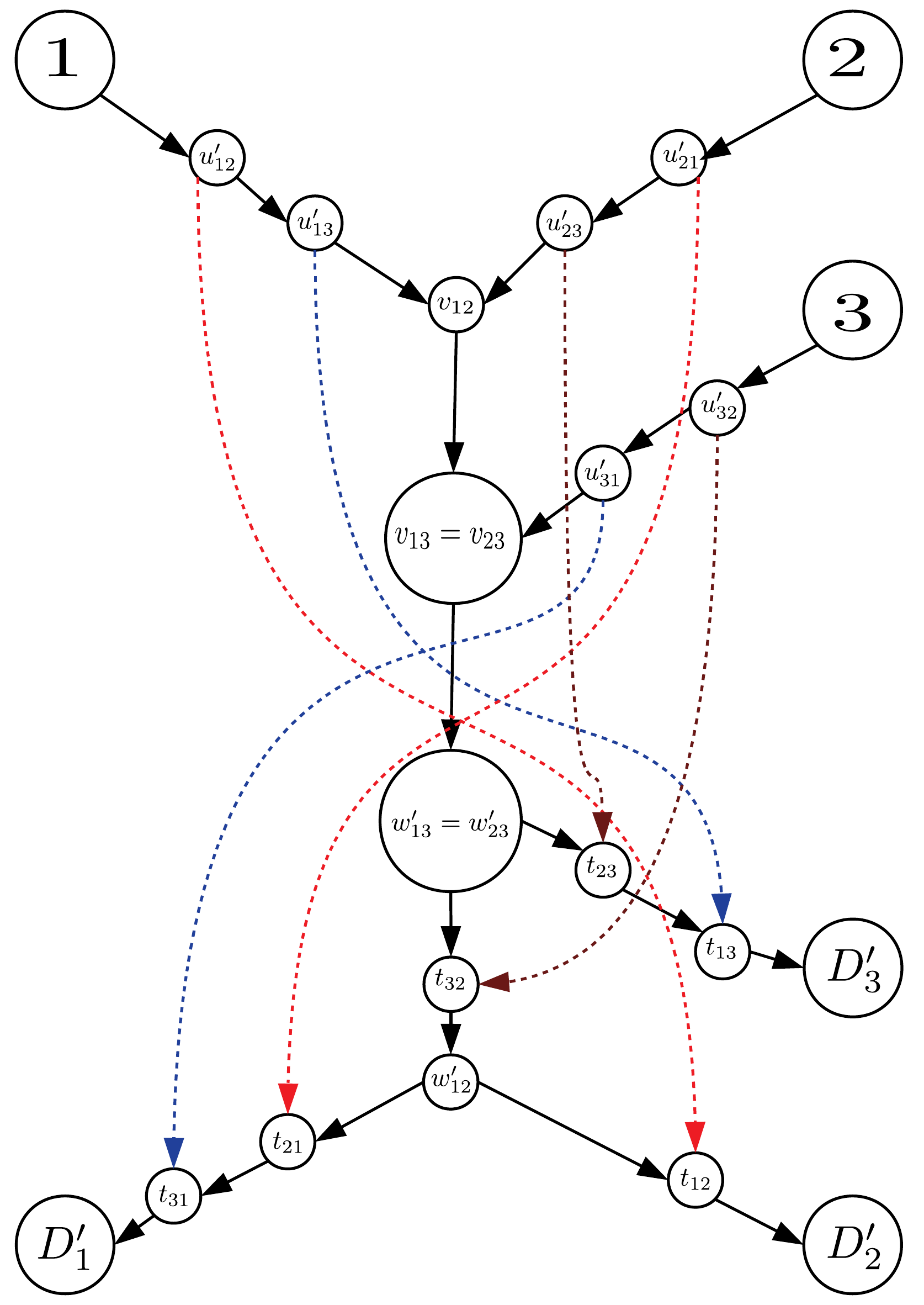}} \hspace{0.2in}
  \subfigure[Network style A original Config  6]{\label{network style A-5}\includegraphics[height=2.0in]{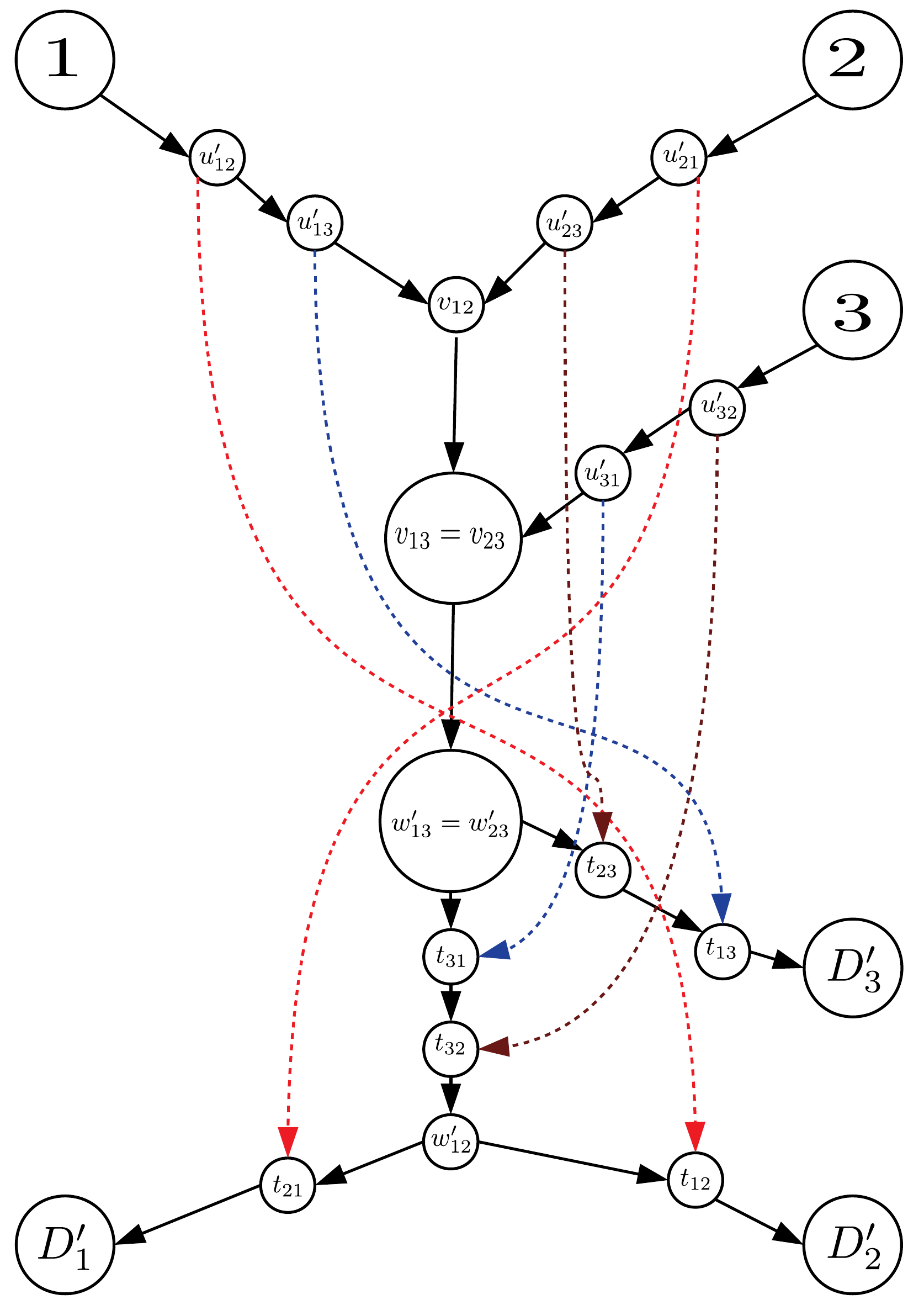}}
  \hspace{0.2in}
  \subfigure[Network style A original Config  5]{\label{network style A-6}\includegraphics[height=2.0in]{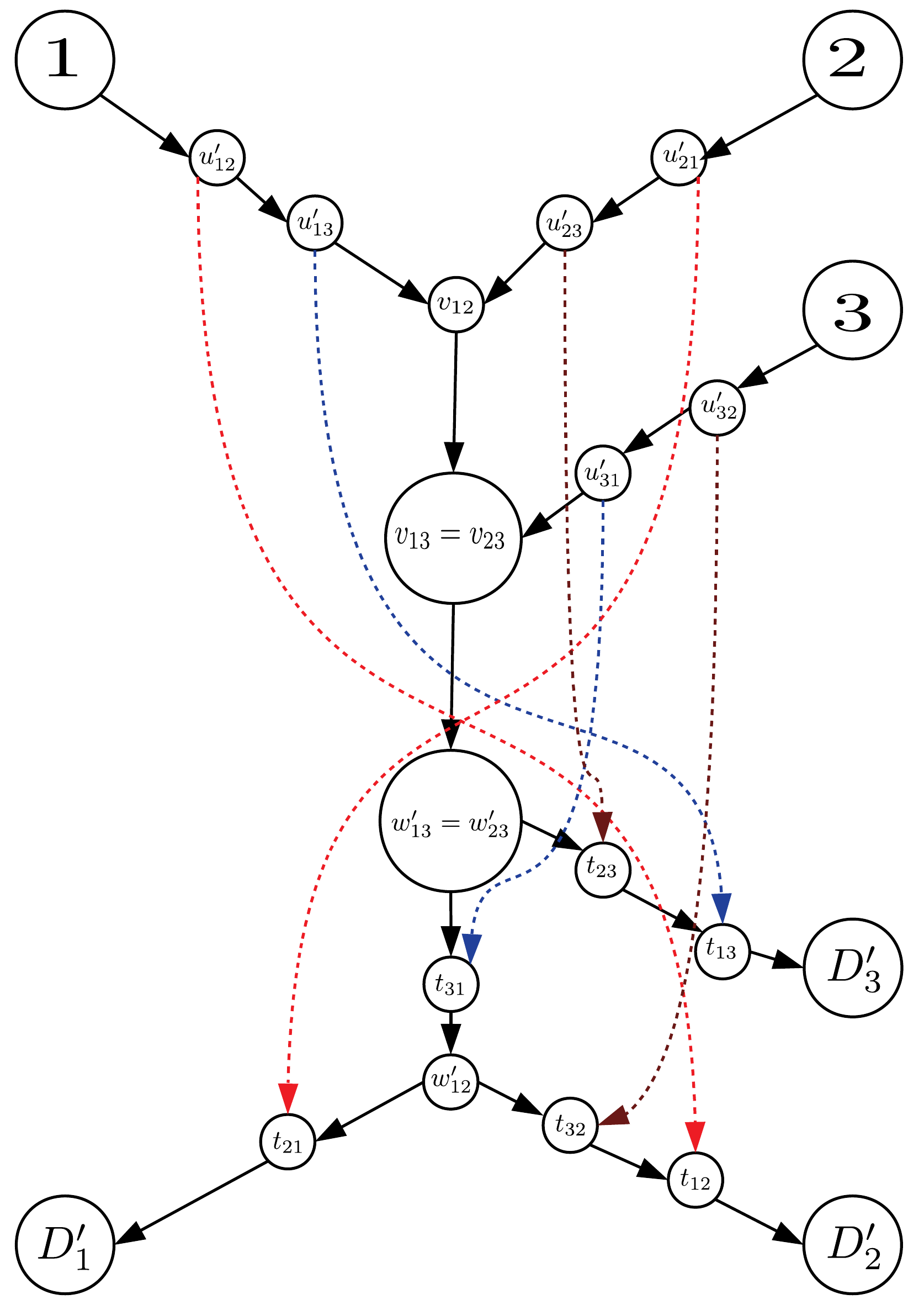}}
  \hspace{0.2in}
  \vspace{0in}
  \hrule
	\caption{}
	 \end{figure*}
	\newpage
\begin{figure*}[htbp]
  \centering
  \vspace{0.0in}
  \subfigure[Network style A stage I reduced Config  S14 for above Configs 2 and 6]{\includegraphics[height=2.5in]{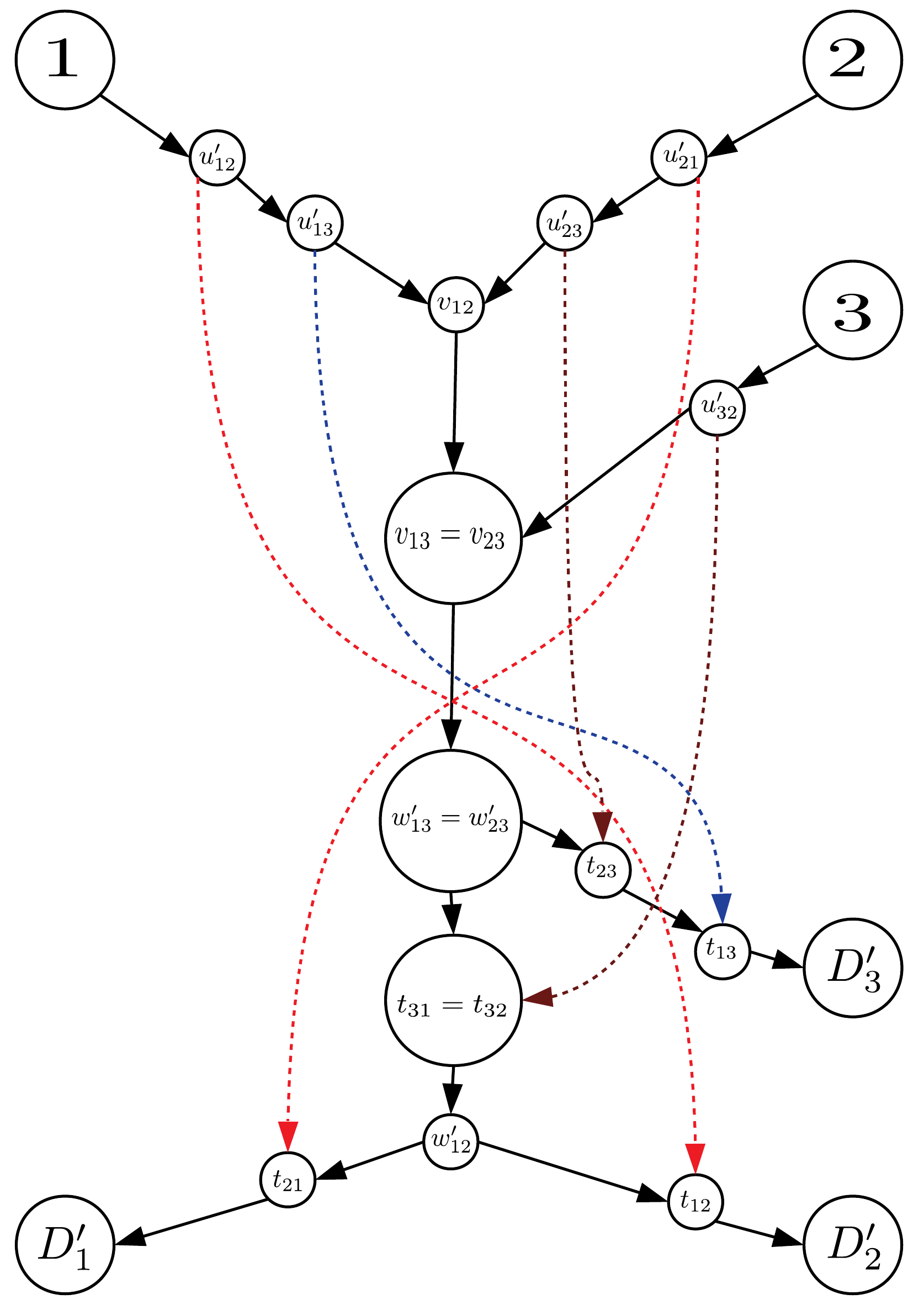}
  \label{network style 1A-2}}
 \hspace{0.0in}
 \subfigure[Network style A stage I reduced Config S15 of above Config 5 ]{\label{network style 1A-5}\includegraphics[height=2.5in]{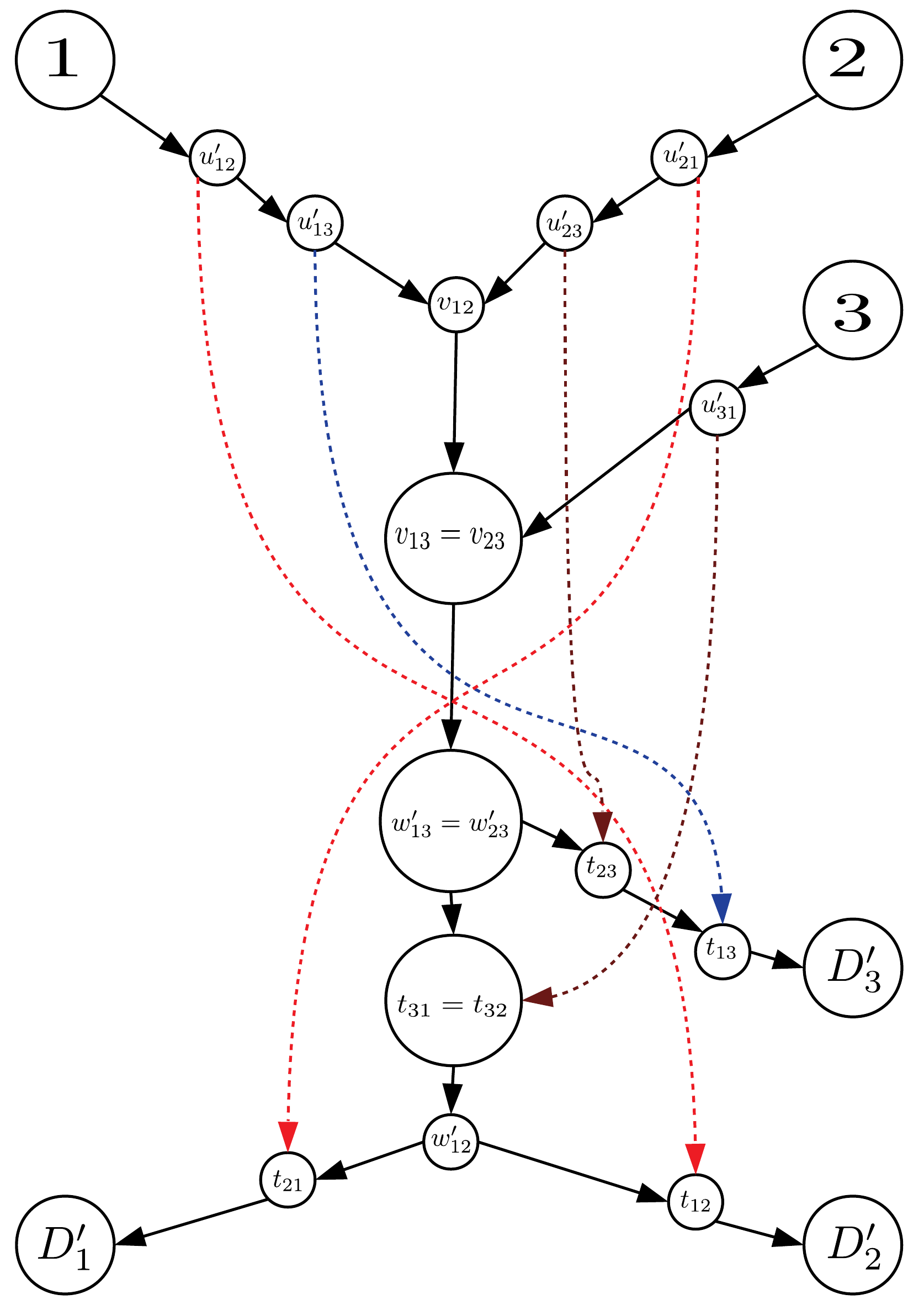}}
   \hspace{0.0in}
   \hrule
	\caption{}
   \end{figure*}
\begin{figure*}[htbp]
  \centering
  \vspace{.0in}
  \subfigure[Network style A stage II reduced Config  S23 for stage I configs (S14 and S15)]{\label{network style 2A-2}\includegraphics[height=3.5in]{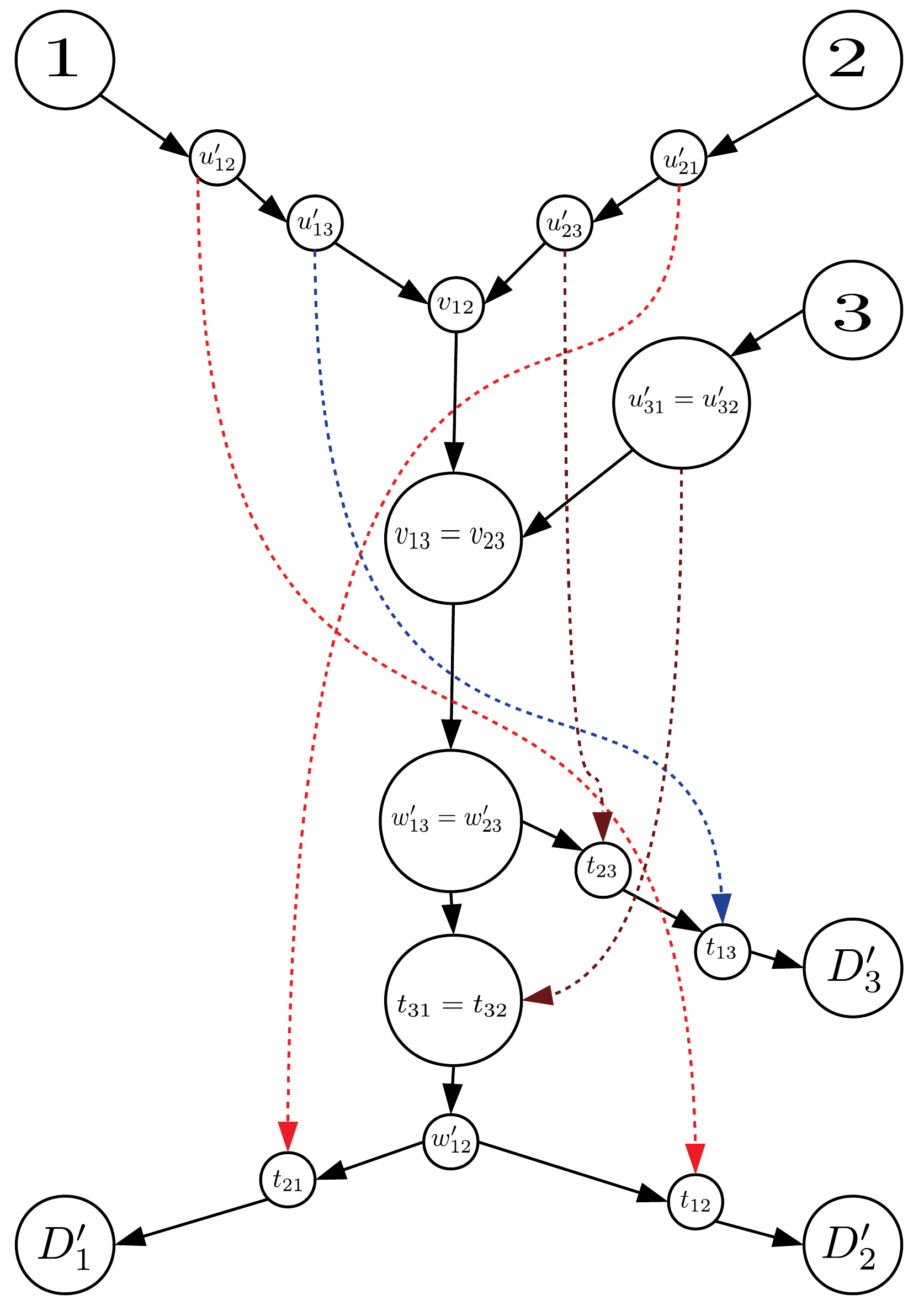}}
  \hrule
	\caption{}
 \end{figure*}
\begin{figure*}[htbp]
  \centering
  \subfigure[Network style A original Config  3]{\label{network style A-3}\includegraphics[height=2.0in]{network_style_A-3.pdf}} \hspace{0.2in}
  \subfigure[Network style A original Config  11]{\label{network style A-11}\includegraphics[height=2.0in]{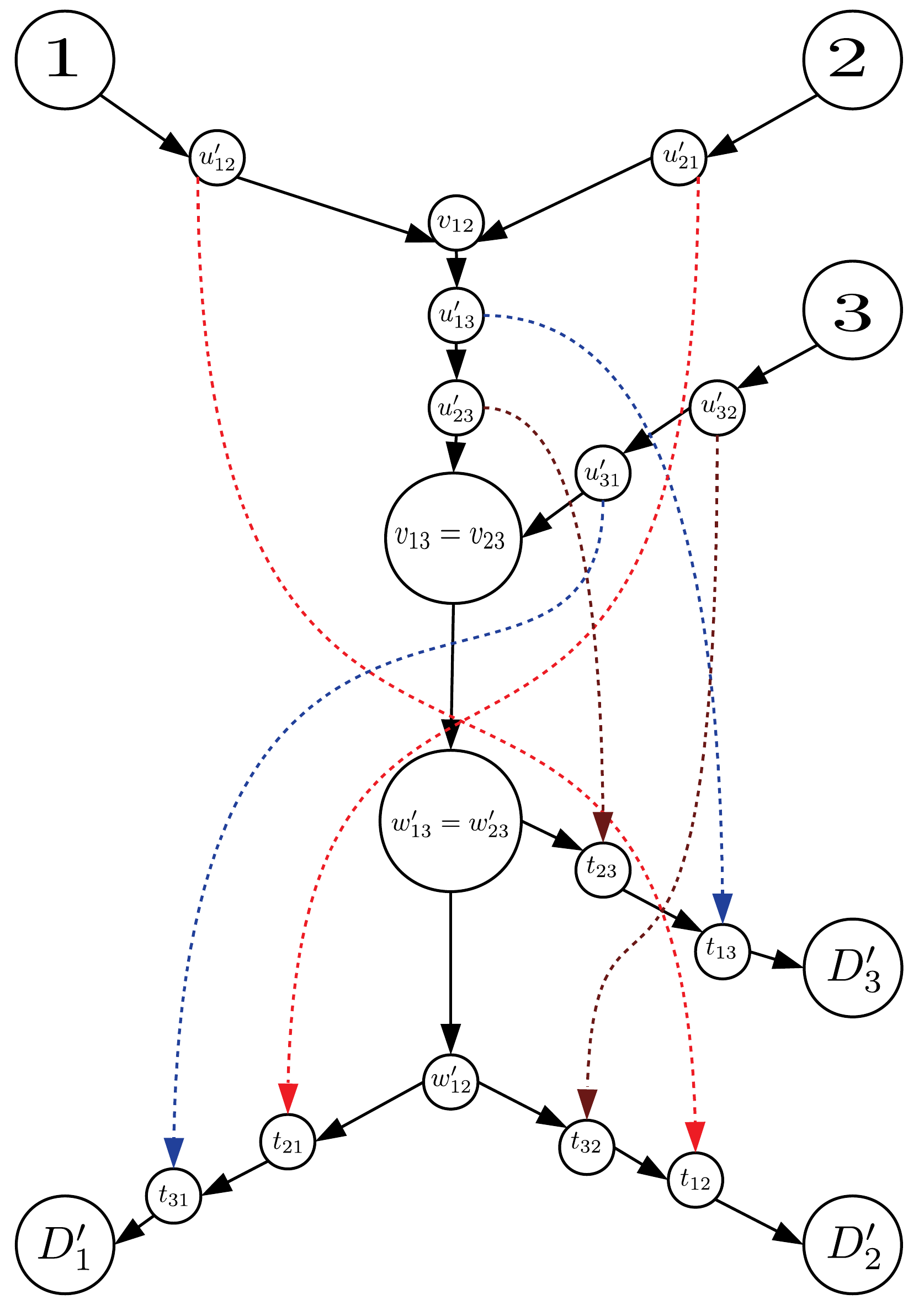}}
  \hspace{0.2in}
  \subfigure[Network style A original Config  9]{\label{network style A-9}\includegraphics[height=2.0in]{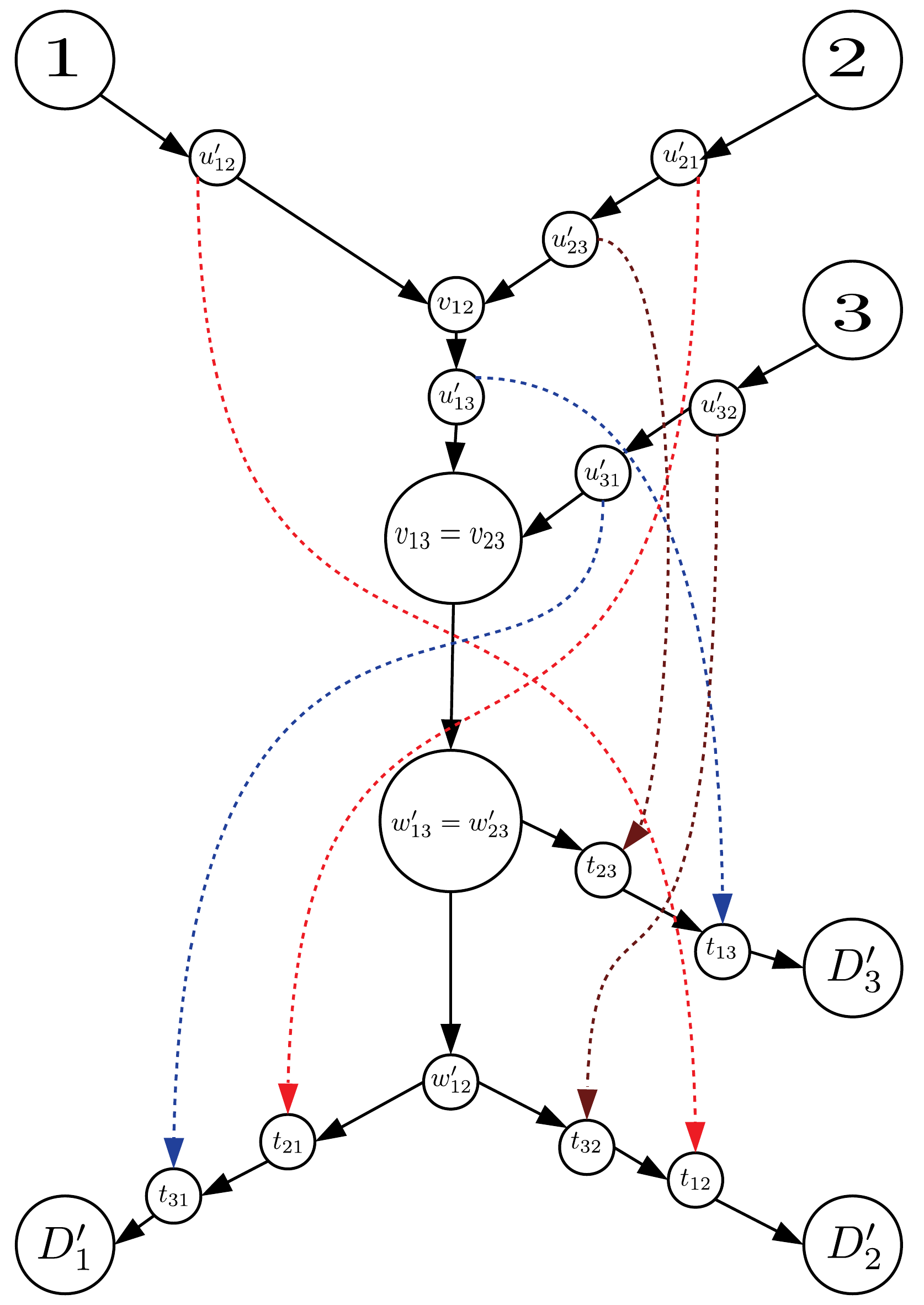}}
  \hspace{0.2in}
  \vspace{0in}
  \hrule
	\caption{}
 \end{figure*}
\begin{figure*}[htbp]
  \centering
  \vspace{0.0in}
  \subfigure[Network style A stage I reduced Config  S12 for above Config  3 and 11]{\includegraphics[height=2.5in]{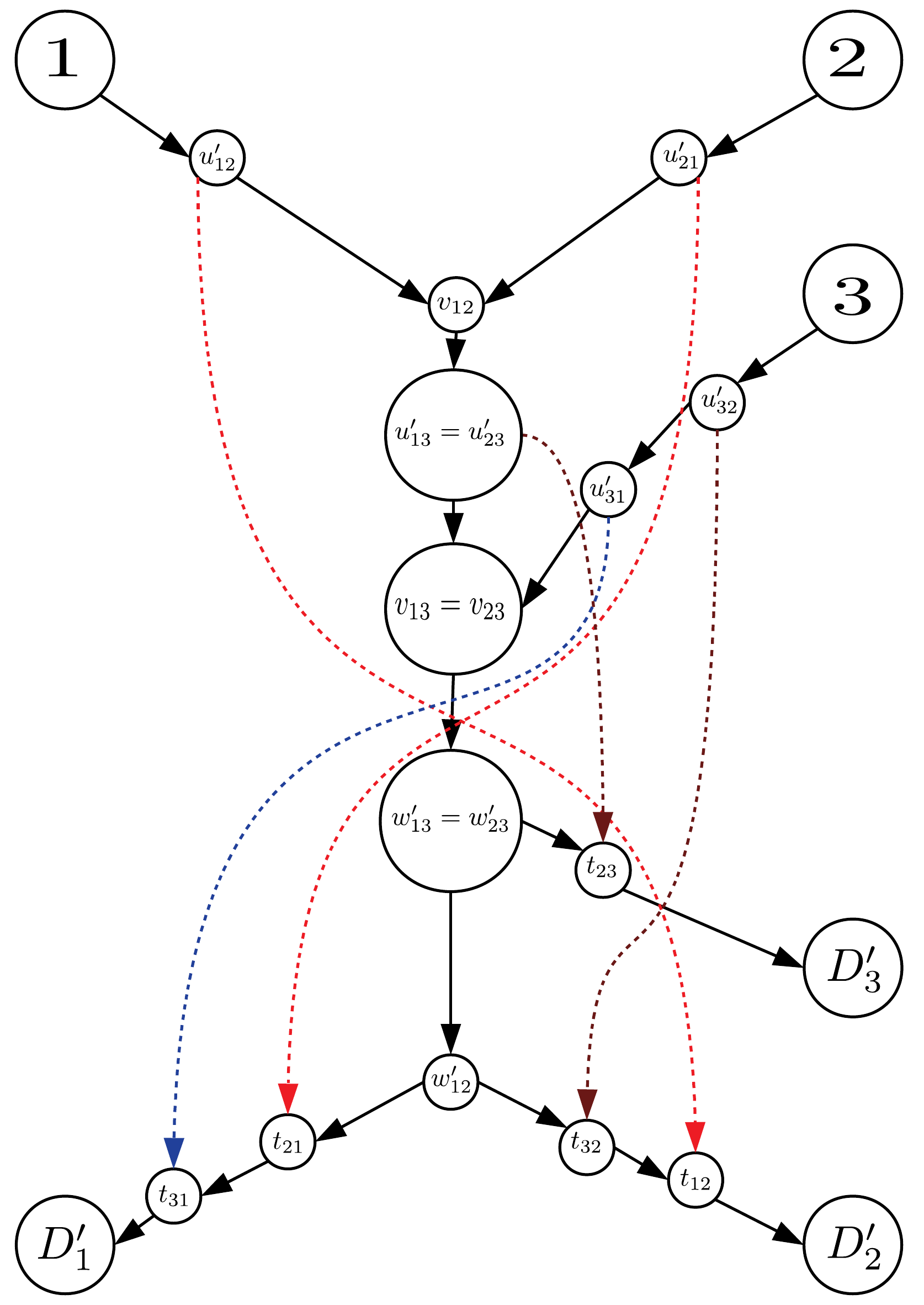}
  \label{network style 1A-3}}
 \hspace{0.0in}
 \subfigure[Network style A stage I reduced Config  S13 of above Config  9 ]{\label{network style 1A-9}\includegraphics[height=2.5in]{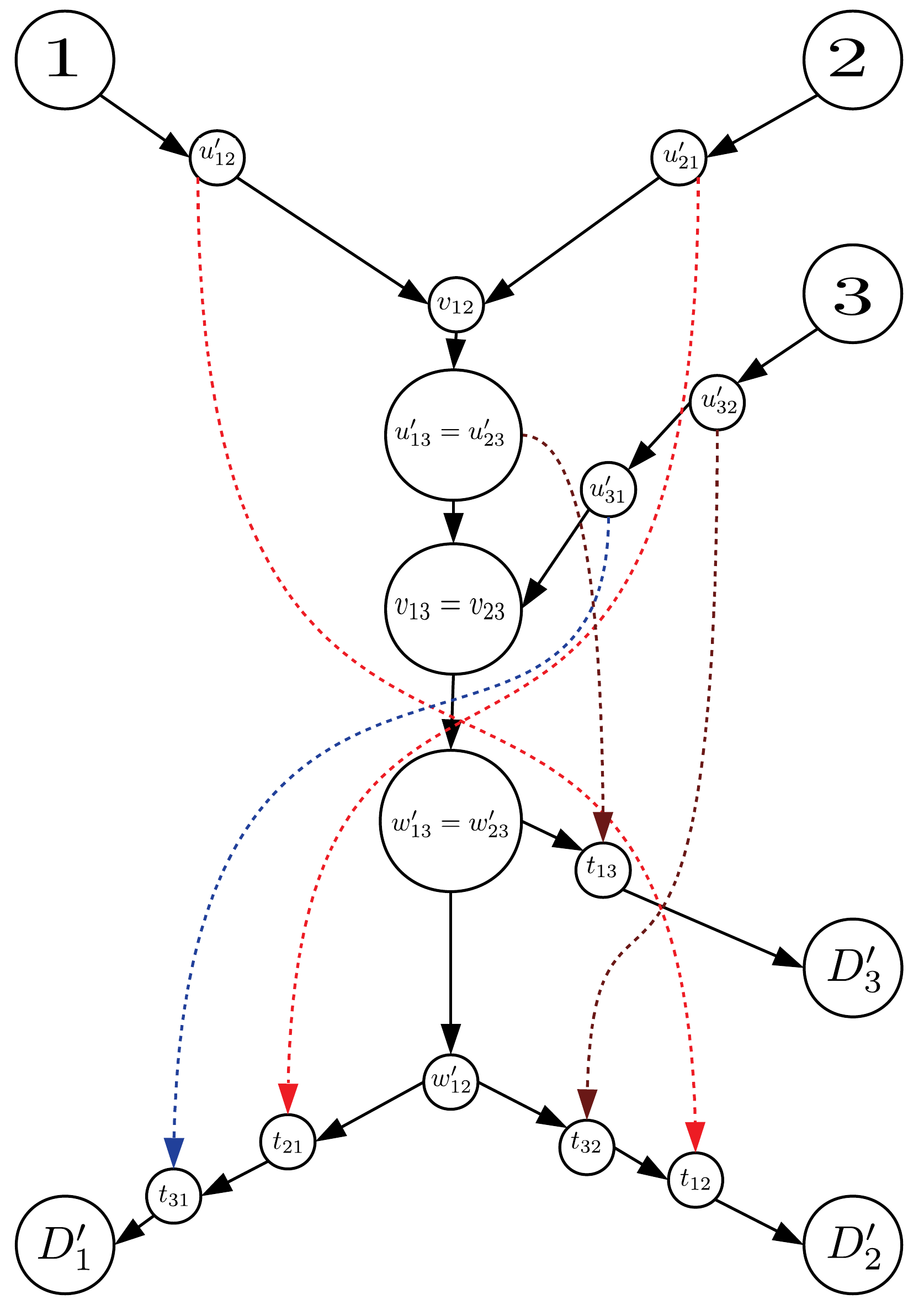}}
   \hspace{0.0in}
   \hrule
	\caption{}
   \end{figure*}
\begin{figure*}[htbp]
  \centering
  \vspace{.0in}
  \subfigure[Network style A stage II reduced Config  S22 of stage I Config  S12 and S13]{\label{network style 2A-3}\includegraphics[height=3.0in]{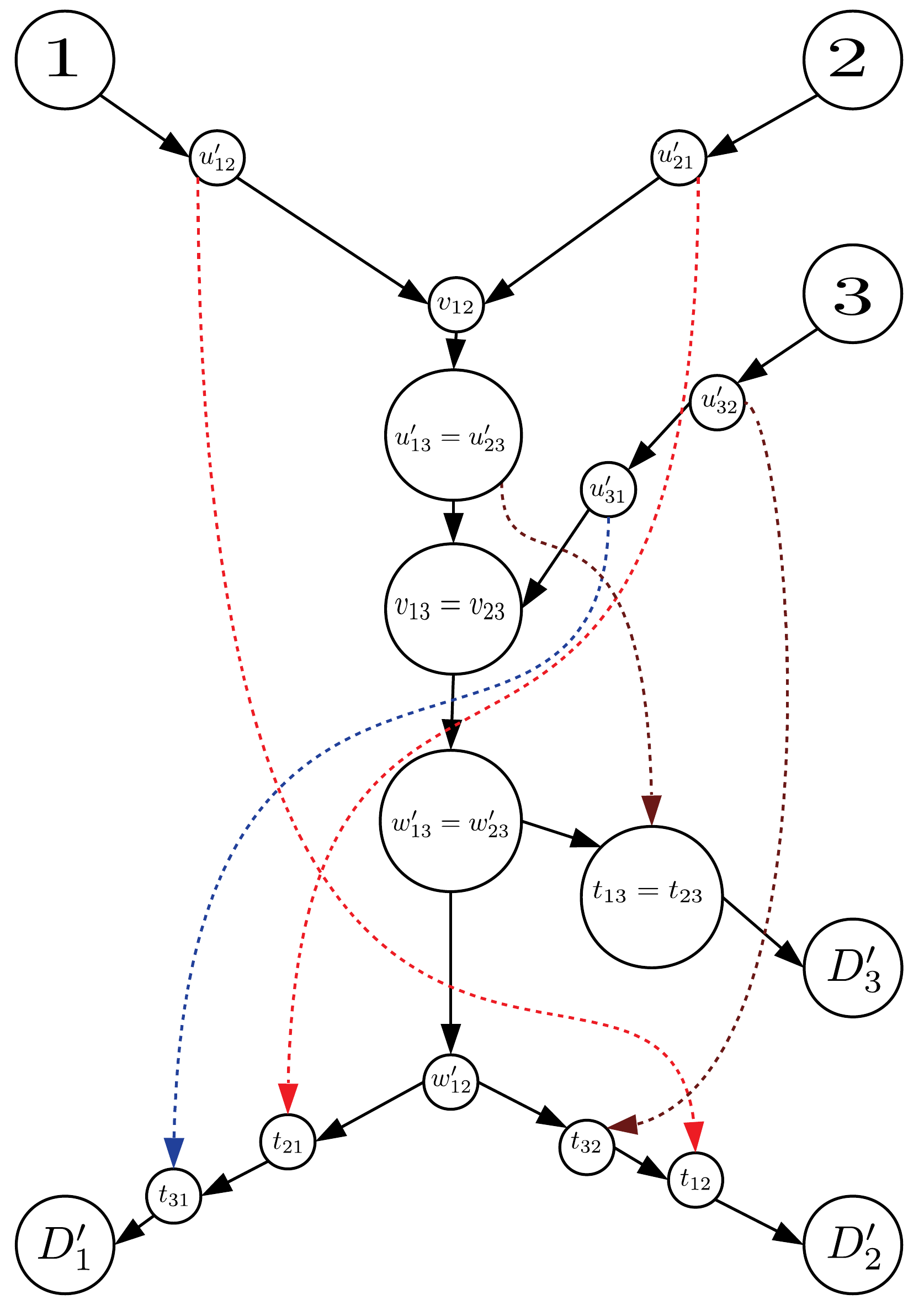}}
  \hspace{0.0in}
	\caption{}
 \end{figure*}
\begin{figure*}[htbp]
  \subfigure[Network style A original Config  4]{\label{network style A-4}\includegraphics[height=2.0in]{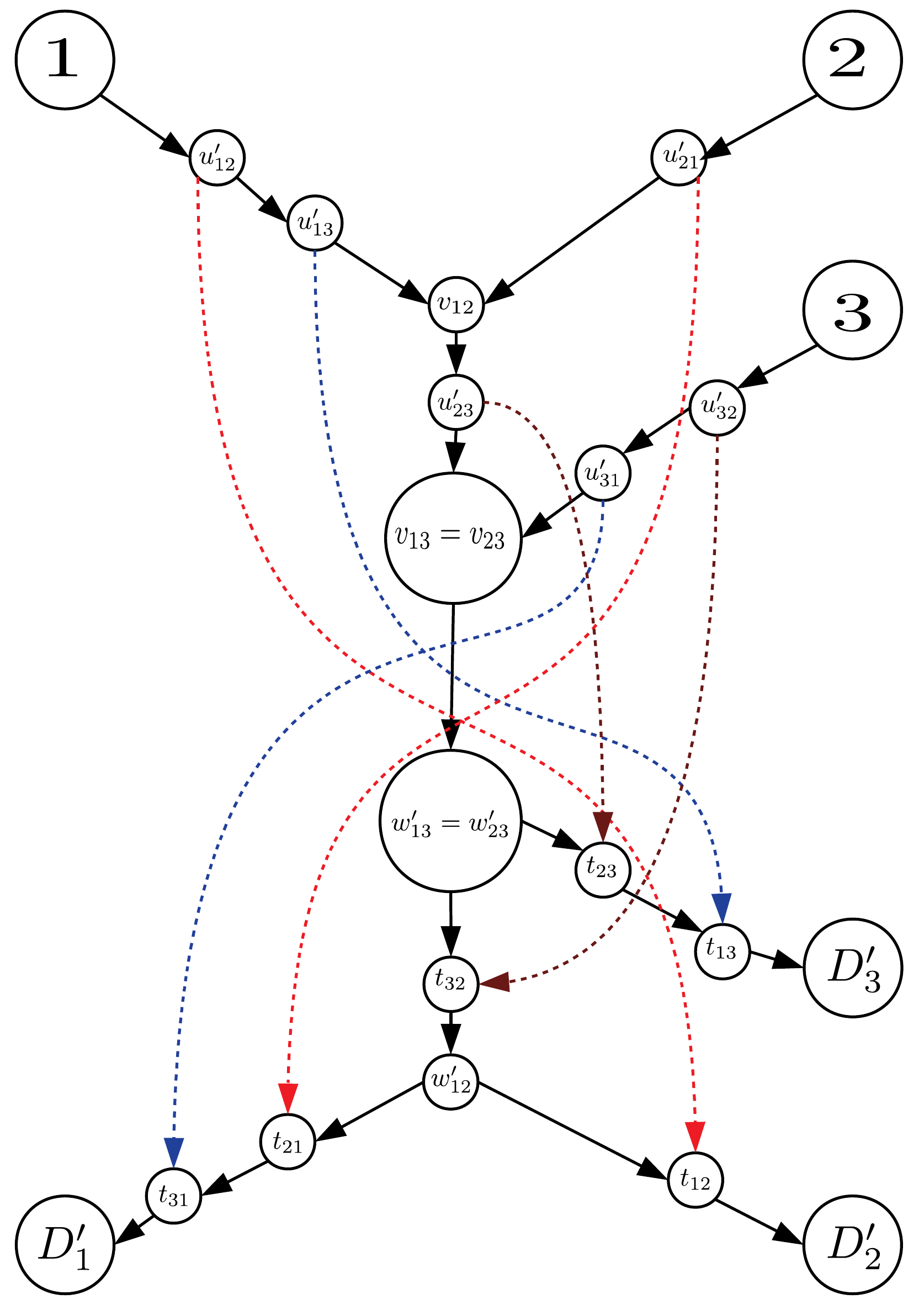}} \hspace{0.0in}
  \subfigure[Network style A original Config  7]{\label{network style A-7}\includegraphics[height=2.0in]{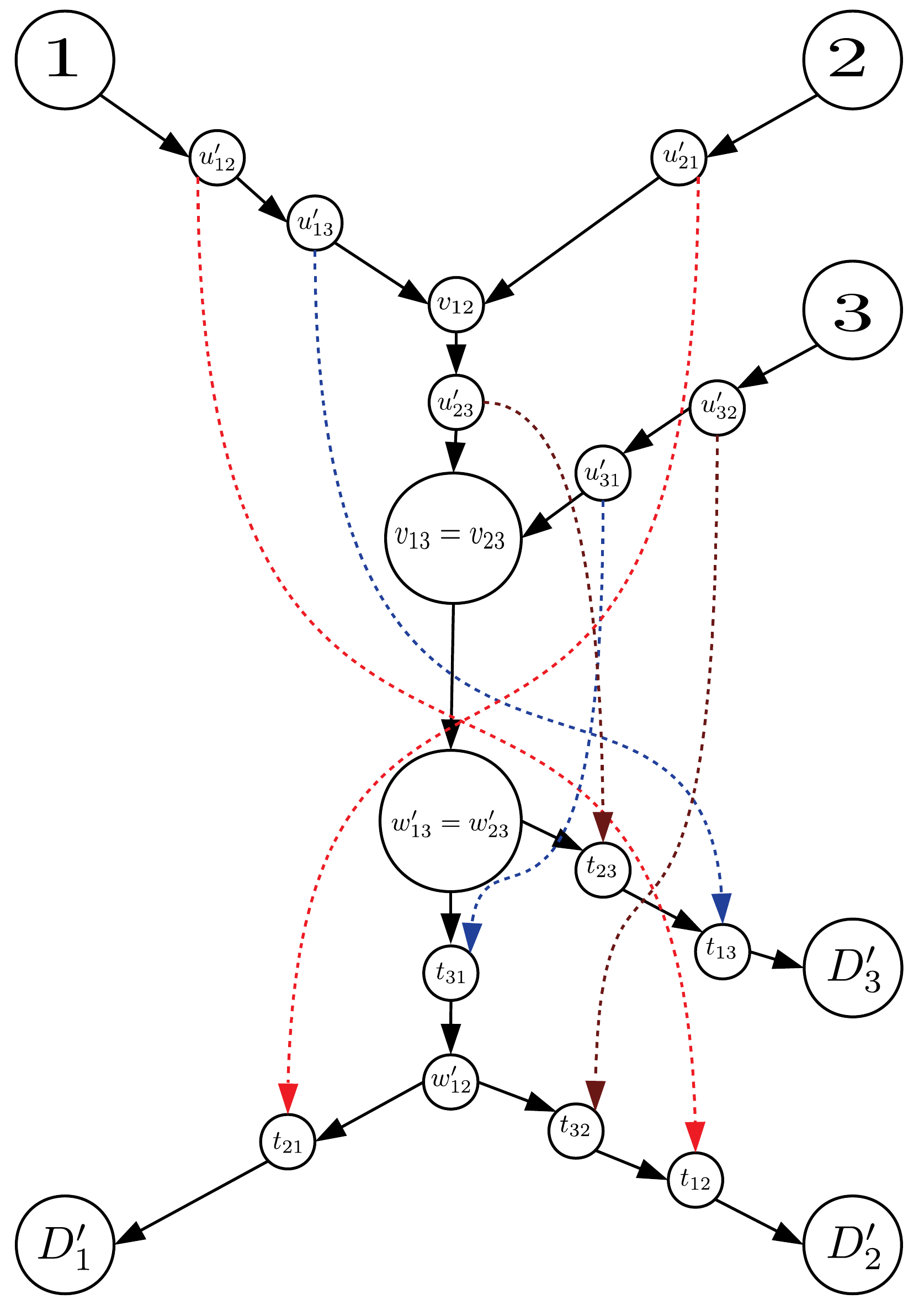}}
  \hspace{0.0in}
  \subfigure[Network style A original Config  8]{\label{network style A-8}\includegraphics[height=2.0in]{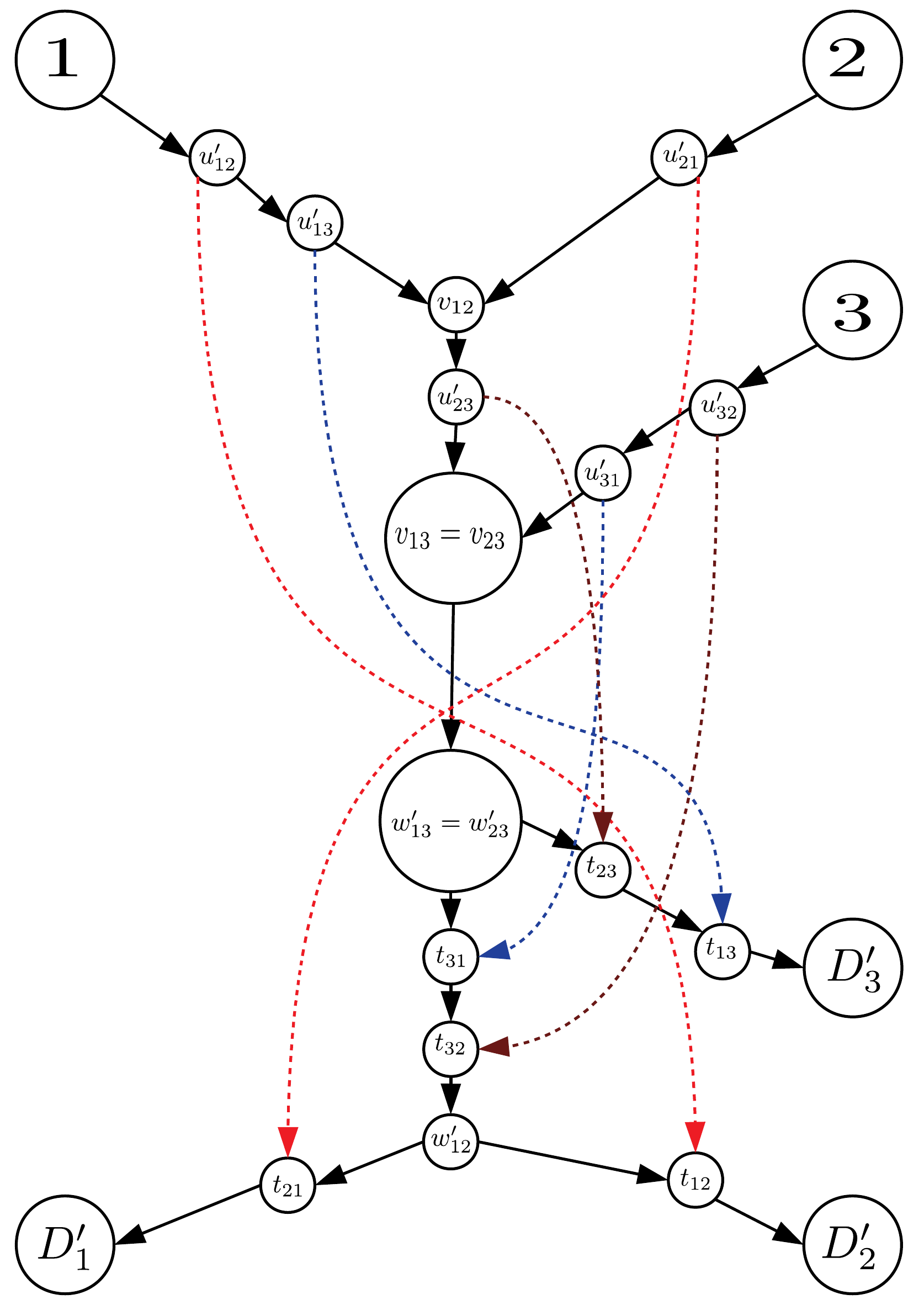}}
  \hspace{0.0in}
  \subfigure[Network style A original Config  15]{\label{network style A-15}\includegraphics[height=2.0in]{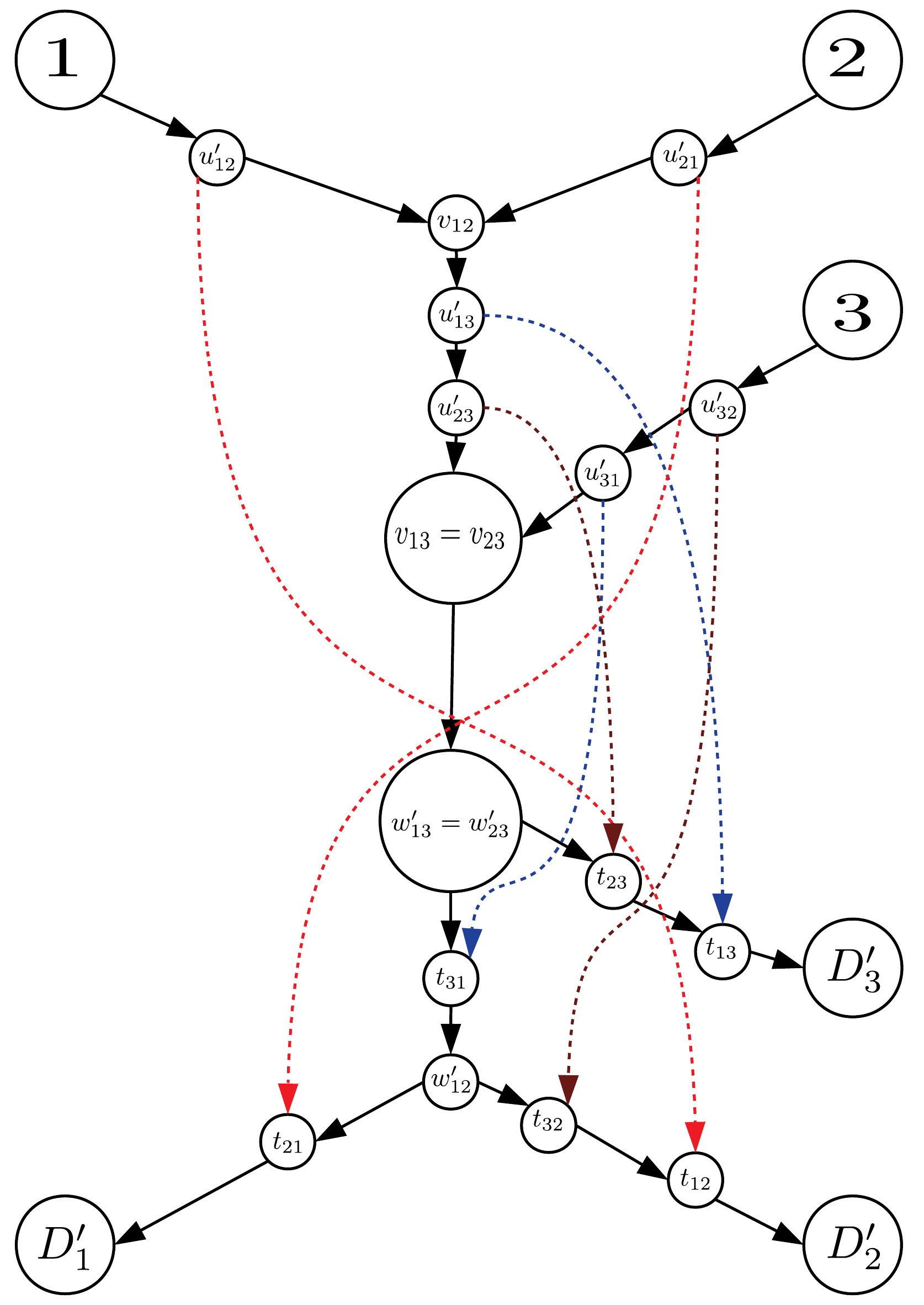}}
   \hspace{0.0in}
  \subfigure[Network style A original Config  16]{\label{network style A-16}\includegraphics[height=2.0in]{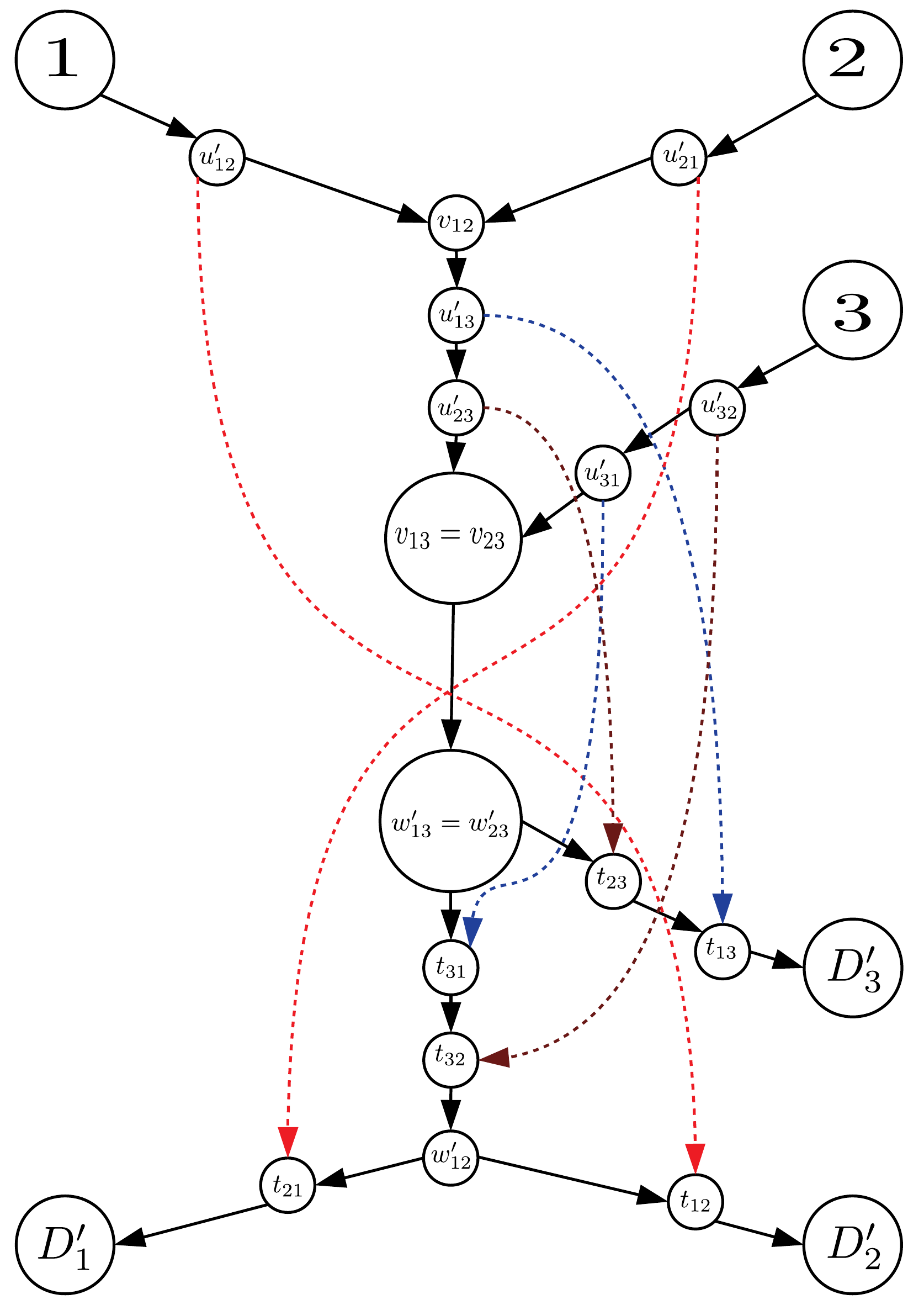}}
  \hspace{0.0in}
  \subfigure[Network style A original Config  10]{\label{network style A-10}\includegraphics[height=2.0in]{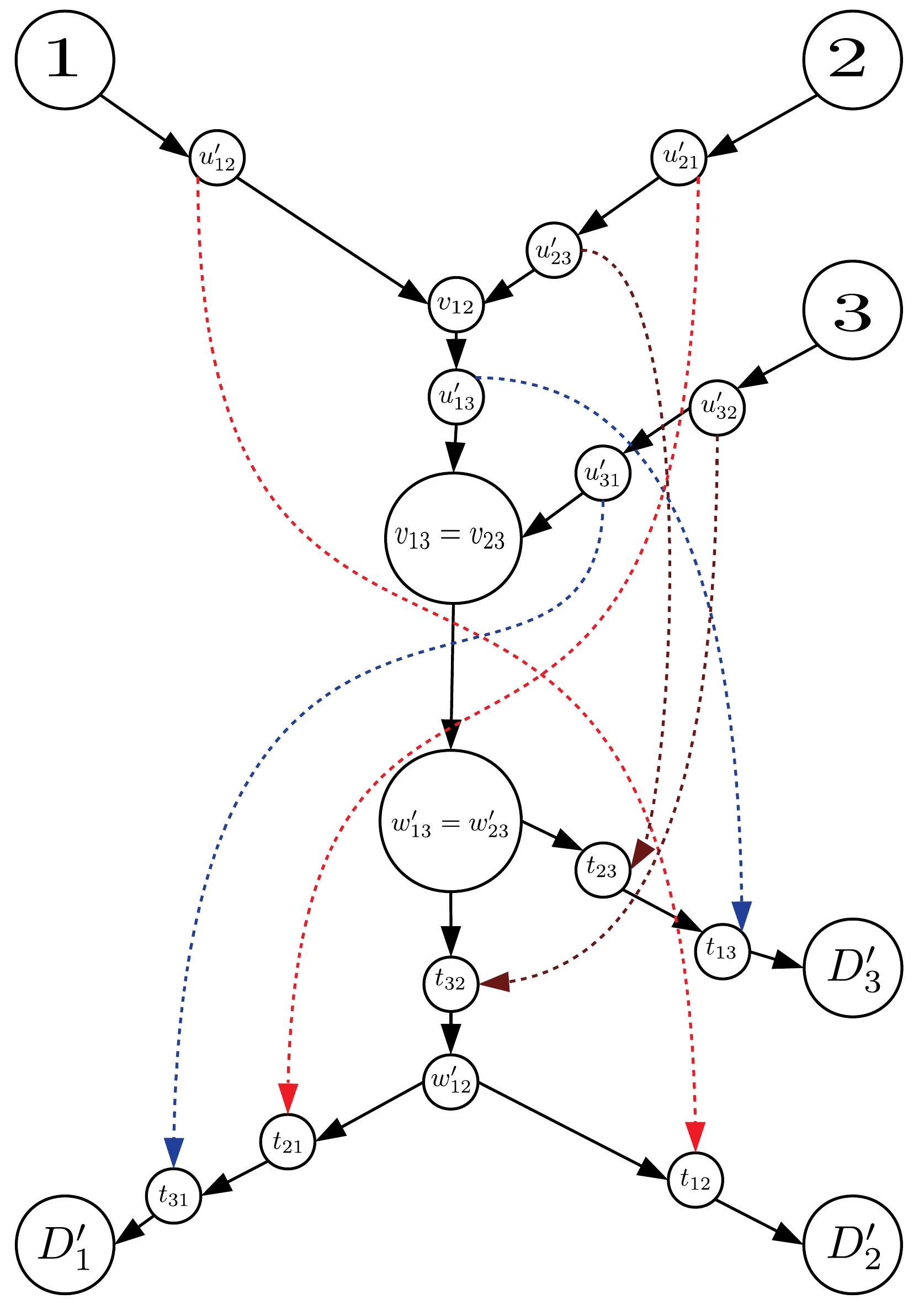}}
  \hspace{0.0in}
  \subfigure[Network style A original Config  12]{\label{network style A-12}\includegraphics[height=2.0in]{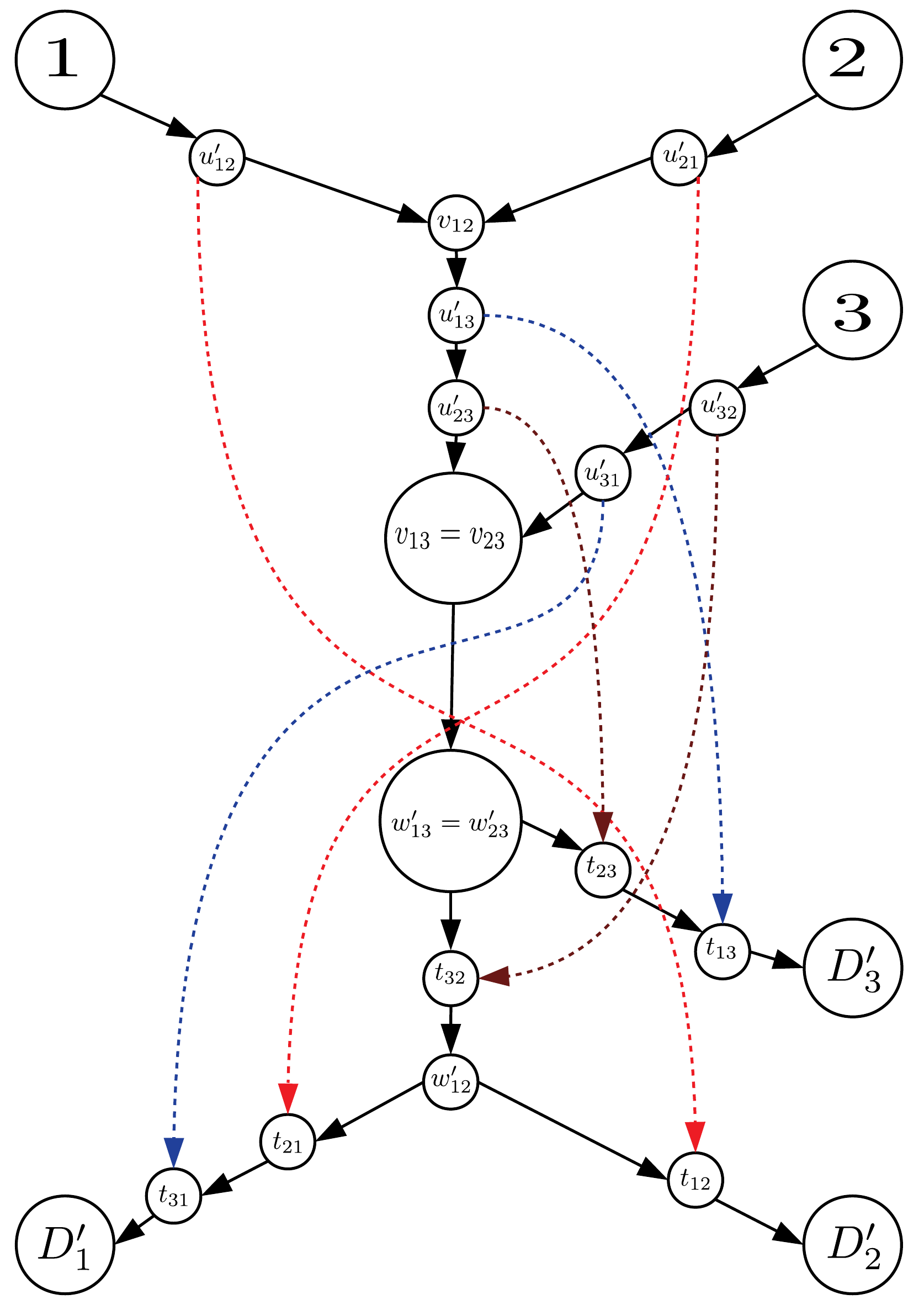}}
  \hspace{0.0in}
  \subfigure[Network style A original Config  14]{\label{network style A-14}\includegraphics[height=2.0in]{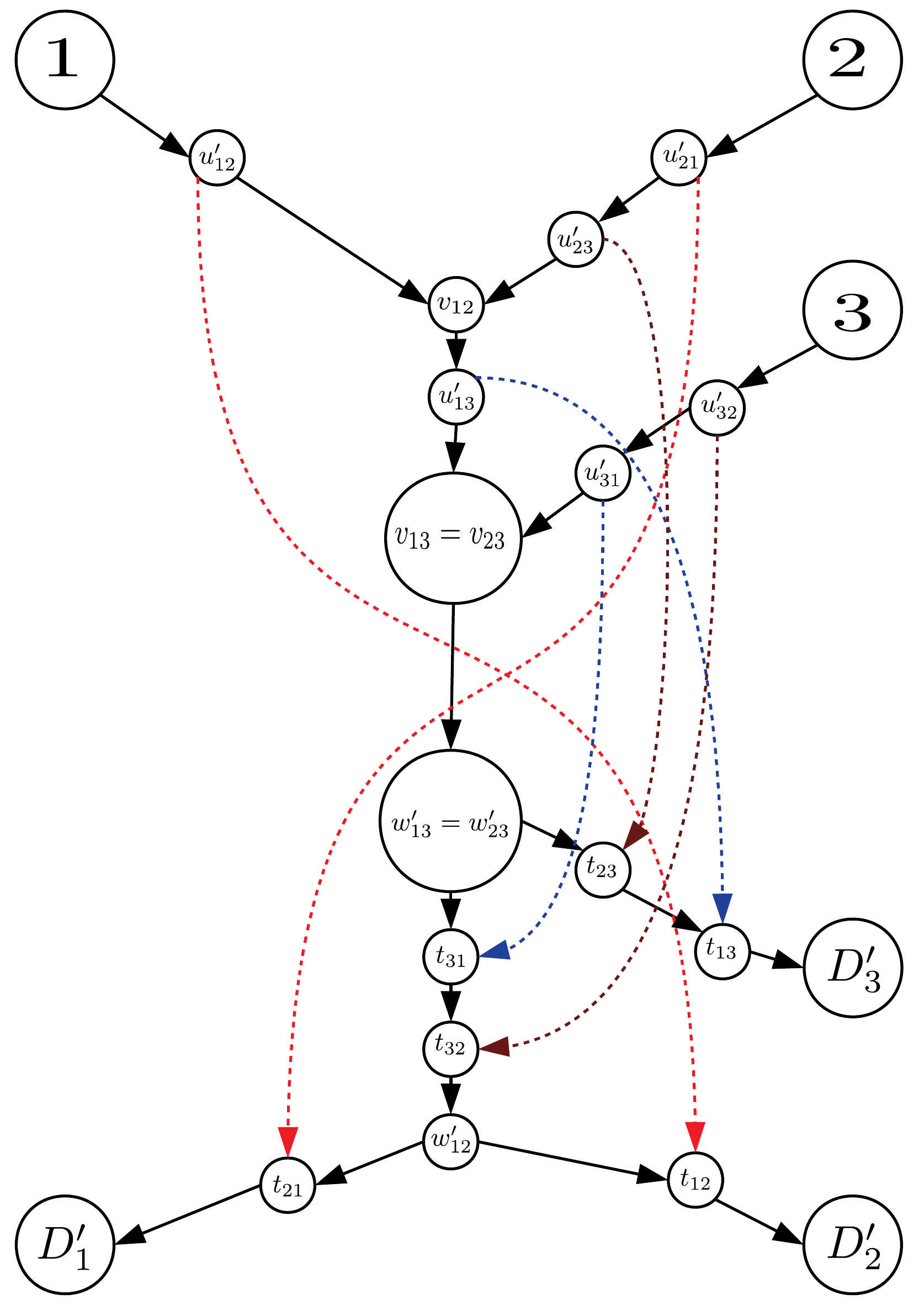}}
   \hspace{0.0in}
  \subfigure[Network style A original Config  13]{\label{network style A-13}\includegraphics[height=2.0in]{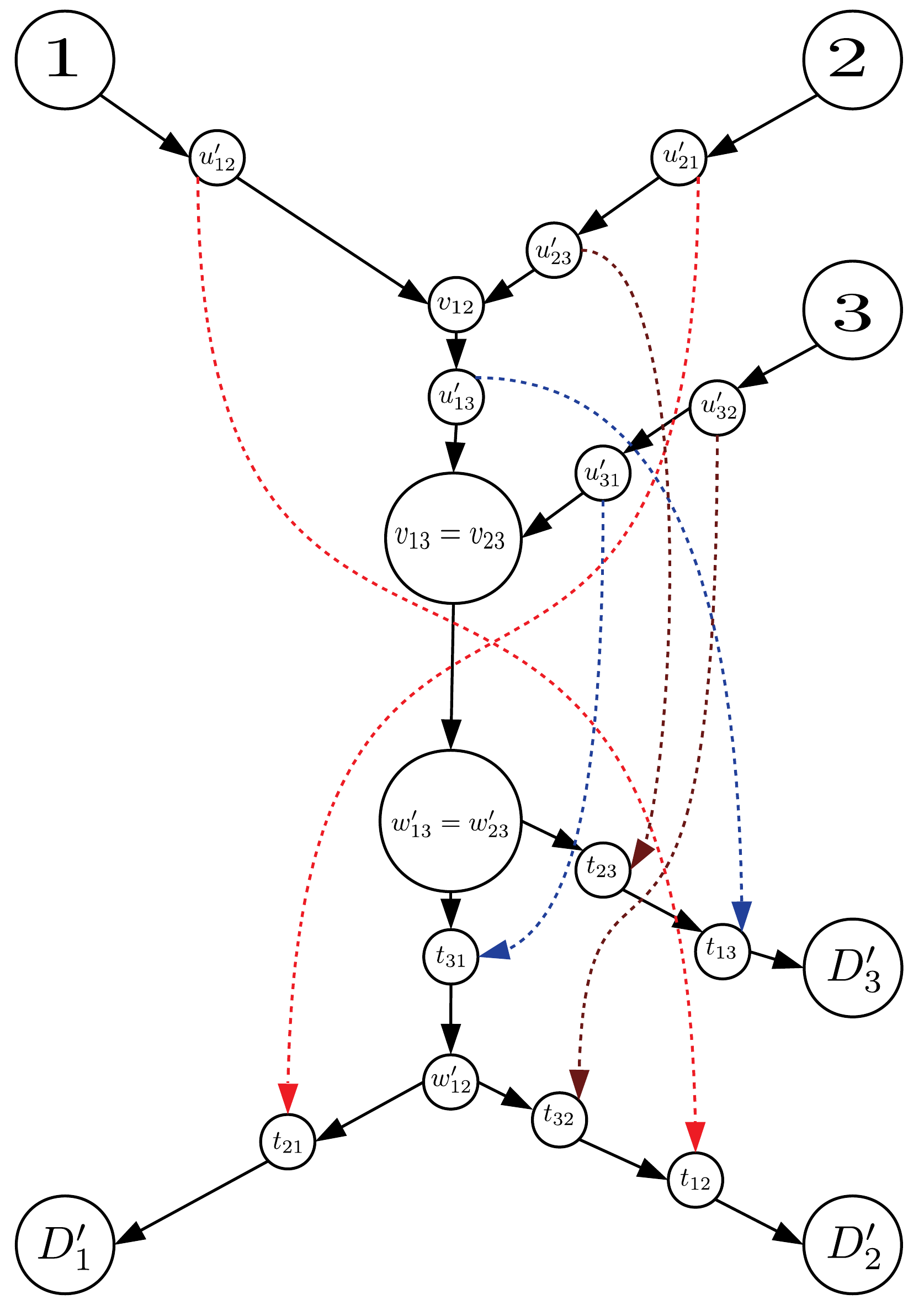}}
    \vspace{0in}
 \hrule
\caption{}
 \end{figure*}
\begin{figure*}[htbp]
  \centering
  \hspace{0.0in}
  \subfigure[Network style A stage I reduced Config  S16 for Config  4]{\includegraphics[height=2.5in]{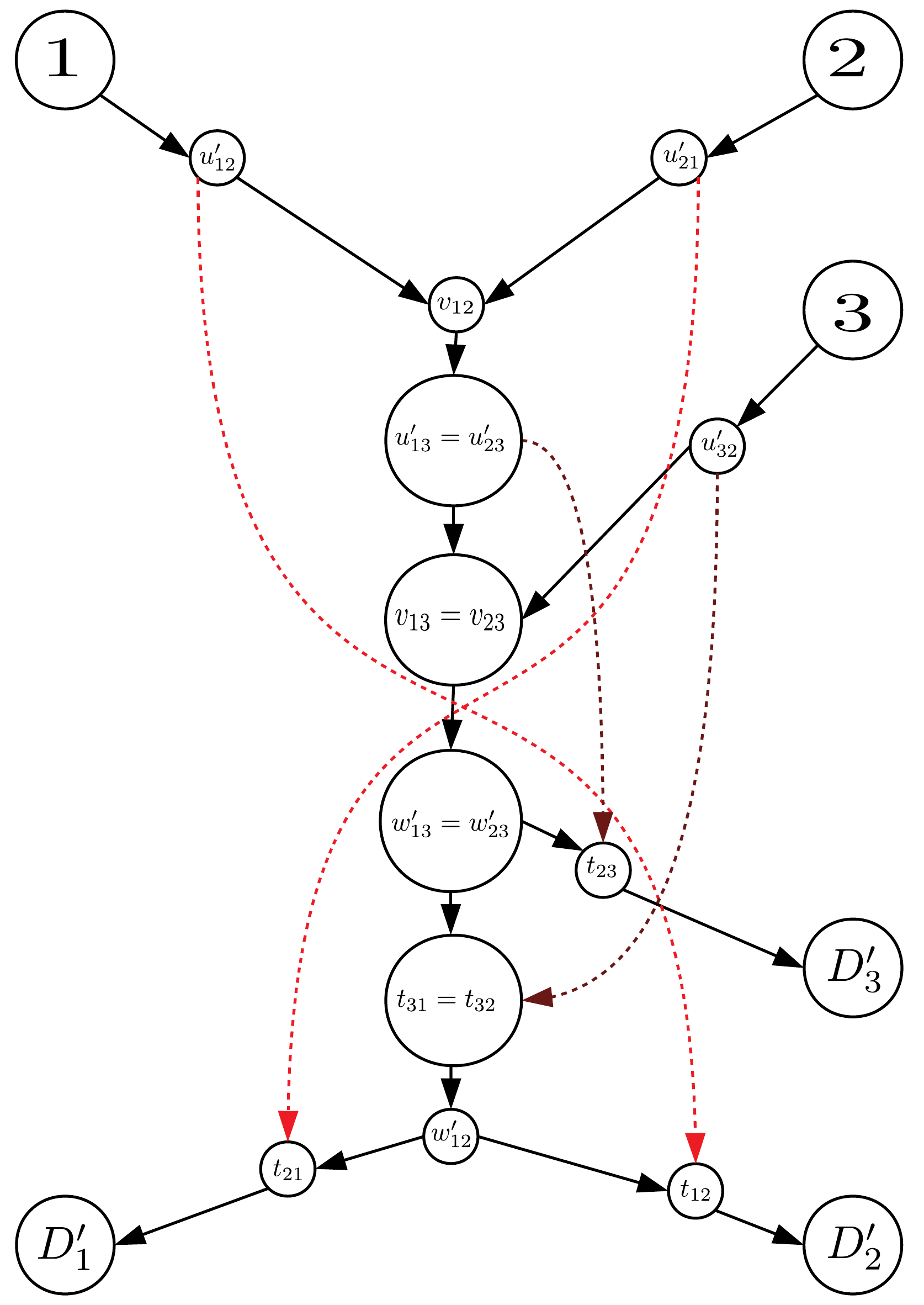}
  \label{network style 1A-4}}
 \hspace{0.0in}
 \subfigure[Network style A stage I reduced Config  S17 for Config  7,8,15 and 16]{\label{network style 1A-7}\includegraphics[height=2.5in]{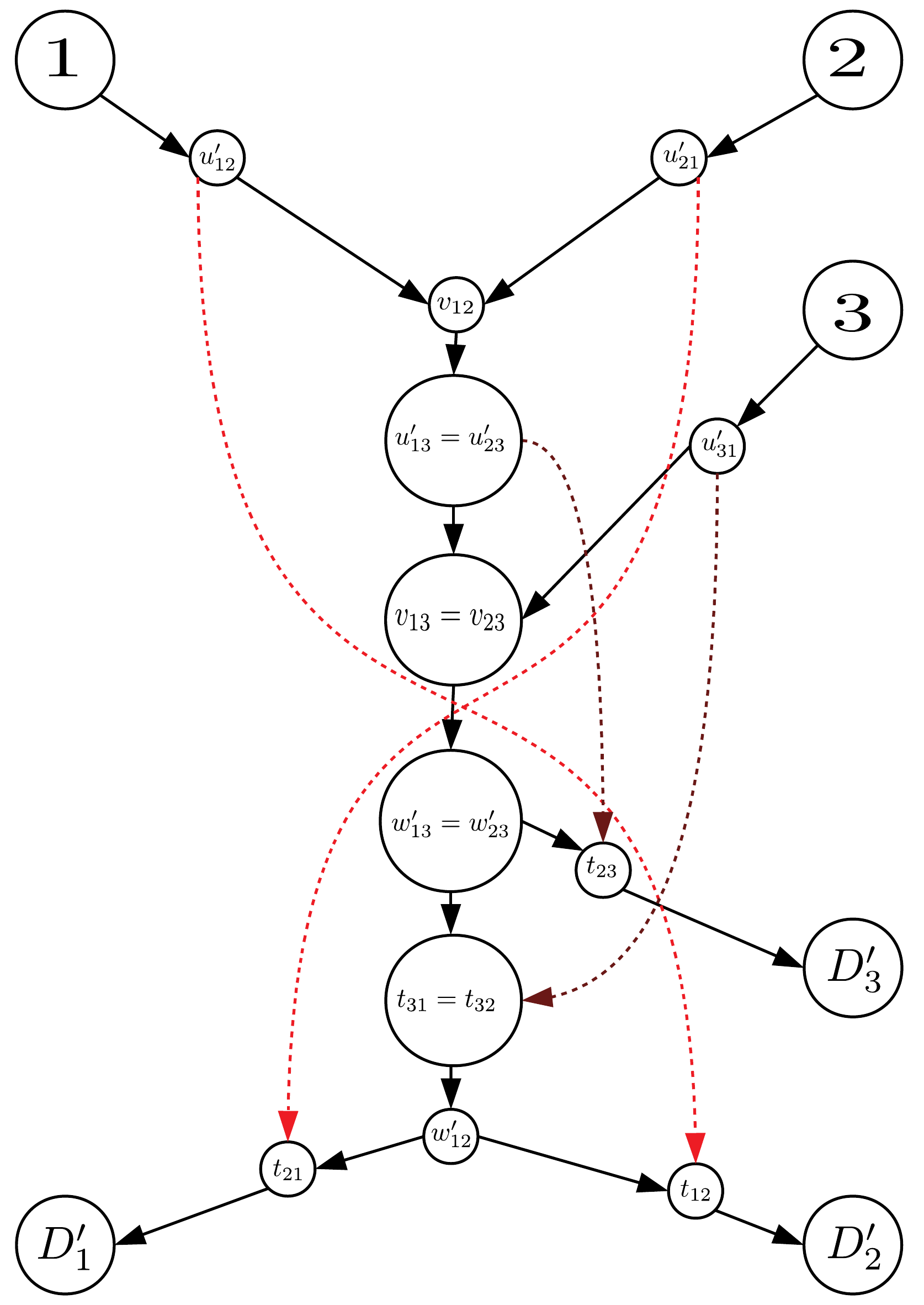}}
   \hspace{0.0in}
   \subfigure[Network style A stage I reduced Config  S18 for Config  10,12 and 14 ]{\label{network style 1A-10}\includegraphics[height=2.5in]{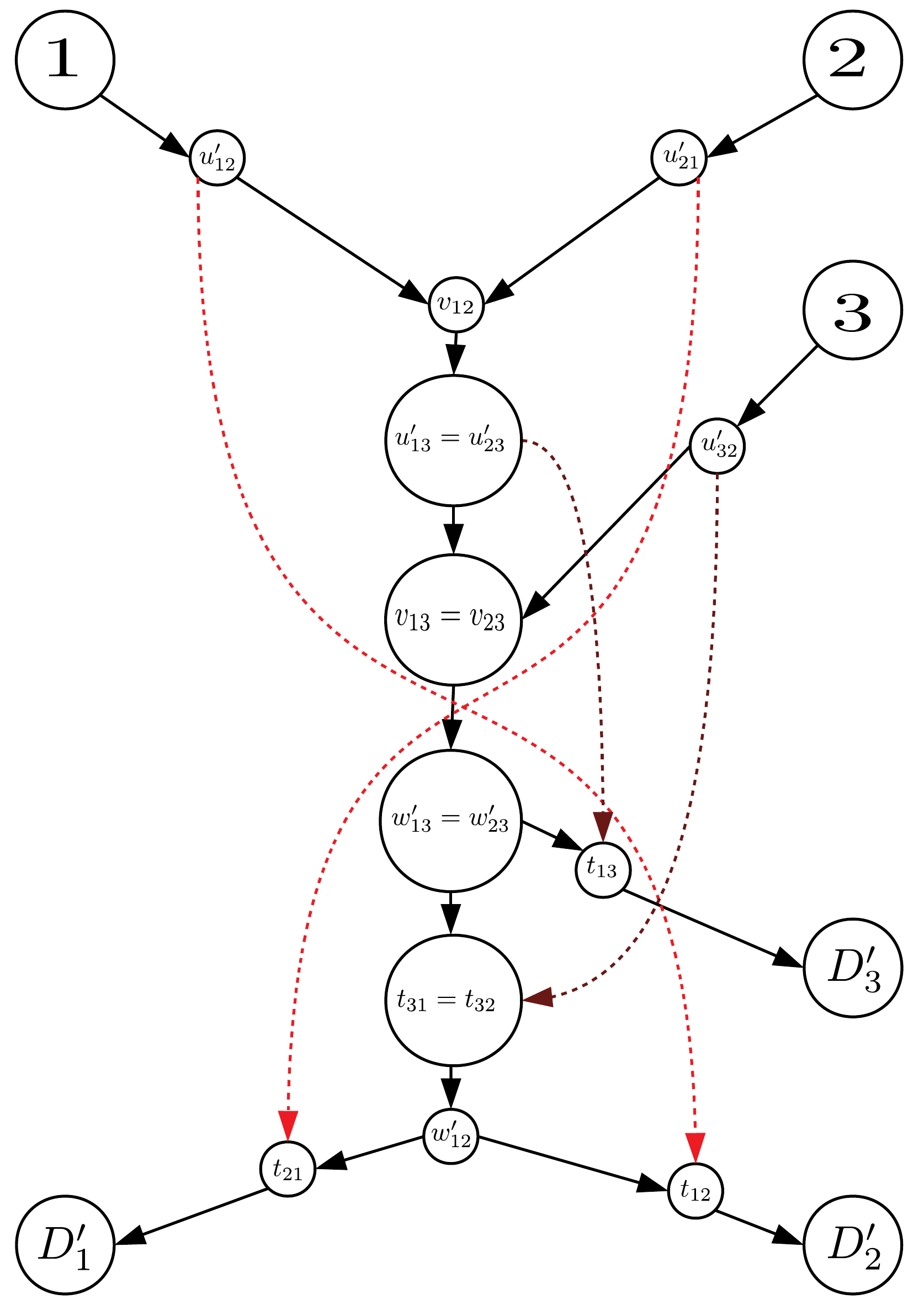}}
   \hspace{0.0in}
   \subfigure[Network style A stage I reduced Config  S19 for Config  13 ]{\label{network style 1A-10}\includegraphics[height=2.5in]{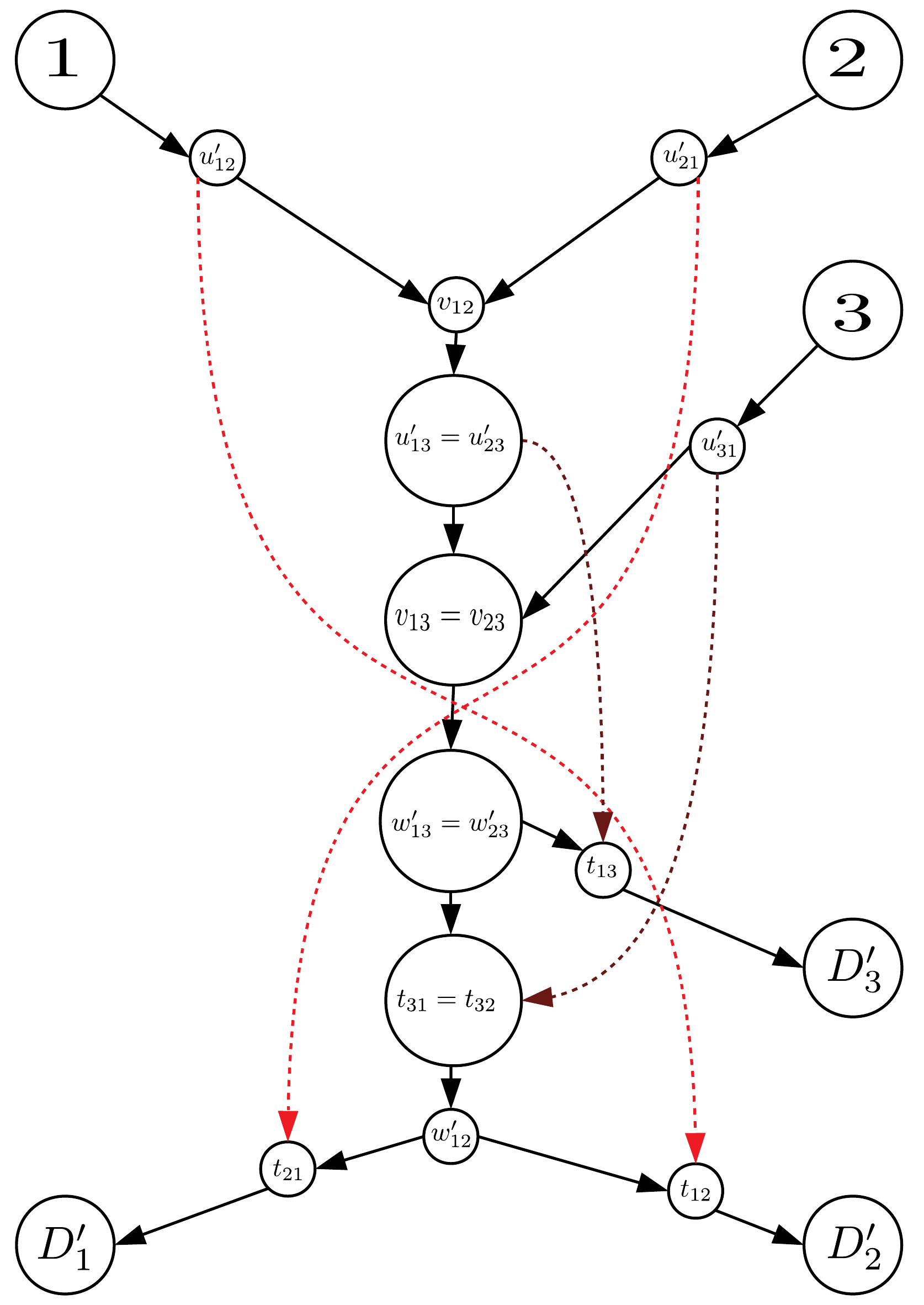}}
      \hrule
			\caption{}
   \end{figure*}
\begin{figure*}[htbp]
  \centering
  \vspace{0.0in}
  \subfigure[Network style A stage II reduced Config  S24 for stage I Config  S16,S17,S18 and S19]{\label{network_style_2A_4}\includegraphics[height=3.2in]{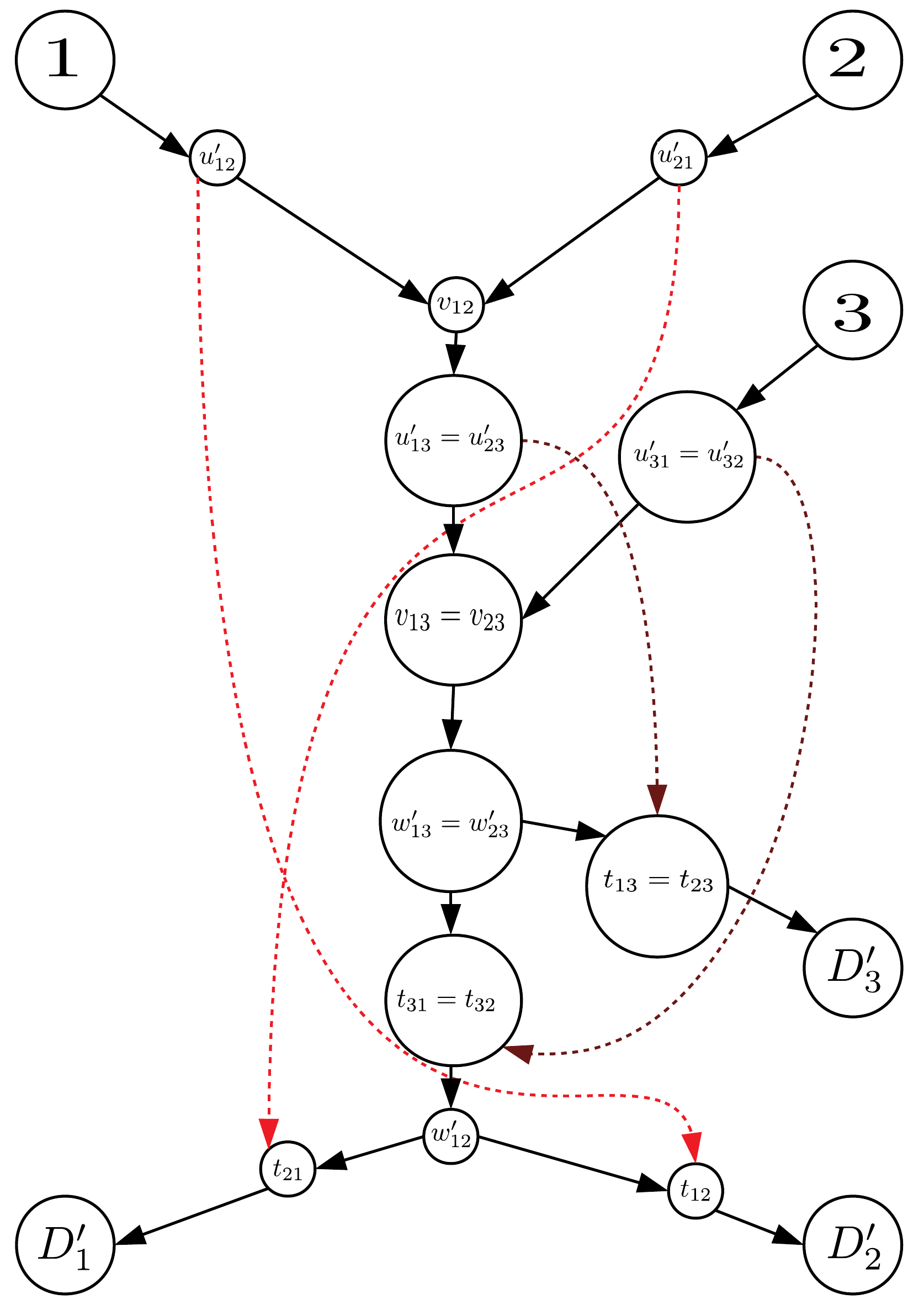}}
  \hspace{0.0in}
	\caption{}
 \end{figure*}

 \begin{figure*}[htbp]
 \centering
  \hspace{0.0in}
  \subfigure[Network style B original Config  1]{\label{network style A-3}\includegraphics[height=2.00in]{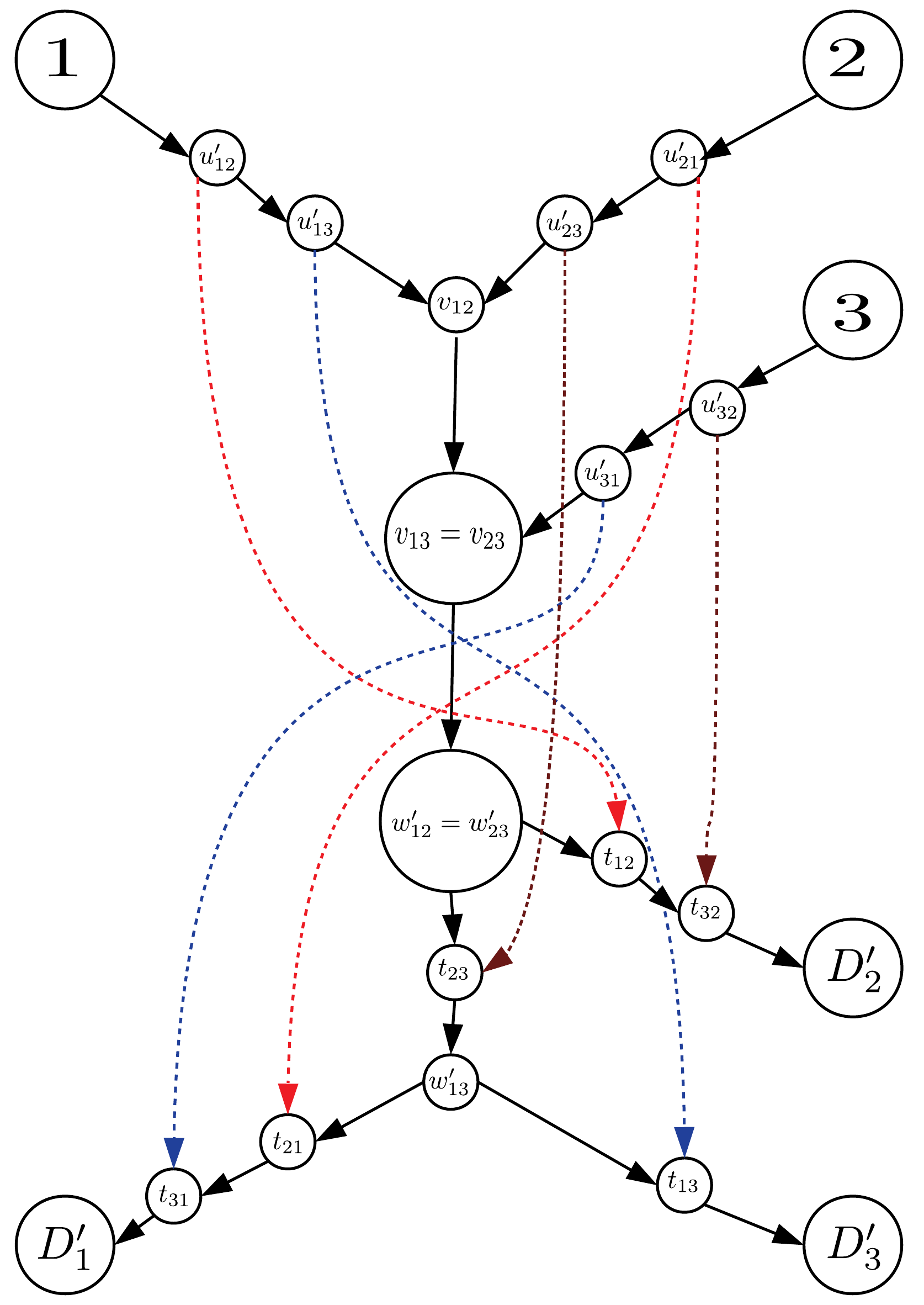}} \hspace{0.0in}
  \subfigure[Network style B original Config  7]{\label{network style A-11}\includegraphics[height=2.0in]{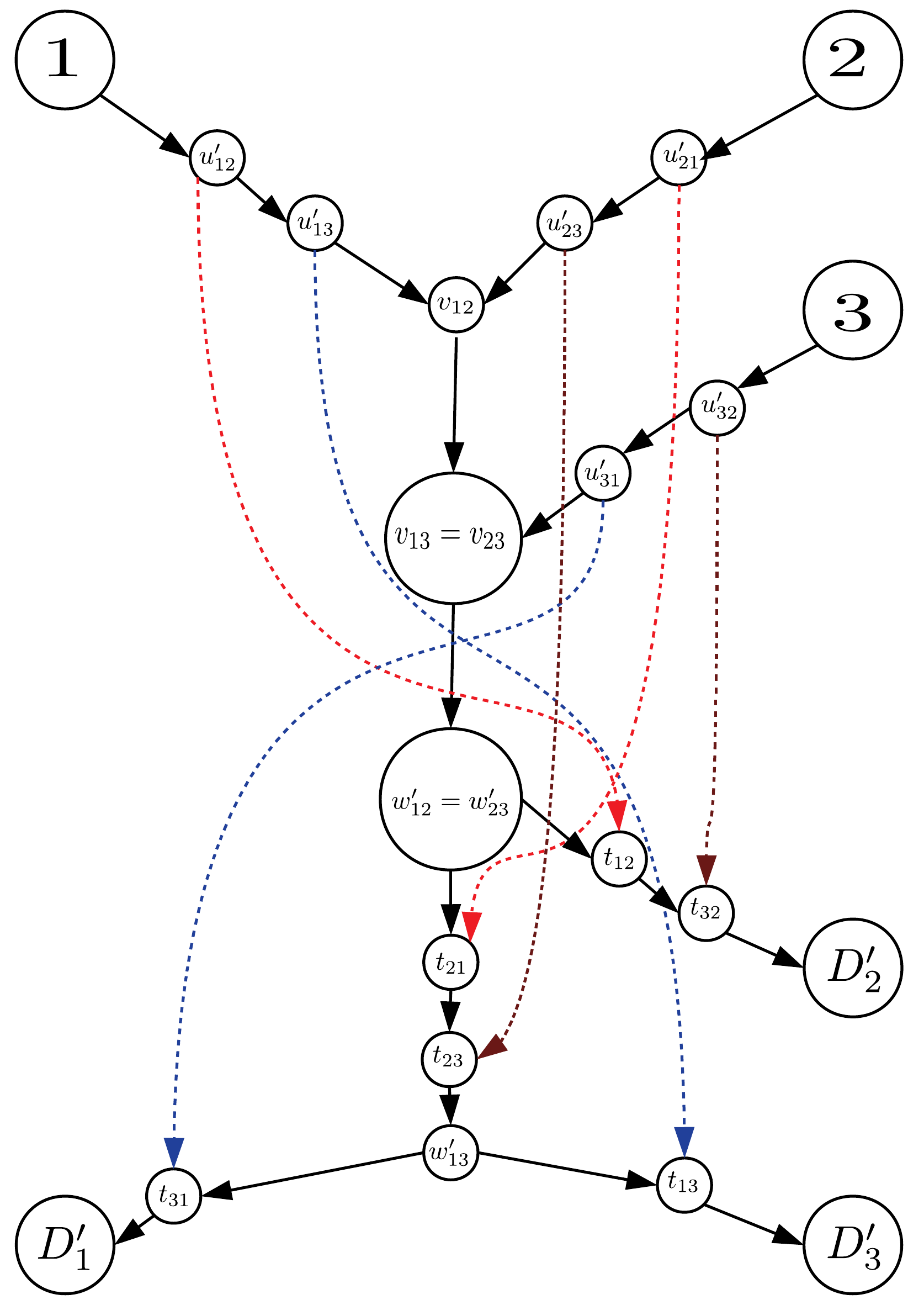}}
  \hspace{0.0in}
  \subfigure[Network style B original Config  8]{\label{network style A-9}\includegraphics[height=2.0in]{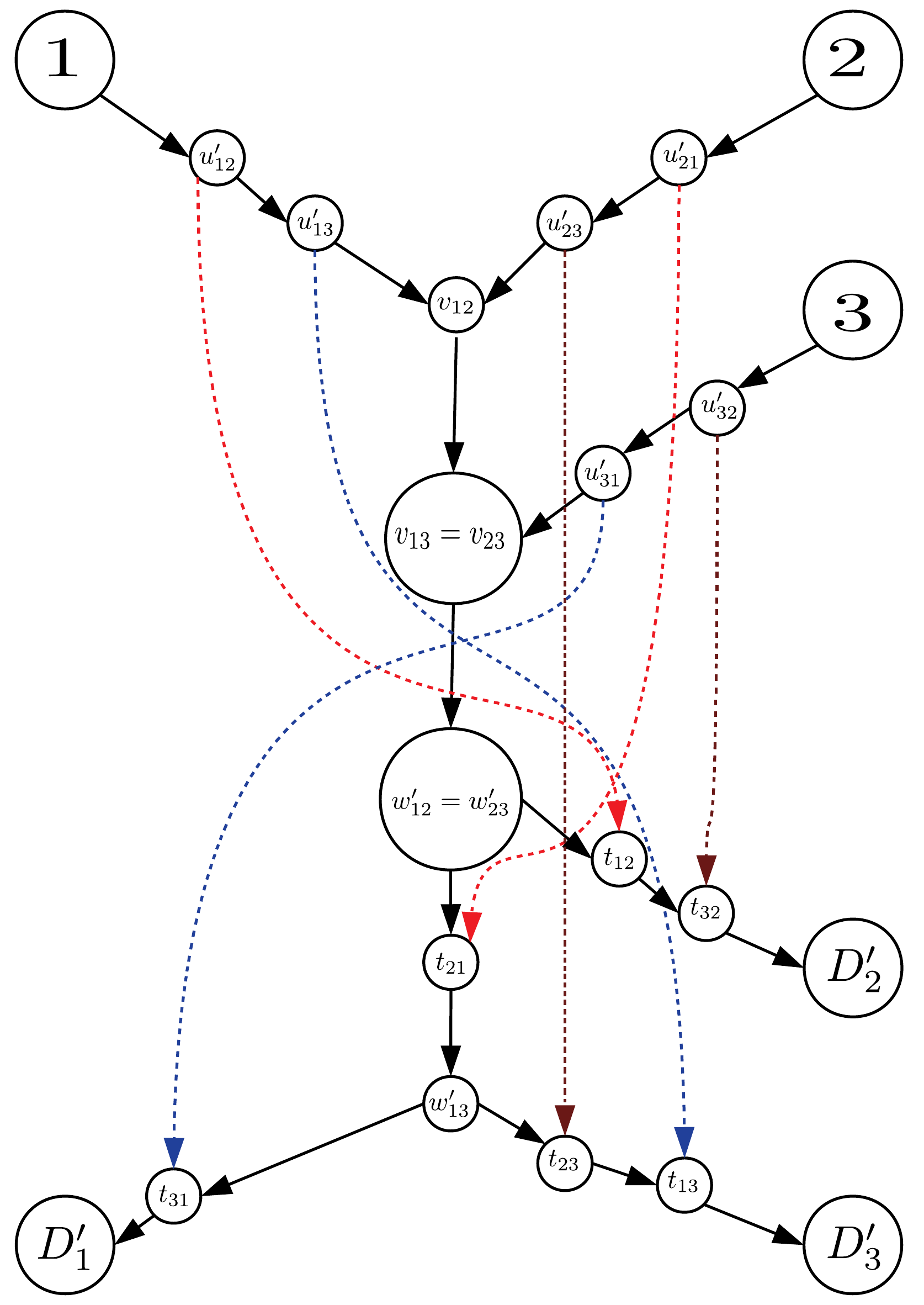}}
  \hspace{0.0in}
  \subfigure[Network style B original Config  9]{\label{network style A-9}\includegraphics[height=2.0in]{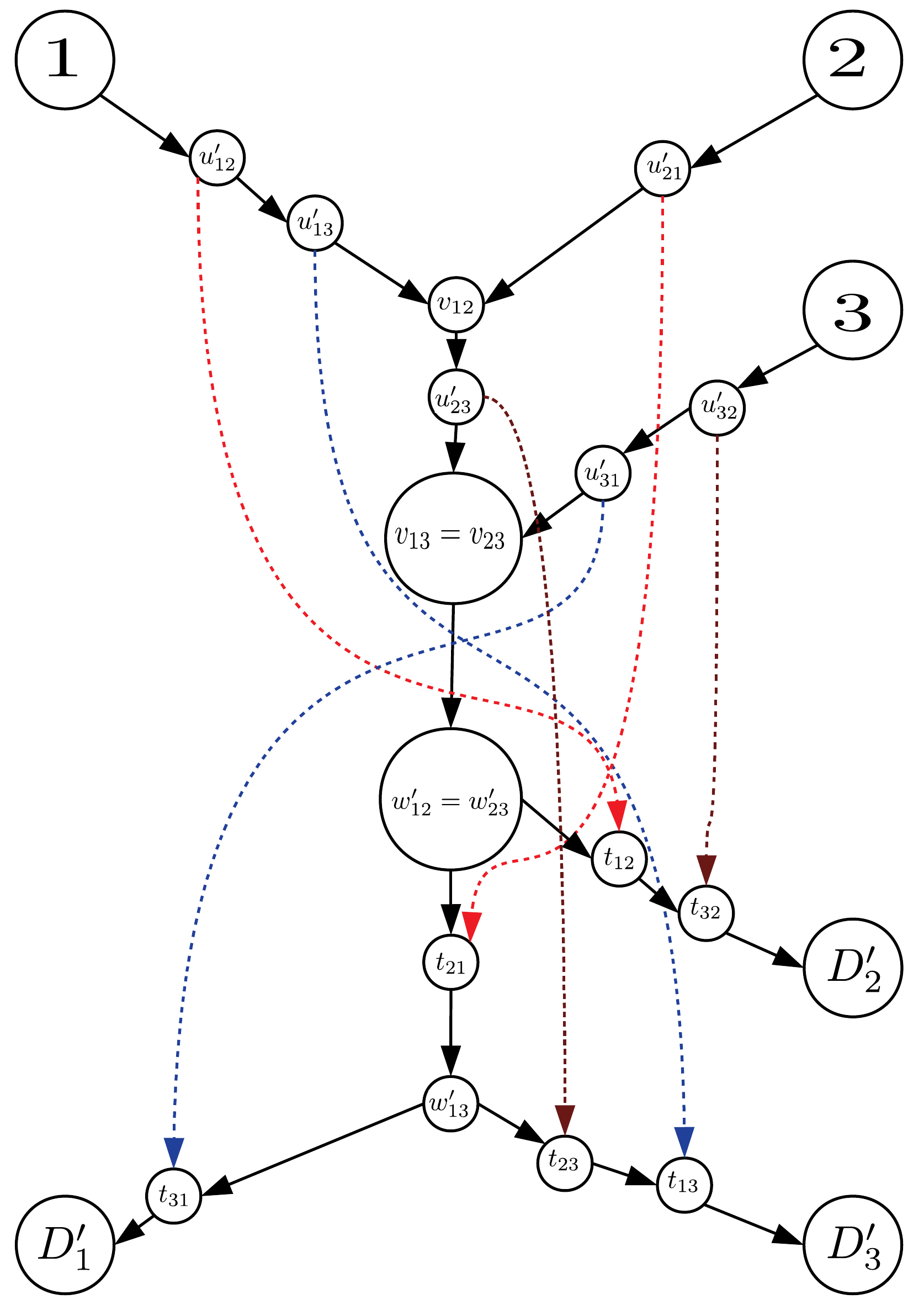}}
  \hspace{0.0in}
  
  \vspace{0in}
  \hrule
  \subfigure[Network style B stage I reduced Config  S11 for above Config  1]{\includegraphics[height=2.5in]{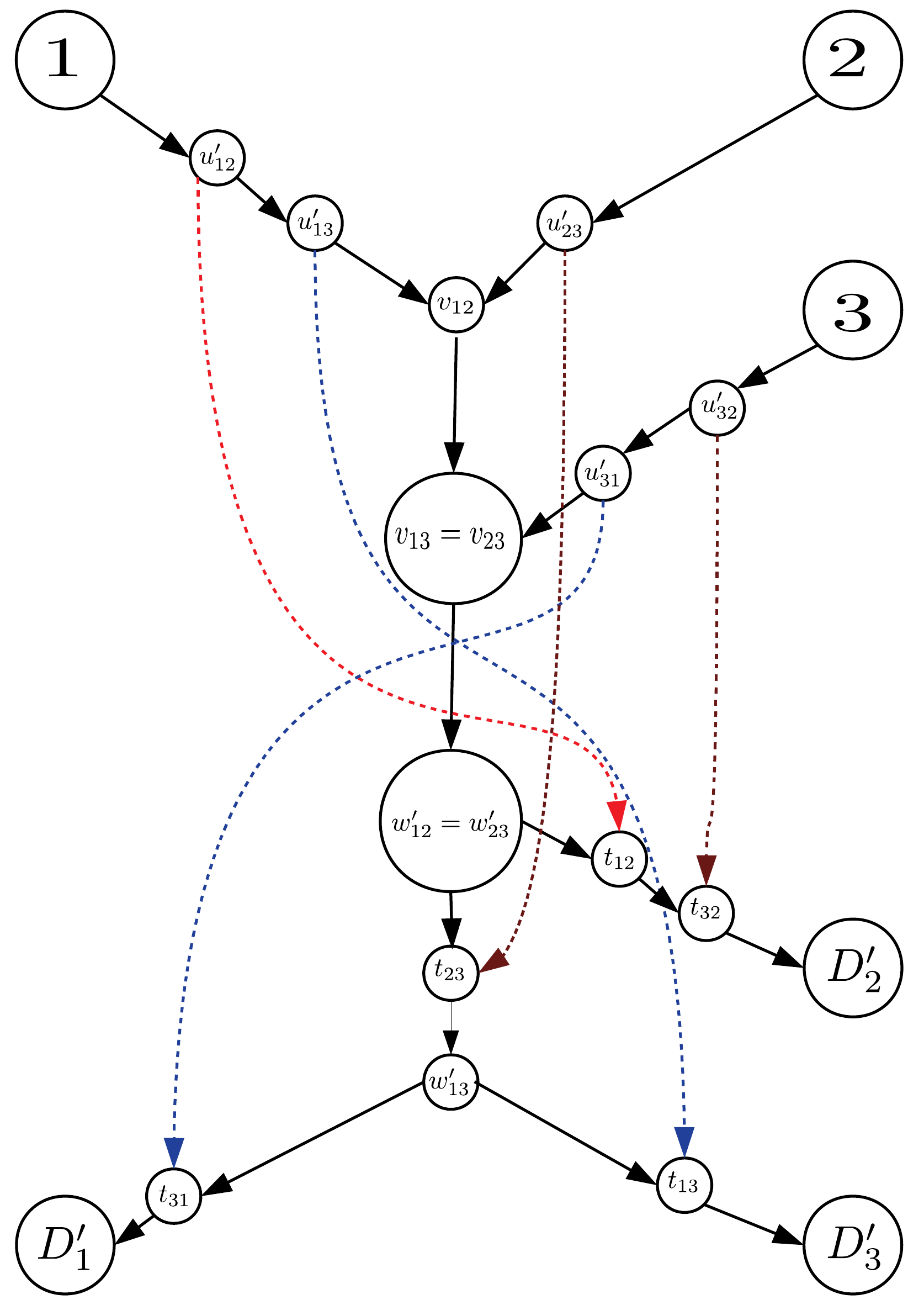}
  \label{network style 1A-3}}
 \hspace{0.0in}
 \subfigure[Network style B stage I reduced Config  S15 for above Config  7,8,9 ]{\label{network style 1A-9}\includegraphics[height=2.5in]{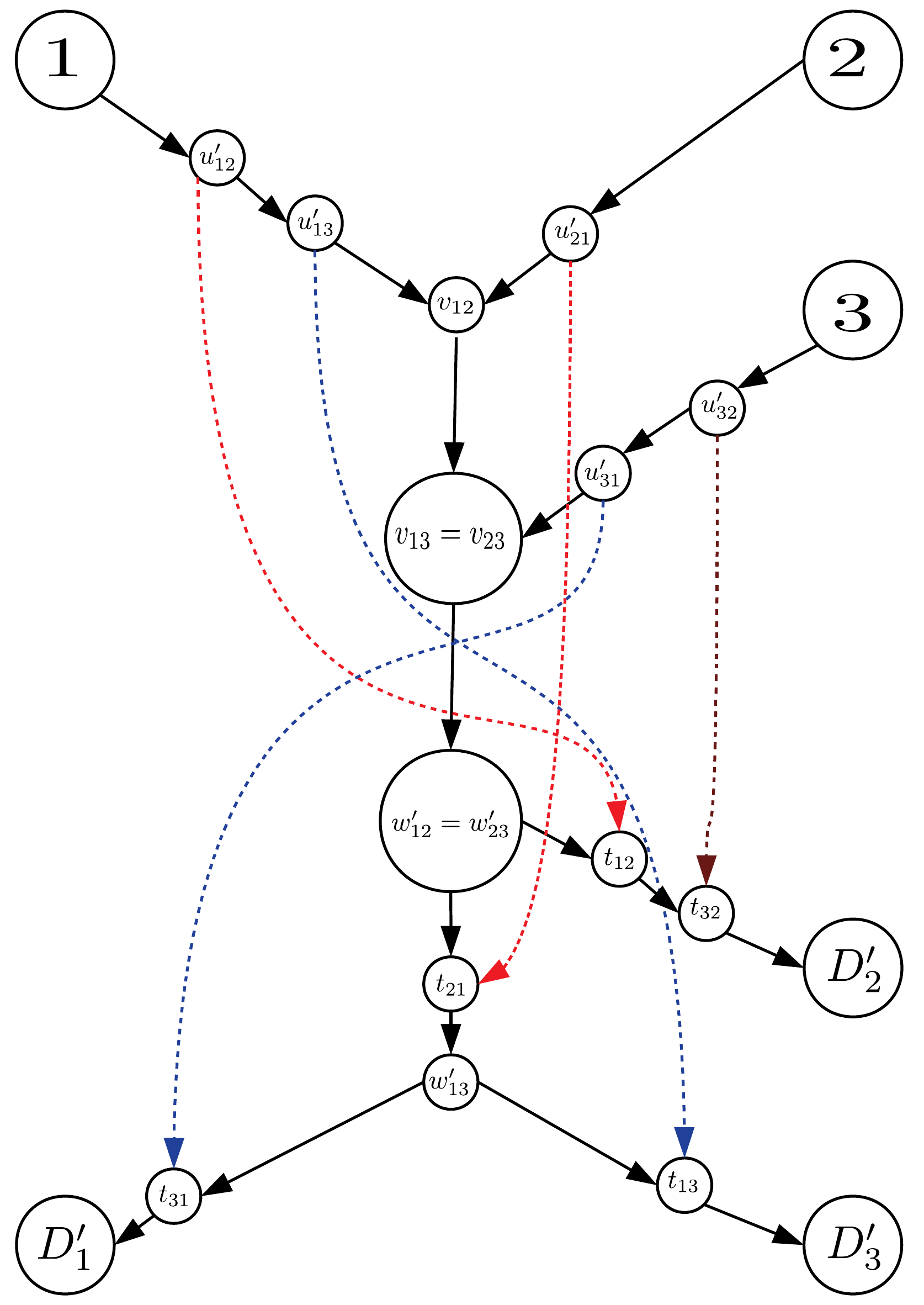}}
   \hspace{0.0in}
   \hrule
   \subfigure[Network style B stage II reduced Config  S21 of stage I Config  S11 and S15]{\label{network style 2A-3}\includegraphics[height=3.2in]{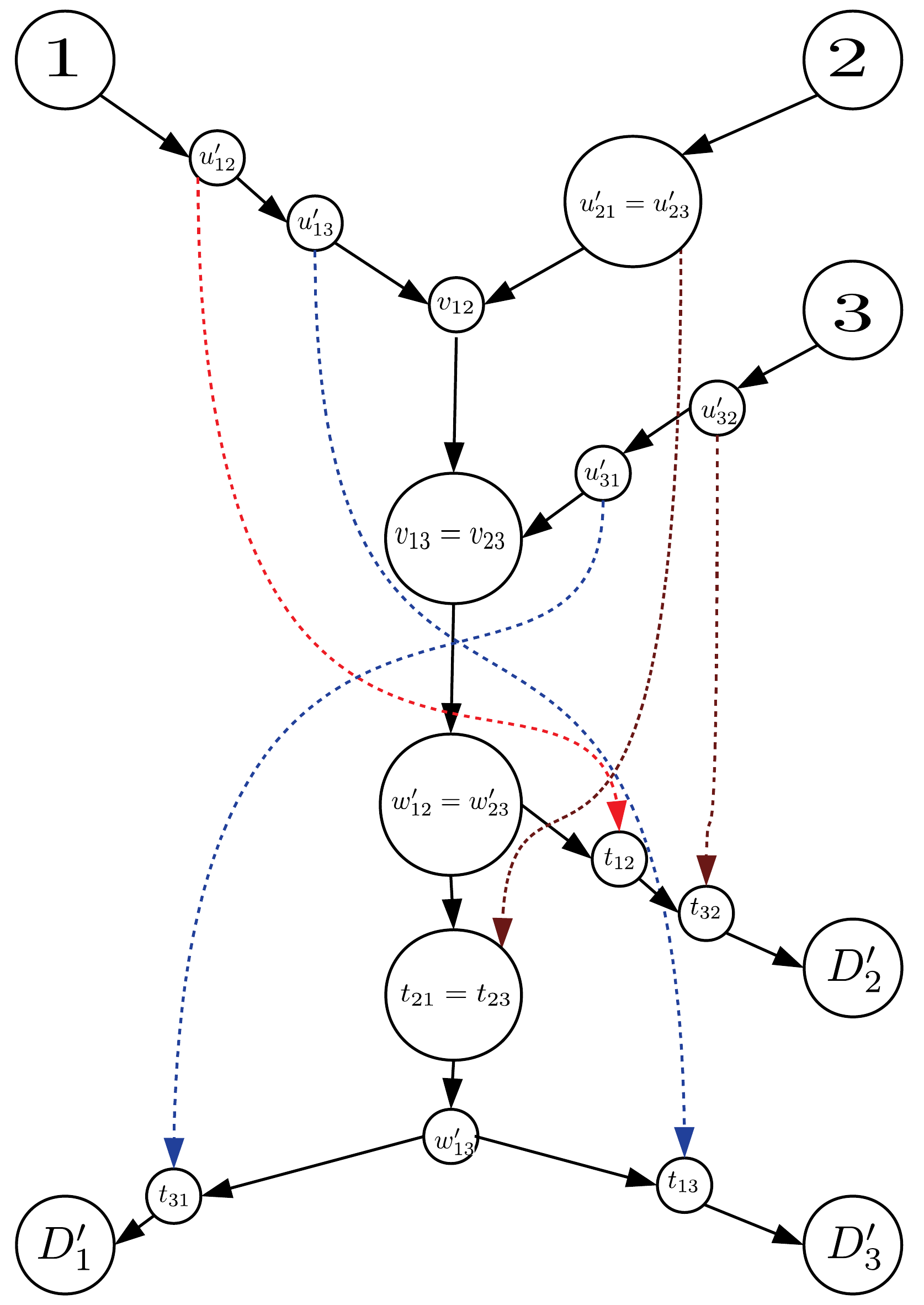}}
  \hspace{0.0in}
   \caption{}
  \end{figure*}
  
  \begin{figure*}[htbp]
  \centering
 \subfigure[Network style B original Config 2 same for reduced Config  S12 and S22]{\label{network style 2A-3}\includegraphics[height=3.2in]{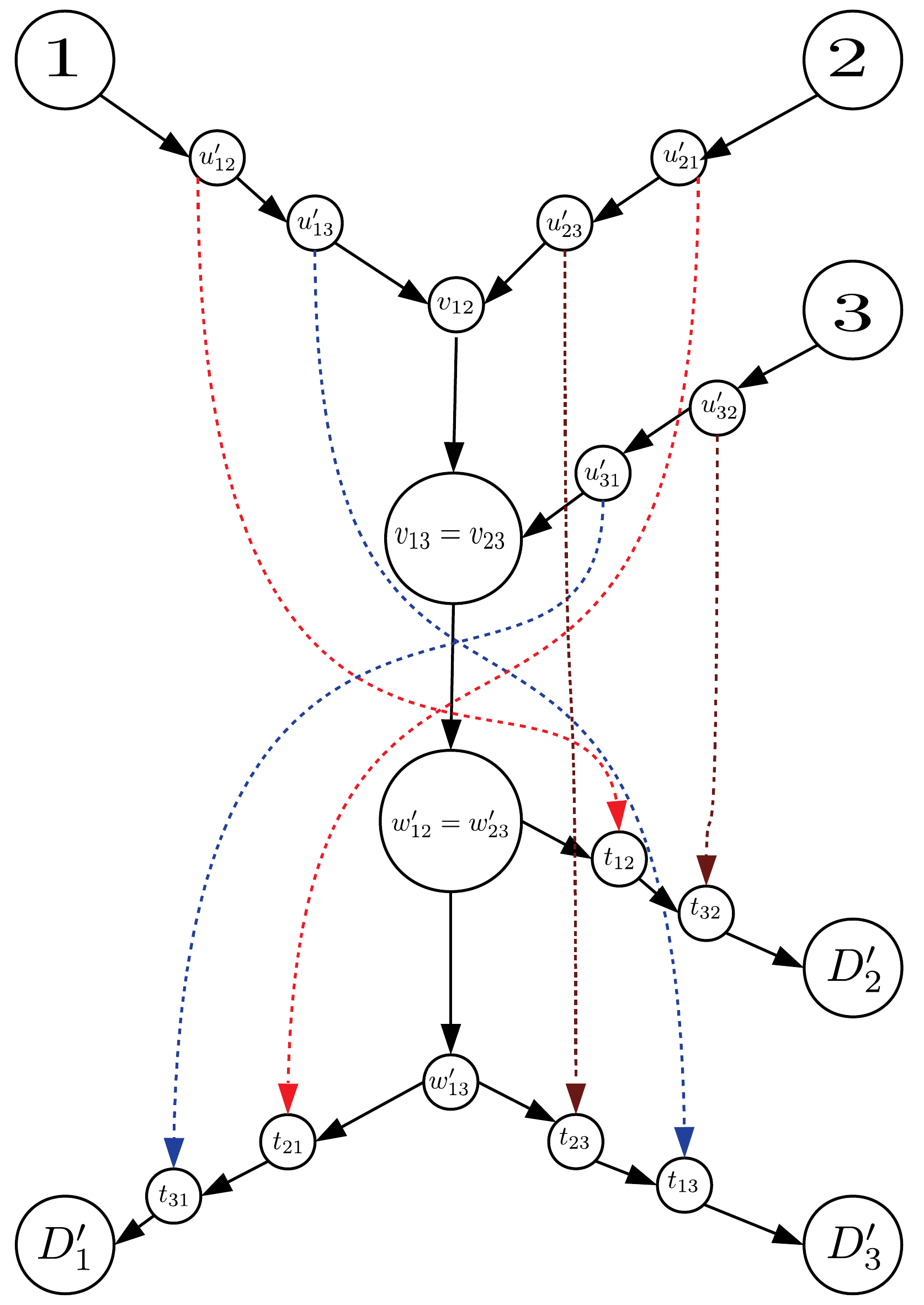}}
  \hspace{0.0in}
   \caption{}
  \end{figure*}
  
   \begin{figure*}[htbp]
 \centering
  \hspace{0.0in}
  \subfigure[Network style B original Config  3]{\label{network style A-3}\includegraphics[height=2in]{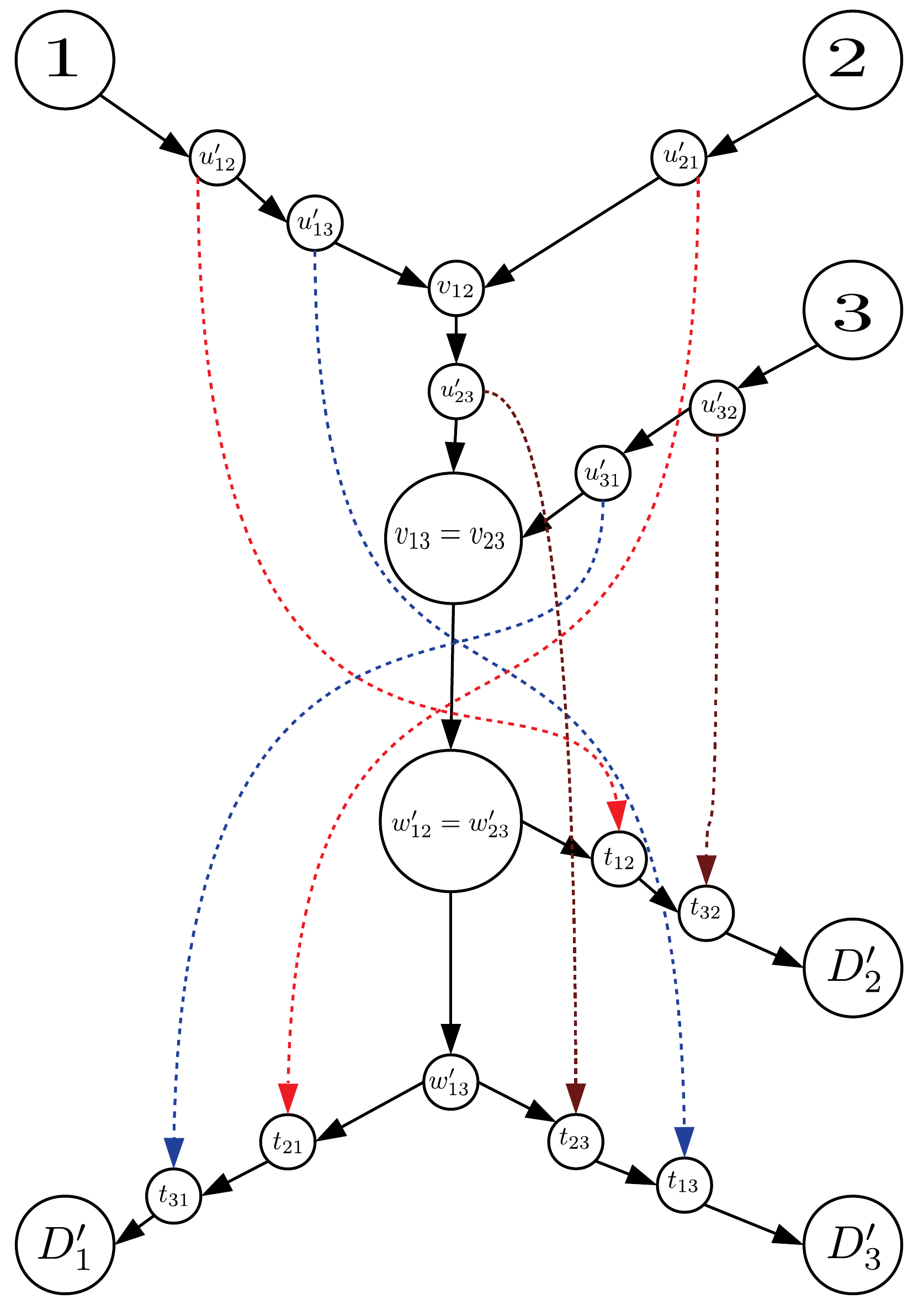}} \hspace{0.0in}
  \subfigure[Network style B original Config  4]{\label{network style A-11}\includegraphics[height=2in]{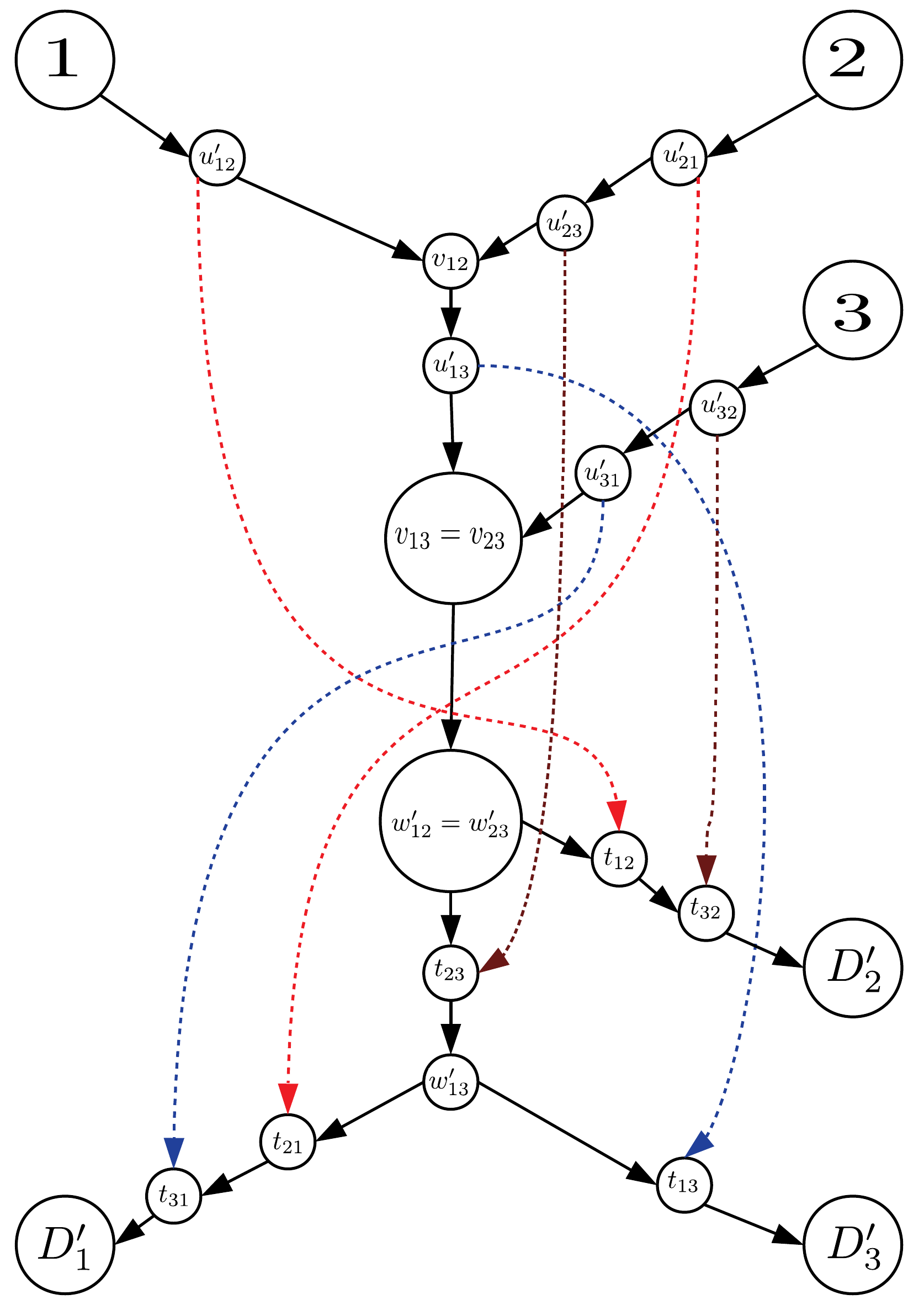}}
  \hspace{0.0in}
  \subfigure[Network style B original Config  5]{\label{network style A-9}\includegraphics[height=2in]{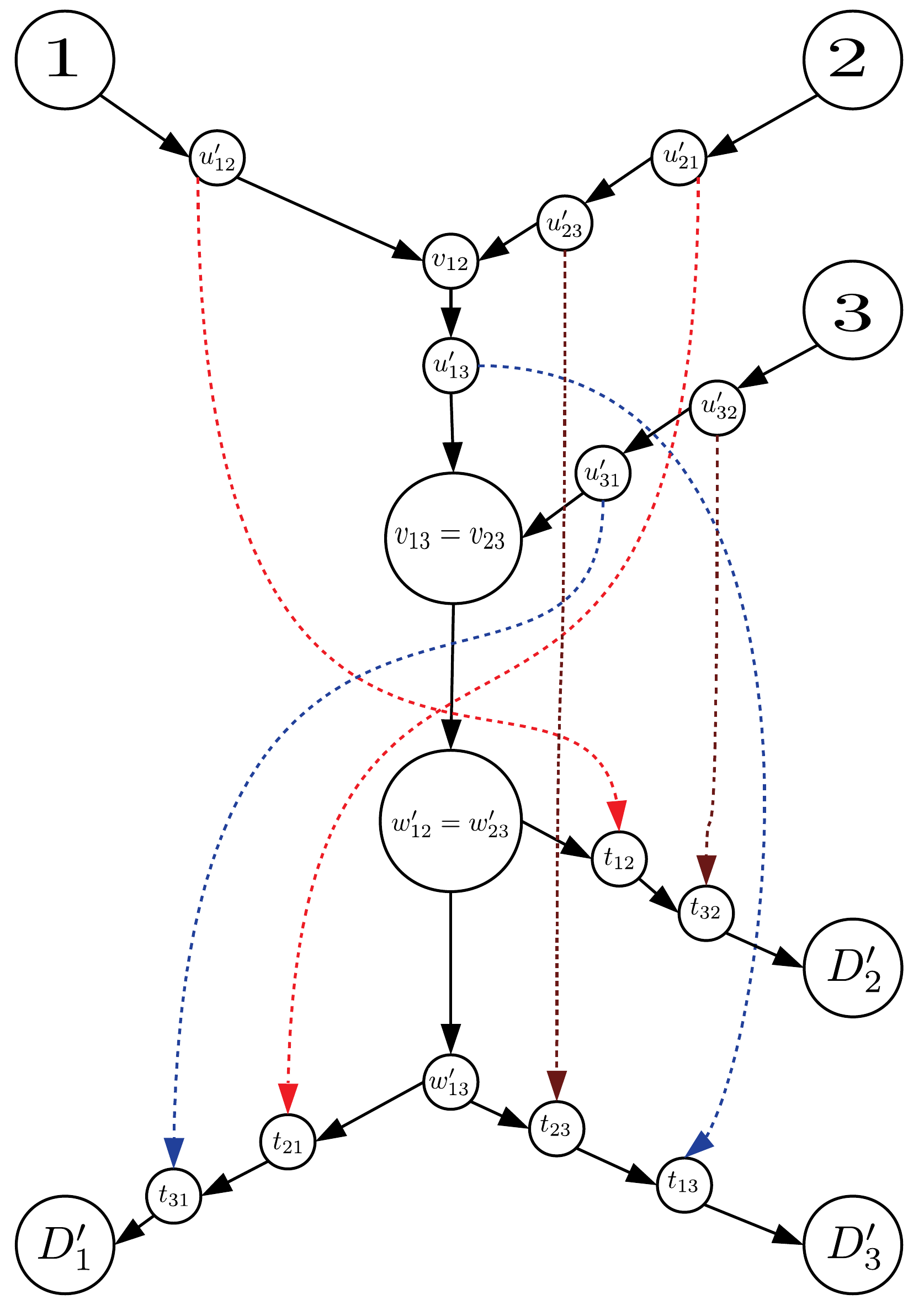}}
  \hspace{0.0in}
  \subfigure[Network style B original Config  6]{\label{network style A-9}\includegraphics[height=2in]{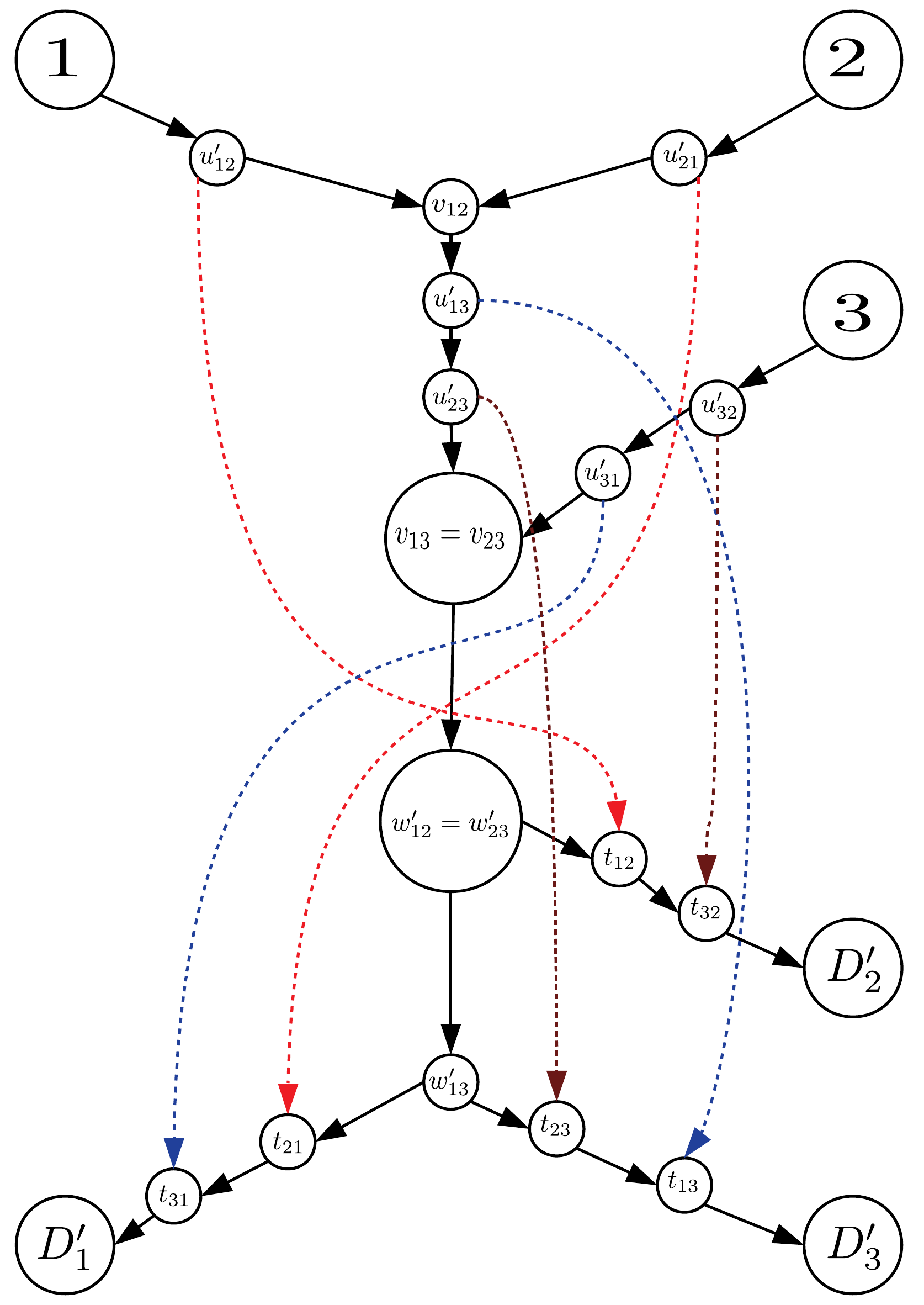}}
  \hspace{0.0in}
  \vspace{0in}
  \hrule
  \subfigure[Network style B stage I reduced Config  S13 for above Config 3]{\includegraphics[height=2.5in]{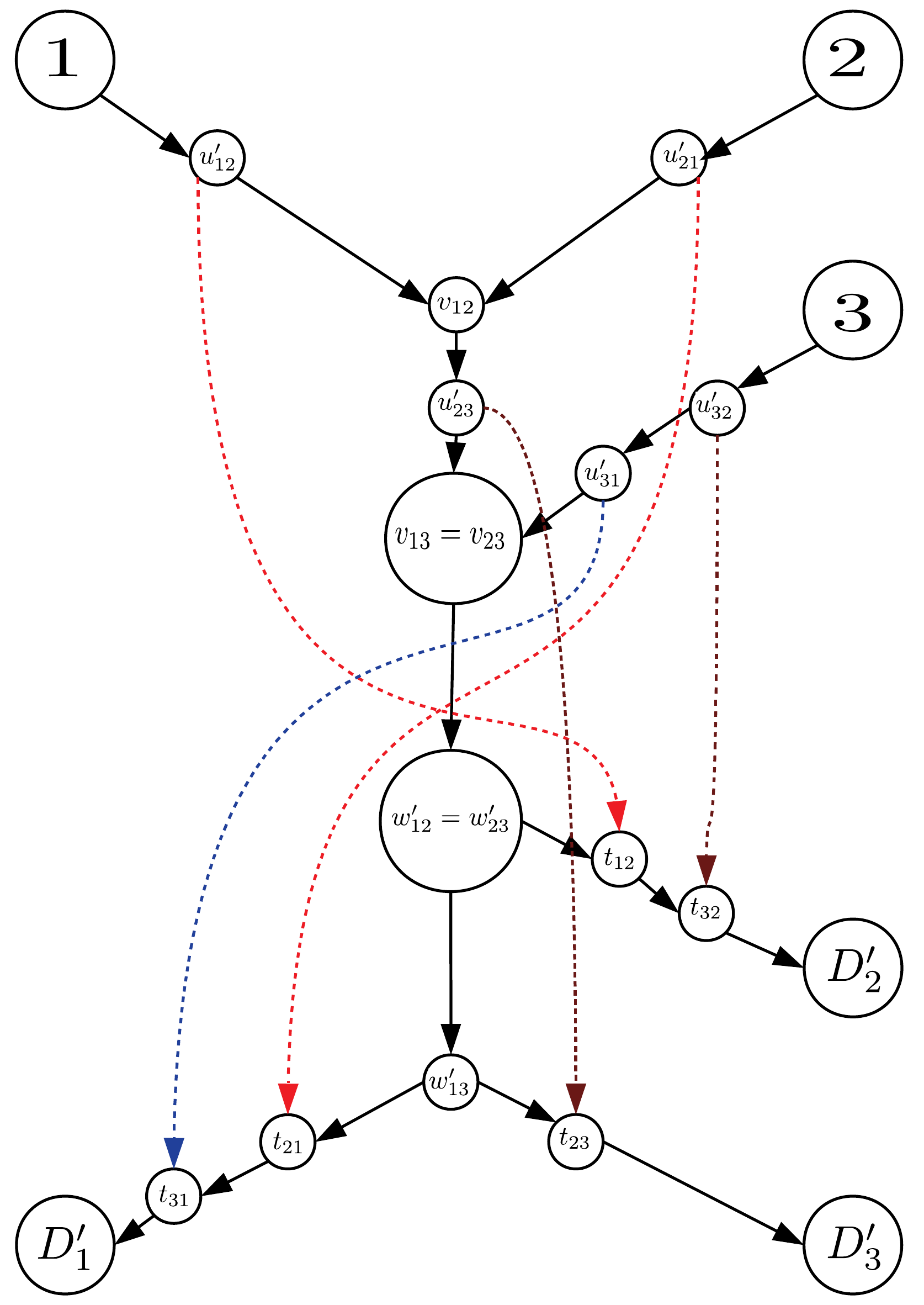}
  \label{network style 1A-3}}
 \hspace{0.0in}
 \subfigure[Network style B stage I reduced Config  S14 for above Config  4,5 and 6 ]{\label{network style 1A-9}\includegraphics[height=2.5in]{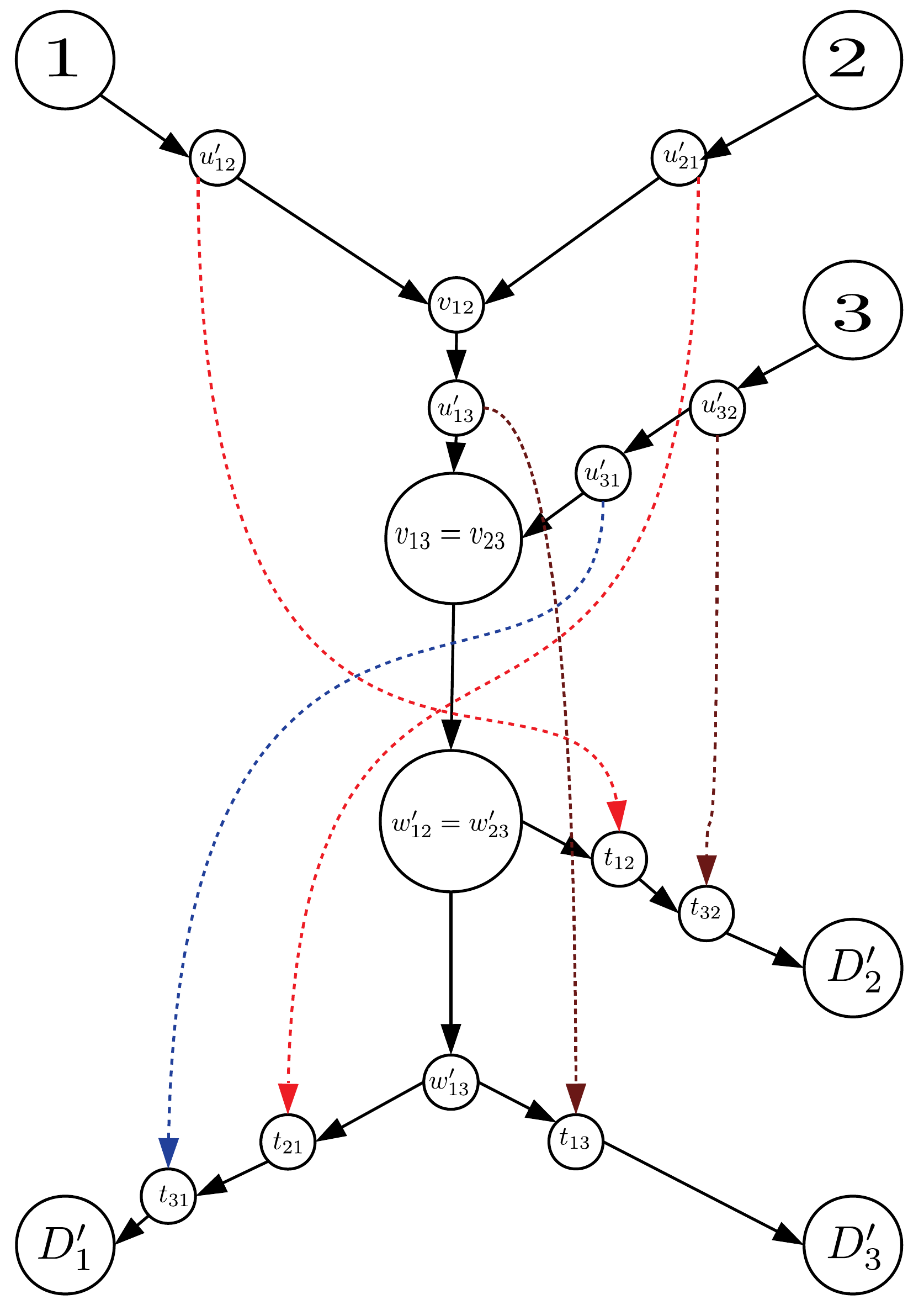}}
  \hrule
  \subfigure[Network style B stage II reduced Config  S21 of stage I Config  S13 and S14]{\label{network style 2A-3}\includegraphics[height=3.2in]{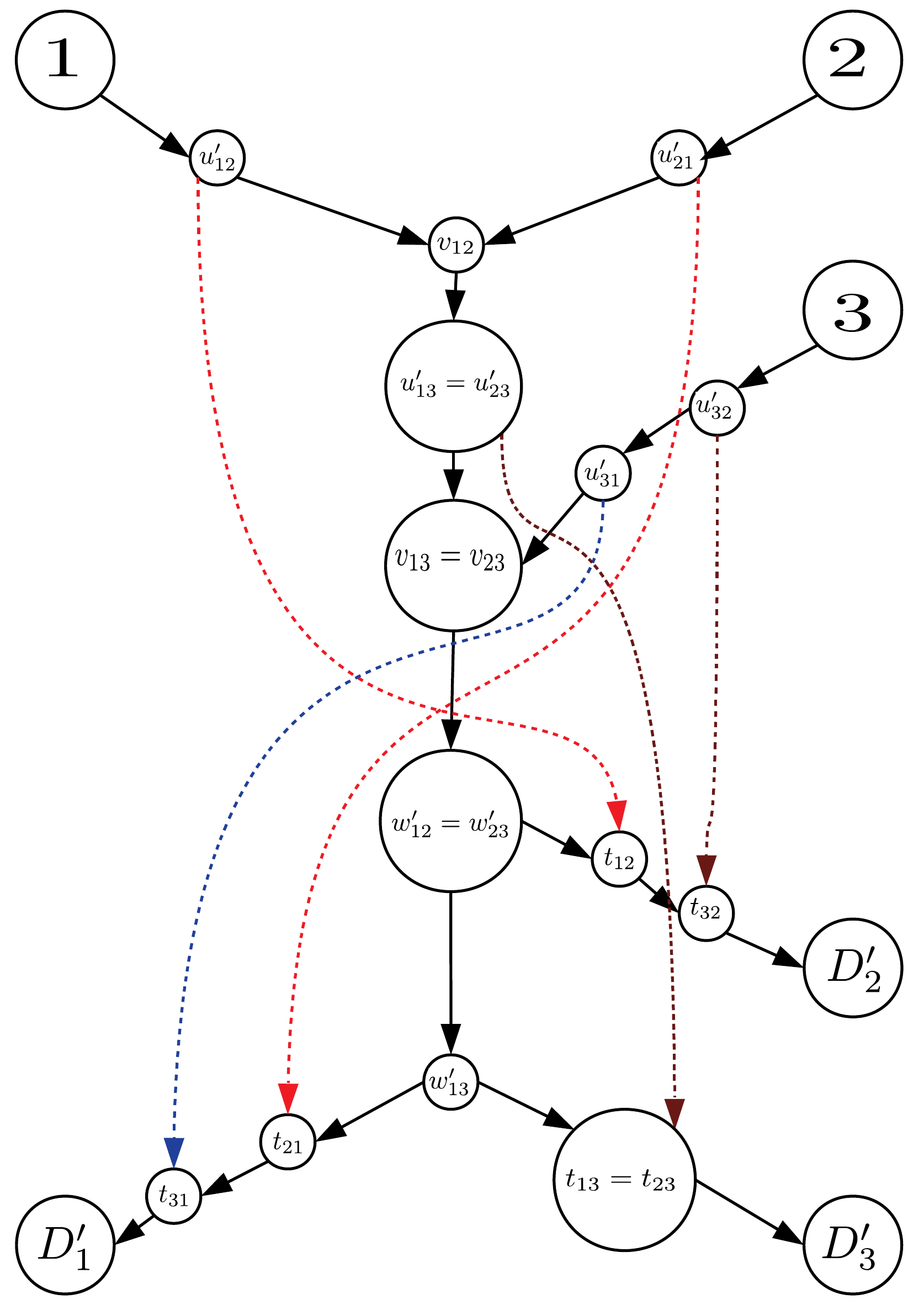}}
  \hspace{0.0in}
   \caption{}
  \end{figure*}
  \begin{figure*}[htbp]
  \centering
 \subfigure[Network style B original Config 11 same for reduced Config  S16 and S24]{\label{network style 2A-3}\includegraphics[height=3.2in]{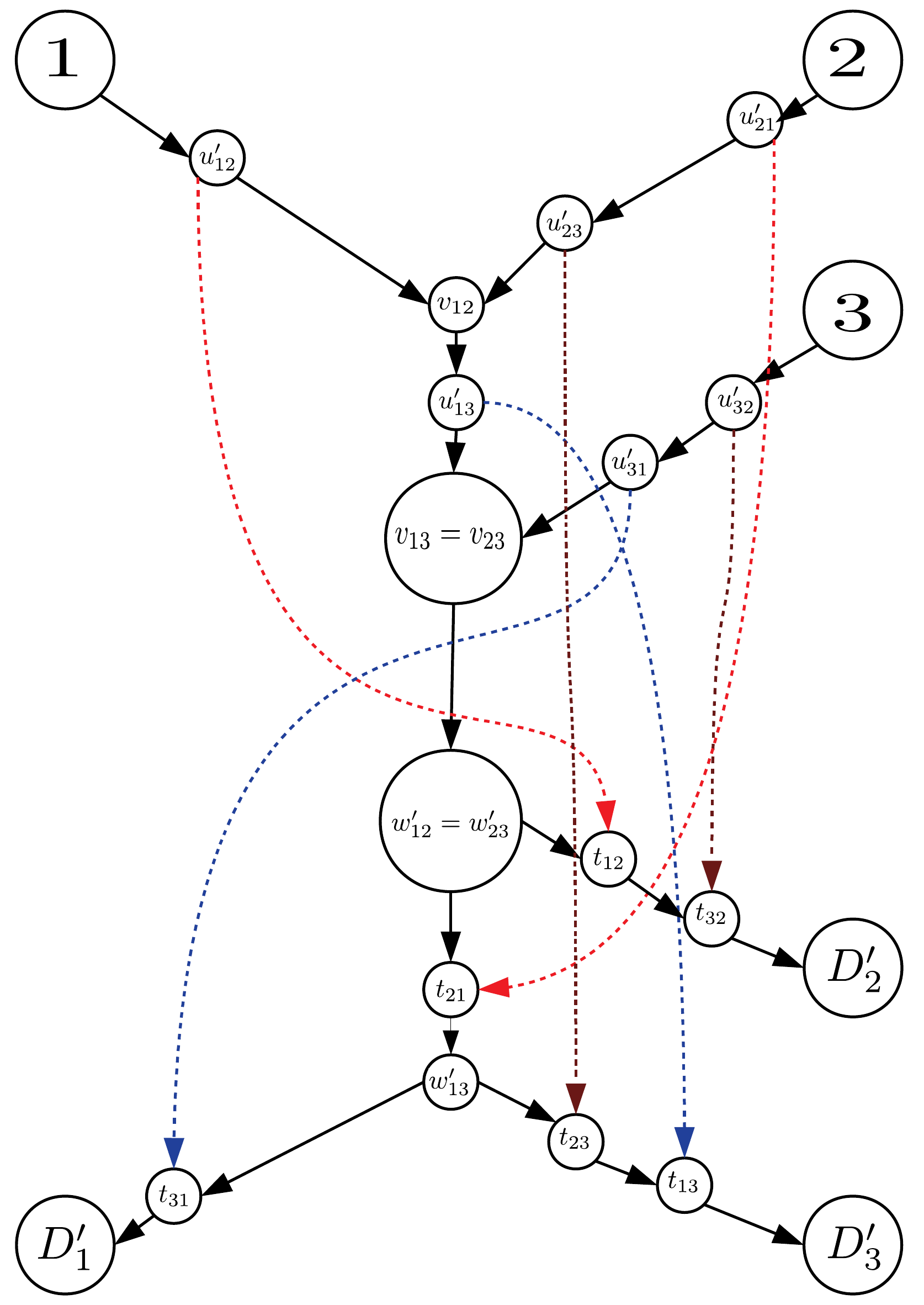}}
  \hspace{0.0in}
   \caption{}
  \end{figure*}

\end{document}